\documentclass{lmcs}

\usepackage[utf8]{inputenc}
\usepackage{amsmath,
	    amsfonts,
	    amstext,
	    amssymb,
	    mathrsfs,
	    mathpartir,
	    stmaryrd,
	    latexsym,
	    enumitem,
	    proof,
	    graphicx,
	    xcolor,
	    hyperref,
	    comment}

\usetikzlibrary{arrows,positioning,arrows.meta}
\tikzset{modal/.style = {>= stealth', shorten >= 0pt, shorten <= 0pt, auto,
			 node distance = 1cm, semithick}, 
	 point/.style = {circle, draw, fill = black, inner sep = 0.5mm}}

\newcommand{\ms}[1]
	{\null\ifmmode\mathord{\mathcode`-="702D\it #1\mathcode`\-="2200}
	\else$\mathord{\mathcode`-="702D\it #1\mathcode`\-="2200}$\fi}

\newcommand{\cws}[2]
	{\\ \centerline{$#2$} \\[-#1pt]}

\newcommand{\fullbox}
	{{\mbox{}\nolinebreak\hfill{$\rule{2.0mm}{2.0mm}$}}}

\newcommand{\bibtrick}[1]
	{}

\newcommand{\blue}[1]
	{{\color{blue}#1}}

\newcommand{\lsp}
	{\llbracket}
\newcommand{\rsp}
	{\rrbracket}

\newcommand{\lmp}
	{\{ \! | \,}
\newcommand{\rmp}
	{\, | \! \}}

\newcommand{\ssfc}
	{\textsf{C}}

\newcommand{\ssfe}
	{\textsf{E}}

\newcommand{\ssff}
	{\textsf{F}}

\newcommand{\ssfn}
	{\textsf{N}}

\newcommand{\cala}
        {\mathcal{A}}

\newcommand{\calb}
        {\mathcal{B}}

\newcommand{\calc}
        {\mathcal{C}}

\newcommand{\cale}
        {\mathcal{E}}

\newcommand{\cali}
        {\mathcal{I}}

\newcommand{\call}
        {\mathcal{L}}

\newcommand{\calp}
        {\mathcal{P}}

\newcommand{\calt}
        {\mathcal{T}}

\newcommand{\natns}
	{\mathbb{N}}

\newcommand{\powerset}
        {\mathscr{P}}

\newcommand{\procs}
	{\mathbb{P}}

\newcommand{\zip}
	{\textit{zip}}

\newcommand{\initial}
	{\textit{init}}

\newcommand{\wf}
	{\textit{wf}}

\newcommand{\toinitial}
	{\textit{to\_init}}

\newcommand{\enr}
	{\textit{enr}}

\newcommand{\brm}
	{\textit{brm}}

\newcommand{\ctx}
	{\textit{ctx}}

\newcommand{\xarrow}[2]
        {\, {\xrightarrow{#1}}_{#2} \,}

\newcommand{\arrow}[2]
        {\, {\auxarrow\limits^{#1}}_{#2} \,}
\newcommand{\auxarrow}
	{\mathop{\longrightarrow}}

\newcommand{\wauxarrow}
	{\mathop{\Longrightarrow}}

\newcommand{\nil}
	{\underline 0}

\newcommand{\parfun}
	{\rightharpoonup}

\newcommand{\restr}
	{\upharpoonright}

\newcommand{\sbis}[1]
	{\sim_{#1}}

\newcommand{\pco}[1]
	{\mathop{\Vert_{#1}}}

\newcommand{\lpar}
	{\rrfloor}

\newcommand{\rpar}
	{\llfloor}

\newcommand{\lplu}
	{\mathop{. \!\! +} \!}

\newcommand{\rplu}
	{\mathop{+ \!\! .} \hspace{-0.04cm}}

\newcommand{\true}
	{\mathsf{true}}

\newcommand{\diam}[2]
	{\langle #1 \rangle_{#2}}

\newcommand{\eildiam}[2]
	{\langle #1 \rangle \! \rangle_{#2}}

\newcommand{\eilbdiam}[2]
	{\langle \! \langle #1 \rangle_{#2}}

\newcommand{\id}
	{\cali}

\newcommand{\fid}
	{\textit{fi}}

\newcommand{\perm}
	{\textit{pe}}

\newcommand{\sbpr}
	{\textit{sp}}

\newcommand{\apt}
	{\textit{apt}}

\newcommand{\depth}
	{\textit{depth}}

\newcommand{\dom}
	{\textit{dom}}

\newcommand{\rge}
	{\textit{rge}}

\newcommand{\proj}
	{\textit{proj}}

\newcommand{\act}
	{\textit{act}}

\newcommand{\supp}
	{\textit{supp}}

\newcommand{\scs}
	{\textit{scs}}

\begin{document}

\title[Characterizations of Hereditary History-Preserving Bisimilarity]
      {Hereditary History-Preserving Bisimilarity: \\
       Characterizations via Backward Ready Multisets}

\author[M.~Bernardo]{Marco Bernardo\lmcsorcid{0000-0003-0267-6170}}[a]
\author[A.~Esposito]{Andrea Esposito\lmcsorcid{0009-0009-2259-902X}}[a]
\author[C.A.~Mezzina]{Claudio A.\ Mezzina\lmcsorcid{0000-0003-1556-2623}}[b]

\address{Dipartimento di Scienze Pure e Applicate, Universit\`a di Urbino, Italy}
\email{marco.bernardo@uniurb.it, andrea.esposito@uniurb.it}

\address{Dipartimento di Informatica, Universit\`a di Bari, Italy}
\email{claudio.mezzina@uniba.it}



\begin{abstract}
We devise two complementary characterizations of hereditary history-preserving bisimilarity (HHPB): a
denotational one, over stable configuration structures, and an operational one, in a reversible process
calculus. Our characterizations rely on forward-reverse bisimilarity \blue{(FRB)} augmented with backward
ready multiset equality, \blue{which is able to distinguish between autoconcurrency and autocausation}.
\blue{With respect to previous work,} we thus shift the emphasis from uniquely identifying events, as done
in other characterizations, to counting occurrences of identically labeled events associated with incoming
transitions. This yields a more lightweight behavioral equivalence than HHPB: \blue{our bisimulations are
still binary relations instead of ternary ones and do not need to incrementally build any labeling- and
causality-preserving bijection for occurred events}. We show that our characterizations are valid \blue{in
the absence of non-local autoconflicts}. We then study the logical foundations of these characterizations by
relating event identifier logic, which captures HHPB, and backward ready multiset logic, developed for our
\blue{extension of FRB}.
\end{abstract}

\maketitle

%
%
\section{Introduction}
\label{sec:intro}
%
%

Understanding and comparing the behavior of concurrent systems is a central problem in semantics and
verification. Since such systems may exhibit nondeterminism, causality, and concurrency in complex ways, it
is not enough to describe what actions they can perform: one must also capture how these actions relate --
whether they are causally dependent, mutually exclusive, or completely independent. To reason about
concurrent systems, bisimulation has emerged as the canonical tool. Bisimilar systems cannot be
distinguished by any experiment that proceeds step by step: whatever one system can do, the other can match,
with the resulting states remaining equivalent. This makes bisimulation a cornerstone of process
equivalence, compositional reasoning, and congruence-based program verification. However, classical
bisimulation -- such as strong bisimilarity~\cite{Par81,Mil89a} -- relies on interleaving semantics:
concurrency is represented as nondeterministic interleaving of actions~\cite{HM85} over labeled transition
systems~\cite{Kel76}. While sufficient for analyzing communication protocols or sequential behaviors,
interleaving semantics combines independence with nondeterminism and cannot distinguish between
fundamentally different concurrent structures. For instance, a process that performs two independent actions
concurrently becomes indistinguishable from a process that performs the same actions sequentially in either
order. Such a loss of information affects verification tasks, refinement reasoning, and expressiveness
results.

This limitation has motivated the development of true concurrency models -- such as Petri nets~\cite{Pet62},
event structures~\cite{Win86}, and reversible variants of process calculi~\cite{DK04,PU07a} -- where
causality, conflict, and concurrency are modeled explicitly. Within these models, bisimulation must be
refined to account for a richer structure: it~is required to match not only the performed actions, but also
the causal histories leading to those actions. This generates finer equivalences constituting the spectrum
of truly concurrent bisimilarities~\cite{GG01,Fec04,PU12}, which includes history-preserving bisimilarity
(HPB)~\cite{RT88} and hereditary history-preserving bisimilarity (HHPB)~\cite{Bed91}. They are particularly
important because they turn out to be the coarsest and the finest equivalences, respectively, that are
preserved under action refinement and are capable of respecting causality, branching, and their interplay
while abstracting from choices between identical alternatives~\cite{GG01}; moreover, HHPB can be obtained as
a special case of a categorical definition of bisimilarity over concurrency models~\cite{JNW96}.  These
equivalences ensure that two systems behave the same not only in terms of which actions they can perform,
but also in terms of how these actions arise from concurrent or dependent pasts. They are crucial in
applications where dependency information matters like, e.g., reversible computing, distributed memory
models, causality-aware verification, and concurrency theory at large.

HPB and HHPB are defined over event structures~\cite{Win86} or their variants akin to labeled transition
systems, which are called configuration structures~\cite{GP09}. A configuration is a finite set of
non-conflicting events, which contains all the events causing the ones in the configuration itself. The
bisimulation game compares transitions between configurations. While HPB considers only outgoing
transitions, HHPB takes into account also incoming transitions. In other words, the former stepwise matches
only forward computations, whereas the latter examines backward computations too. Both equivalences rely on
ternary bisimulation relations, where the third component is a labeling- and causality-preserving bijection
between the sets of events that occurred so far in the two structures, with the bijection being built
incrementally during the bisimulation game. This makes HPB and HHPB hard to reason about and mechanize.

Logical characterizations of both equivalences have been provided in~\cite{PU14,BC14}. Furthermore, an
axiomatization for HHPB has been developed over forward-only processes in~\cite{FL05}. Finally, HPB is known
to coincide with causal bisimilarity~\cite{DD89,DD90}, hence the latter offers a characterization, as well
as an axiomatization via~\cite{DP92}, for the former. In this paper, we concentrate on characterizing HHPB,
\blue{denotationally and operationally}.

%
\subsection{\blue{Previous Work}}
\label{sec:previous_work}
%

The first alternative characterization of HHPB appeared in~\cite{Bed91} for configuration graphs of prime
event structures. The characterizing equivalence is called back-and-forth bisimilarity -- not to be confused
with the homonymous one in~\cite{DMV90}, which retrieves an interleaving semantics by constraining backward
computations to take place along the corresponding forward computations even in the presence of concurrency.
The main difference between HHPB and back-and-forth bisimilarity is that the latter relies on binary
bisimulation relations as it does not \blue{need to} stepwise build any labeling- and causality-preserving
bijection during the bisimulation game. The characterization result holds under the assumption of no
autoconcurrency, i.e., the absence of configurations from which it is possible to execute two identically
labeled, distinct events that are not in conflict with each other.

	\begin{figure}[t]

\centerline{\includegraphics{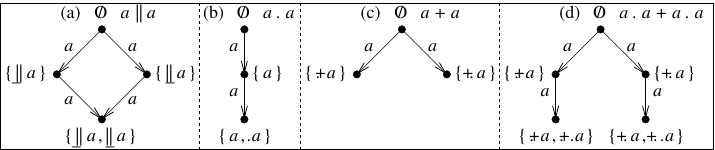}}
\caption{Autoconcurrency (a), autocausation (b, d), and autoconflict (c, d)}
\label{fig:auto_conc_caus_confl}

	\end{figure}

In Figures~\ref{fig:auto_conc_caus_confl}(a) and~(b) we show the configuration graphs respectively
associated with the following two processes for a given action $a$:

	\begin{itemize}

\item Autoconcurrency on $a$, which we express as $a \pco{} a$ where $\pco{}$ stands for parallel
composition. There are two equally labeled, non-conflicting events, which we denote by $\lpar a$ and $\rpar
a$, that can be executed in any order.

\item Autocausation on $a$, which we express as $a \, . \, a$ where dot represents action prefix. There are
two equally labeled, non-conflicting events, which we denote by $a$ and $. a$, such that the former has to
be executed before the latter.

	\end{itemize}

\noindent
These two configuration graphs are back-and-forth bisimilar as witnessed by the symmetric binary relation
that contains the pairs of configurations $(\emptyset, \emptyset)$,  $(\{ \lpar a \}, \{ a \})$,
$(\{ \rpar a \}, \{ a \})$, and $(\{ \lpar a, \rpar a \}, \{ a, . a \})$. However, they are not hereditary
history-preserving bisimilar because, with respect to the last pair, there is no (labeling- and)
causality-preserving bijection that maps the two independent events $\lpar a$ and $\rpar a$ to the two
causally-related events $a$ and $. a$.

The second alternative characterization of HHPB was given in~\cite{PU07b}. The characterizing equivalence is
the forward-reverse bisimilarity -- very close in spirit to the back-and-forth bisimilarity of~\cite{Bed91}
-- originally defined in~\cite{PU07a} for a reversible variant of CCS~\cite{Mil89a} called CCSK. The
operational semantics of CCSK produces labeled transition systems based on a forward transition relation and
a backward one ensuring the loop property~\cite{DK04}. Each transition label comprises an action and a
communication key; the latter is necessary when building backward transitions so as to know who synchronized
with whom in the forward direction. In~\cite{PU07b} forward-reverse bisimilarity was generalized to
configuration graphs of prime event structures and shown to coincide with HHPB in the absence of repeated,
identically labeled events along forward computations, which implies the absence of autoconcurrency (and
autocausation), i.e., the assumption made in~\cite{Bed91}.

In~\cite{PU12} it was shown, by working on stable configuration structures, how to relax the conditions
under which the two characterization results of~\cite{Bed91} and~\cite{PU07b} hold. Specifically, it is
sufficient to require the absence of equidepth autoconcurrency, i.e., the absence of identically labeled
events occurring at the same depth within a configuration. The depth of an event is defined as the length of
the longest causal chain of events up to and including the considered event.

The third alternative characterization of HHPB was provided in~\cite{AC20} and, unlike the previous two,
does not need any restrictive assumption. Based on earlier work~\cite{AC17} -- in which HHPB was shown to
coincide with back-and-forth barbed bisimulation congruence over singly-labeled processes, i.e., processes
with no autoconcurrency and autoconflict (see Figure~\ref{fig:auto_conc_caus_confl}(c)) -- it refers to the
setting of a different reversible variant of CCS called RCCS~\cite{DK04,DK05,Kri12}. While in CCSK all
executed actions and discarded alternative subprocesses are kept within the syntax of processes so as to
enable reversibility, in RCCS the same information is stored into stack-based memories attached to
processes; the two approaches have been proven to be equivalent in~\cite{LMM21}. The idea in~\cite{AC20} is
to import HHPB in the RCCS setting by encoding memories, i.e., the past behavior, as identified
configuration structures. These are stable configuration structures enriched with unique event identifiers,
used in transition labels and exploited when undoing synchronizations. The characterizing equivalence,
called back-and-forth bisimilarity and defined over RCCS processes, relies on ternary bisimulation relations
in which the third component is a bijection between the sets of identifiers of the actions executed so far
in the two processes.

Having to reintroduce a third component in the bisimulation relations in order to characterize HHPB exactly,
i.e., also in the presence of equidepth autoconcurrency, amounts to certifying that ``reversibility is not
just back and forth''~\cite{AC20}, i.e., the forward and backward bisimulation games alone are not enough.
The question then becomes whether and to what extent a systematic event identification is really necessary
to reach HHPB.

This question also arises from the fact that, in the aforementioned bisimulation games, CCSK transition
labels such as $a[i]$ and $a[j]$ are deemed to be different when the two keys $i$ and $j$ are
different~\cite{PU07a} -- which results in the absence of repeated, identically labeled events along forward
computations~\cite{PU07b} -- while identified RCCS transition labels like $i$:$a$ and $j$:$a$ are viewed as
compatible even if $i$ and $j$ are different~\cite{AC20}. On the one hand, in CCSK the two processes $a
\pco{} a$ and $a \, . \, a$ are told apart by forward-reverse bisimilarity because the former evolves to
$a[i] \pco{} a[j]$, which can undo $a[i]$ and $a[j]$ in any order, while the latter evolves to $a[i] \, . \,
a[j]$, from which only $a[j]$ can be undone, hence undoing $a[i]$ \blue{in $a[i] \pco{} a[j]$} cannot be
matched by undoing $a[j]$ \blue{in $a[i] \, . \, a[j]$}. On the other hand, in identified RCCS the same two
processes are distinguished by back-and-forth bisimilarity because, although undoing $i$:$a$ can be matched
by undoing $j$:$a$, it is not possible to establish a suitable bijection from a distributed memory
containing $i$:$a$ in a location and $j$:$a$ in another location \linebreak to a centralized memory
containing $j$:$a$ on top of $i$:$a$.

%
\subsection{\blue{Our Approach}}
\label{sec:our_approach}
%

In this paper we pursue an approach different from event identification. We observe that, in many
distinguishing examples, what matters is not the identity of individual events, but the number of incoming
transitions of a configuration that are labeled with a given action. This suggests that instead of uniquely
identifying events, it may suffice to count them. Building on this idea, we introduce backward ready
multisets, a multiset-based abstraction of the past behavior of configurations; they are extensions of the
backward ready sets exploited in~\cite{BEM24} to axiomatize forward-reverse bisimilarity over reversible
concurrent processes. These multisets record in how many distinct ways an action can be undone, thus
capturing essential concurrency information without relying on global event identifiers or labeling- and
causality-preserving bijections.

Let us consider again Figures~\ref{fig:auto_conc_caus_confl}(a) and~(b). If we look at the two top (resp.\
bottom) configurations, we note that the one on the left has two outgoing (resp.\ incoming) transitions,
while the one on the right has only one. As for the top configuration on the left, in principle we may not
know whether the branch is due to the fact that the two events are concurrent or conflicting. However, for
the bottom configuration on the left we can certainly say that the two events are concurrent, as the models
we are considering are truly concurrent and hence the configuration graph of process $a \, . \, a + a \, .
\, a$ where $+$ stands for nondeterministic choice (see Figure~\ref{fig:auto_conc_caus_confl}(d)) cannot be
isomorphic to the one of $a \pco{} a$ because it must have two different bottom configurations ($\{ \lplu a,
\lplu . a \}$ and $\{ \rplu a, \rplu . a \}$) instead of a single one.

\pagebreak

This paper is an extended version of~\cite{BEM25} \blue{in which we rectify the main result of that paper:
our characterizations, which overcome any limitation related to the absence of autoconcurrency, are valid
when all possible conflicts among identically labeled events are local}. After recalling in
Section~\ref{sec:config_struct_hhp_bisim} the definitions of stable configuration structure~\cite{GP09} and
HHPB~\cite{Bed91}, we provide the following contributions:

	\begin{itemize}

\item In Section~\ref{sec:alt_char_config_struct} we exhibit a denotational characterization over stable
configuration structures. HHPB turns out to coincide with forward-reverse bisimilarity extended with a
clause for checking the equality of the backward ready multisets of matching configurations. The result
holds in the absence of non-local autoconflicts, i.e., \blue{under the assumption that, whenever two
identically labeled events are in conflict with each other, then there is no third event with which one or
both conflicting events are concurrent}.

\item In Section~\ref{sec:alt_char_operational} we exhibit an operational characterization based on a
variant of \blue{PRPC}, the \blue{proved} reversible process calculus of~\cite{BEM24,Esp25}, where executed
action identification is limited to synchronizations. After revising its proved operational semantics
inspired by~\cite{DP92} in order to faithfully account for causality and concurrency, we set up a
backward-ready-multiset variant of forward-reverse bisimilarity. Then we define a denotational semantics
based on stable configuration structures in which events are formalized as proof terms~\cite{BC88b,BC94}, so
as to import the notion of HHPB. We show that, in the absence of non-local autoconflicts, the stable
configuration structures associated with two processes are hereditary history-preserving bisimilar iff the
two processes are equated by the backward-ready-multiset variant of forward-reverse bisimilarity.

\item In Section~\ref{sec:rel_eil_brml} we investigate the relationships between the event identifier logic
of~\cite{PU14}, which captures HHPB, and \blue{a novel} backward ready multiset logic, which we show to
capture forward-reverse bisimilarity extended with backward ready multiset equality.

	\end{itemize}

\noindent
Section~\ref{sec:concl} closes the paper with directions for future work. 

\blue{
We conclude this introduction by mentioning that our results are in line with those of~\cite{APU25},
appeared shortly after~\cite{BEM25}, where alternative characterizations of HPB and HHPB are given in terms
of just forward and backward trasitions, without resorting to ternary relations. Their approach is based on
CCSK and on keyed event structures, where each event is uniquely identified by a key. More recent
work~\cite{AK25} introduces a symmetric residuation operation on configuration structures to model
reversibility in a purely denotational setting. Stable configuration structures turn out to be closed under
this residuation. A semantics is derived in terms of prime event structures, where a switch operation
dualizes causality and conflict. Their approach is closely aligned with our own goals: we also aim to
faithfully represent reversible process behavior in a true concurrency model, preserving causality and
independence relationships. However, our approach differs in how we encode memory and backward moves: rather
than relying on residuation, we build memory syntactically -- via proof terms -- which gives us a more
operational perspective. Furthermore, our backward-ready-multiset forward-reverse bisimilarity is proven to
coincide with HHPB in a way that is tailored to the structure of our extension of PRPC.
}

%
%
\section{\blue{Configuration Structures and} Hereditary History-Preserving Bisimilarity}
\label{sec:config_struct_hhp_bisim}
%
%

In this section we recall the stable version of configuration structures~\cite{GP09}
(Section~\ref{sec:config_struct}) as well as hereditary history-preserving bisimilarity~\cite{Bed91} over
them (Section~\ref{sec:hhpb}).

%
\subsection{\blue{Stable Configuration Structures}}
\label{sec:config_struct}
%

Configuration structures~\cite{GP09} are variants of event structures~\cite{Win86} expressing their
underlying labeled transition system view. A configuration, which represents a state, is the set of
non-conflicting events \blue{that} occurred so far. Every configuration is finite and left-closed with
respect to causality, i.e., it contains all the events causing the ones in the configuration itself.
Therefore, given two arbitrary events in a configuration, they are either causally related, \blue{i.e., can
occur only in a specific order}, or independent of each other, \blue{i.e., can occur in any order}.

In the following definitions taken from~\cite{GG01}, we assume a countable set $\cala$ of action labels.
Furthermore, we denote by $\powerset_{\rm fin}(\cale)$ the set of finite subsets of set $\cale$.

	\begin{defi}\label{def:config_struct}

A \emph{configuration structure} is a triple $\emph{\ssfc} = (\cale, \calc, \ell)$ where:

		\begin{itemize}

\item $\cale$ is a set of \emph{events}.

\item $\calc \subseteq \powerset_{\rm fin}(\cale)$ is a set of \emph{configurations}.

\item $\ell : \bigcup_{X \in \calc} X \rightarrow \cala$ is a \emph{labeling function} \blue{for events
occurring in configurations}.
\fullbox

		\end{itemize}

	\end{defi}

As shown in~\cite{GG01}, the configurations of a prime event structure~\cite{Win86} \blue{-- based on
causality and conflict relations --} form a configuration structure, while those of a stable event
structure~\cite{Win86} \blue{-- based on consistency and enabling notions --} form a configuration structure
that is stable in the following sense.

	\begin{defi}\label{def:stable_config_struct}

A configuration structure $\emph{\ssfc} = (\cale, \calc, \ell)$ is said to be \emph{stable} iff it is:

		\begin{itemize}

\item Rooted: $\emptyset \in \calc$.

\item Connected: $\forall X \in \calc \setminus \{ \emptyset \} \ldotp \exists e \in X \ldotp X \setminus \{
e \} \in \calc$.

\item Closed under bounded unions and intersections: $\forall X, Y, Z \in \calc \ldotp X \cup Y \subseteq Z
\Longrightarrow X \cup Y, \linebreak X \cap Y \in \calc$.
\fullbox

		\end{itemize}

	\end{defi}

The notions of causality and concurrency for events can be defined locally to configurations, while the
notion of conflict has to be defined globally, i.e., by examining all configurations in the considered
structure.

	\begin{defi}\label{def:caus_conc_confl}

Let $\emph{\ssfc} = (\cale, \calc, \ell)$ be a stable configuration structure and $X \in \calc$ be one of
its configurations:

		\begin{itemize}

\item The \emph{causality relation} over $X$ is defined by letting $e_{1} \le_{X} e_{2}$, for two events
$e_{1}, e_{2} \in X$, iff $e_{2} \in Y$ implies $e_{1} \in Y$ for all $Y \in \calc$ such that $Y \subseteq
X$. We write $e_{1} <_{X} e_{2}$ when $e_{1} \le_{X} e_{2}$ and $e_{1} \neq e_{2}$.

\item Two distinct events $e_{1}, e_{2} \in X$ are \emph{concurrent} in $X$, written $e_{1} \,
\textit{co}_{X} \, e_{2}$, iff $e_{1} \not<_{X} e_{2}$ and $e_{2} \not<_{X} e_{1}$.

\item Two distinct events $e_{1}, e_{2} \in \cale$ are \emph{conflicting}, written $e_{1} \, \# \; e_{2}$,
iff there is no $Z \in \calc$ such that $e_{1}, e_{2} \in Z$.
\fullbox

		\end{itemize}

	\end{defi}

A notion of transition between configurations can be naturally introduced.

	\begin{defi}\label{def:trans}

Let $\emph{\ssfc} = (\cale, \calc, \ell)$ be a stable configuration structure, $a \in \cala$, and $X, X' \in
\calc$. We say that there exists a \emph{transition} from $X$ to $X'$ \blue{labeled with $a$}, written $X
\arrow{a}{\emph{\ssfc}} X'$, iff $X \subseteq X'$, $X' \setminus X = \{ e \}$, and $\ell(e) = a$.
\fullbox

	\end{defi}

%
\subsection{\blue{Hereditary History-Preserving Bisimilarity}}
\label{sec:hhpb}
%

HHPB~\cite{Bed91} performs the bisimulation game both in the forward direction and in the backward one.
Since there is a single transition relation, in the bisimulation game a distinction is made between the
outgoing transitions of two matching configurations $X_{1}$ and $X_{2}$ (i.e., $X_{1}
\arrow{a}{\emph{\textsf{C}}_{1}} X'_{1}$ and $X_{2} \arrow{a}{\emph{\textsf{C}}_{2}} X'_{2}$ in the forward
direction) and their incoming transitions (i.e., $X'_{1} \arrow{a}{\emph{\textsf{C}}_{1}} X_{1}$ and $X'_{2}
\arrow{a}{\emph{\textsf{C}}_{2}} X_{2}$ in the backward direction). Below, $f \restr X$ denotes the
restriction of function $f$ to set $X$.

	\begin{defi}\label{def:hhp_bisim}

We say that two stable configuration structures $\emph{\ssfc}_{i} = (\cale_{i}, \calc_{i}, \ell_{i})$, $i
\in \{ 1, 2 \}$, are \emph{hereditary history-preserving bisimilar}, written $\emph{\ssfc}_{1} \sbis{\rm
HHPB} \emph{\ssfc}_{2}$, iff there exists a hereditary history-preserving bisimulation between
$\emph{\ssfc}_{1}$ and~$\emph{\ssfc}_{2}$, i.e., a relation $\calb \subseteq \calc_{1} \times \calc_{2}
\times \powerset(\cale_{1} \times \cale_{2})$ such that $(\emptyset, \emptyset, \emptyset) \in \calb$ and,
whenever $(X_{1}, X_{2}, f) \in \calb$, then:

		\begin{itemize}

\item $f$ is a bijection from $X_{1}$ to $X_{2}$ that preserves:

			\begin{itemize}

\item Labeling: $\ell_{1}(e) = \ell_{2}(f(e))$ for all $e \in X_{1}$.

\item Causality: $e \le_{X_{1}} e' \Longleftrightarrow f(e) \le_{X_{2}} f(e')$ for all $e, e' \in X_{1}$.

			\end{itemize}

\item For each $X_{1} \arrow{a}{\emph{\ssfc}_{1}} X'_{1}$ there exist $X_{2} \arrow{a}{\emph{\ssfc}_{2}}
X'_{2}$ and $f' \subseteq \cale_{1} \times \cale_{2}$ such that $(X'_{1}, X'_{2}, f') \in \calb$ and $f'
\upharpoonright X_{1} = f$, and vice versa.

\item For each $X'_{1} \arrow{a}{\emph{\ssfc}_{1}} X_{1}$ there exist $X'_{2} \arrow{a}{\emph{\ssfc}_{2}}
X_{2}$ and $f' \subseteq \cale_{1} \times \cale_{2}$ such that $(X'_{1}, X'_{2}, f') \in \calb$ and $f
\upharpoonright X'_{1} = f'$, and vice versa.
\fullbox

		\end{itemize}

	\end{defi}

%
%
\section{Characterization on Stable Configuration Structures}
\label{sec:alt_char_config_struct}
%
%

The first characterization that we provide for $\sbis{\rm HHPB}$ is on stable configuration structures.
\blue{The idea is to move from a ternary bisimulation relation to a binary one.} Instead of stepwise
building a labeling- and causality-preserving bijection between the events of matching configurations --
which are the events occurred so far in both stable configuration structures -- we just count the
identically labeled incoming transitions of matching configurations. \linebreak In the following, $\lmp$ and
$\rmp$ are used as multiset delimiters.

	\begin{defi}\label{def:brm_cs}

\blue{
Let $\emph{\ssfc} = (\cale, \calc, \ell)$ be a stable configuration structure. The \emph{backward ready
multiset} of $X \in \calc$ is $\brm(X) = \lmp a \in \cala \mid X' \arrow{a}{\emph{\textsf{C}}} X \rmp$.
\fullbox
}

	\end{defi}


	\begin{defi}\label{def:frb_brm_config_struct}

We say that two stable configuration structures $\emph{\ssfc}_{i} = (\cale_{i}, \calc_{i}, \ell_{i})$, $i
\in \{ 1, 2 \}$, are \emph{brm-forward-reverse bisimilar}, written $\emph{\ssfc}_{1} \sbis{\rm FRB:brm}
\emph{\ssfc}_{2}$, iff there exists a brm-forward-reverse bisimulation between $\emph{\ssfc}_{1}$ and
$\emph{\ssfc}_{2}$, i.e., a relation $\calb \subseteq \calc_{1} \times \calc_{2}$ such that $(\emptyset,
\emptyset) \in \calb$ and, whenever $(X_{1}, X_{2}) \in \calb$, then:

		\begin{itemize}

\item \blue{$\brm(X_{1}) = \brm(X_{2})$.}

\item For each $X_{1} \arrow{a}{\emph{\ssfc}_{1}} X'_{1}$ there exists $X_{2} 
\arrow{a}{\emph{\ssfc}_{2}} X'_{2}$ such that $(X'_{1}, X'_{2}) \in \calb$, and vice versa.

\item For each $X'_{1} \arrow{a}{\emph{\ssfc}_{1}} X_{1}$ there exists $X'_{2} 
\arrow{a}{\emph{\ssfc}_{2}} X_{2}$ such that $(X'_{1}, X'_{2}) \in \calb$, and vice versa.
\fullbox

		\end{itemize}

	\end{defi}

We emphasize that $\sbis{\rm FRB:brm}$ is capable of discriminating between autoconcurrency and
autocausation like $\sbis{\rm HHPB}$. For example, the two configuration structures in
Figure~\ref{fig:auto_conc_caus_confl}(a) and~(b) are \blue{distinguished} by $\sbis{\rm FRB:brm}$ because
the backward ready multiset of the bottom configuration of the former is $\lmp a, a \rmp$ while the one of
the bottom configuration of the latter is $\lmp a \rmp$. This makes the binary-relation-based $\sbis{\rm
FRB:brm}$ a good candidate to characterize the ternary-relation-based $\sbis{\rm HHPB}$ without the
limitations of previous results~\cite{Bed91,PU07b,PU12}. However, not even our characterization result --
originally stated in~\cite{BEM25} without any constraint -- holds in general, because a restriction
different from the absence of equidepth autoconcurrency comes into play.

	\begin{figure}[t]

\begin{tikzpicture}[modal]

\node[] (ea1) [label = 135: {$\emph{\ssfe}$}] {$a_{1}$};
\node[] (ea2) [right = of ea1]                {$a_{2}$};
\node[] (ea3) [right = of ea2]                {$a_{3}$};
\node[] (eb1) [below = of ea1]                {$b_{1}$};
\node[] (eb2) [below = of ea2]                {$b_{2}$};
\node[] (eb3) [below = of ea3]                {$b_{3}$};

\path[->] (ea1) edge                            (eb1);
\path[->] (ea2) edge                            (eb2);
\path[->] (ea3) edge                            (eb3);
\path[-]  (ea1) edge[bend left, loosely dotted] (ea3);
\path[-]  (eb1) edge[loosely dotted]            (eb2);
\path[-]  (eb2) edge[loosely dotted]            (eb3);

\node[] (fa1) [right = 3cm of ea3, label = 135: {$\emph{\ssff}$}] {$a_{1}$};
\node[] (fa2) [right = of fa1]                                    {$a_{2}$};
\node[] (fa3) [right = of fa2]                                    {$a_{3}$};
\node[] (fb1) [below = of fa1]                                    {$b_{1}$};
\node[] (fb2) [below = of fa2]                                    {$b_{2}$};
\node[] (fb3) [below = of fa3]                                    {$b_{3}$};

\path[->] (fa1) edge                            (fb1);
\path[->] (fa2) edge                            (fb2);
\path[->] (fa3) edge                            (fb3);
\path[-]  (fa1) edge[bend left, loosely dotted] (fa3);
\path[-]  (fb1) edge[loosely dotted]            (fb2);
\path[-]  (fb2) edge[loosely dotted]            (fa3);

\end{tikzpicture}

\caption{\blue{Event structures $\emph{\ssfe}$ and $\emph{\ssff}$: arrows denote causality, dotted lines
denote conflict}}
\label{fig:cex_frbbrm_not_hhpb_es}

	\end{figure}

	\begin{figure}[t]

\begin{tikzpicture}[modal]

\node[] (e1)  [label = 90: {$\emph{\ssfe}$}]                                     {1};
\node[] (e2)  [below left = of e1, xshift = 0pt, yshift = 0pt, label = 180: {}]  {2};
\node[] (e3)  [below = of e1, xshift = 0pt, yshift = 0pt, label = 180: {}]       {3};
\node[] (e4)  [below right = of e1, xshift = 0pt, yshift = 0pt, label = 180: {}] {4};
\node[] (e5)  [below = of e2, xshift = 0pt, yshift = 0pt, label = 180: {}]       {5};
\node[] (e6)  [below right = of e3, xshift = 0pt, yshift = 0pt, label = 180: {}] {6};
\node[] (e7)  [below left = of e2, xshift = 0pt, yshift = 0pt, label = 180: {}]  {7};
\node[] (e8)  [below = of e3, xshift = 0pt, yshift = 0pt, label = 180: {}]       {8};
\node[] (e9)  [below right = of e4, xshift = 0pt, yshift = 0pt, label = 180: {}] {9};
\node[] (e10) [below = of e7, xshift = 0pt, yshift = 0pt, label = 180: {}]       {10};
\node[] (e11) [below = of e5, xshift = 0pt, yshift = 0pt, label = 180: {}]       {11};
\node[] (e12) [below = of e9, xshift = 0pt, yshift = 0pt, label = 180: {}]       {12};
\node[] (e13) [below = of e6, xshift = 0pt, yshift = 0pt, label = 180: {}]       {13};

\path[->] (e1) edge node[above left]  {$a_{1}$} (e2);
\path[->] (e1) edge node[left]        {$a_{2}$} (e3);
\path[->] (e1) edge node[above right] {$a_{3}$} (e4);
\path[->] (e2) edge node[above left]  {$b_{1}$} (e7);
\path[->] (e2) edge node[left]        {}        (e5);
\path[->] (e3) edge node[left]        {}        (e5);
\path[->] (e3) edge node[left]        {$b_{2}$} (e8);
\path[->] (e3) edge node[left]        {}        (e6);
\path[->] (e4) edge node[left]        {}        (e6);
\path[->] (e4) edge node[above right] {$b_{3}$} (e9);
\path[->] (e7) edge node[left]        {}        (e10);
\path[->] (e5) edge node[left]        {}        (e10);
\path[->] (e5) edge node[left]        {}        (e11);
\path[->] (e8) edge node[left]        {}        (e11);
\path[->] (e8) edge node[left]        {}        (e13);
\path[->] (e6) edge node[left]        {}        (e13);
\path[->] (e6) edge node[left]        {}        (e12);
\path[->] (e9) edge node[left]        {}        (e12);

\node[] (f1)  [right = 5.5cm of e1, label = 90: {$\emph{\ssff}$}]                {1};
\node[] (f2)  [below left = of f1, xshift = 0pt, yshift = 0pt, label = 180: {}]  {2};
\node[] (f3)  [below = of f1, xshift = 0pt, yshift = 0pt, label = 180: {}]       {3};
\node[] (f4)  [below right = of f1, xshift = 0pt, yshift = 0pt, label = 180: {}] {4};
\node[] (f5)  [below = of f2, xshift = 0pt, yshift = 0pt, label = 180: {}]       {5};
\node[] (f6)  [below right = of f3, xshift = 0pt, yshift = 0pt, label = 180: {}] {6};
\node[] (f7)  [below left = of f2, xshift = 0pt, yshift = 0pt, label = 180: {}]  {7};
\node[] (f8)  [below = of f3, xshift = 0pt, yshift = 0pt, label = 180: {}]       {8};
\node[] (f9)  [below right = of f4, xshift = 0pt, yshift = 0pt, label = 180: {}] {9};
\node[] (f10) [below = of f7, xshift = 0pt, yshift = 0pt, label = 180: {}]       {10};
\node[] (f11) [below = of f5, xshift = 0pt, yshift = 0pt, label = 180: {}]       {11};
\node[] (f12) [below = of f9, xshift = 0pt, yshift = 0pt, label = 180: {}]       {12};

\path[->] (f1) edge node[above left]  {$a_{1}$} (f2);
\path[->] (f1) edge node[left]        {$a_{2}$} (f3);
\path[->] (f1) edge node[above right] {$a_{3}$} (f4);
\path[->] (f2) edge node[above left]  {$b_{1}$} (f7);
\path[->] (f2) edge node[left]        {}        (f5);
\path[->] (f3) edge node[left]        {}        (f5);
\path[->] (f3) edge node[left]        {$b_{2}$} (f8);
\path[->] (f3) edge node[left]        {}        (f6);
\path[->] (f4) edge node[left]        {}        (f6);
\path[->] (f4) edge node[above right] {$b_{3}$} (f9);
\path[->] (f7) edge node[left]        {}        (f10);
\path[->] (f5) edge node[left]        {}        (f10);
\path[->] (f5) edge node[left]        {}        (f11);
\path[->] (f8) edge node[left]        {}        (f11);
\path[->] (f6) edge node[left]        {}        (f12);
\path[->] (f9) edge node[left]        {}        (f12);

\end{tikzpicture}

\caption{$\emph{\ssfe}$ and $\emph{\ssff}$ are equated by $\sbis{\rm FRB:brm}$ but not by $\sbis{\rm HHPB}$
due to non-local autoconflicts}
\label{fig:cex_frbbrm_not_hhpb_cs}

	\end{figure}

\blue{
Consider the two event structures $\emph{\ssfe}$ and $\emph{\ssff}$ in
Figure~\ref{fig:cex_frbbrm_not_hhpb_es}, which are taken from~\cite[Example~4.7]{PU12}. Both of them feature
six events, of which three are labeled with action~$a$ and three are labeled with action $b$. To preserve
uniqueness among identically labeled events, each action has a distinct subscript. In both event structures,
every event $a_{i}$ causes the corresponding event $b_{i}$ for $i \in \{ 1, 2, 3 \}$ as arrows denote
causality, with event $a_{1}$ being in conflict with event $a_{3}$ as denoted by the dotted line between
them. Moreover, in event structure $\emph{\ssfe}$ event $b_{2}$ is in conflict with events $b_{1}$ and
$b_{3}$, while in event structure $\emph{\ssff}$ event $b_{2}$ is in conflict with events $b_{1}$ and
$a_{3}$ (and also $b_{3}$ due to conflict inheritance via causality).
}

The two corresponding configuration structures, still denoted by $\emph{\ssfe}$ and $\emph{\ssff}$ and taken
from~\cite[Example~4.7]{PU12}, are shown in Figure~\ref{fig:cex_frbbrm_not_hhpb_cs}, where configurations
are represented by numbers and transitions parallel to each other are labeled with the same subscripted
action. As pointed out by I.~Ulidowski, I.~Phillips, and C.~Aubert in a personal communication to the
authors, the two configuration structures $\emph{\ssfe}$ and $\emph{\ssff}$ constitute a counterexample to
the characterization result in~\cite{BEM25}. It turns out that $\emph{\ssfe} \not\sbis{\rm HHPB}
\emph{\ssff}$ because, starting the bisimulation game from the two initial configurations $1_{\emph{\ssfe}}$
and $1_{\emph{\ssff}}$, we have that:

	\begin{itemize}

\item If $1_{\emph{\ssff}}$ performs $a_{3}$ thus evolving to $4_{\emph{\ssff}}$, then $1_{\emph{\ssfe}}$
can respond by performing $a_{3}$ (or equivalently~$a_{1}$) thus evolving to $4_{\emph{\ssfe}}$ (or
equivalently $2_{\emph{\ssfe}}$). The initial bijection is $\{ (a_{3}, a_{3}) \}$.

\item If $4_{\emph{\ssff}}$ performs $a_{2}$ thus evolving to $6_{\emph{\ssff}}$, then $4_{\emph{\ssfe}}$
can respond by performing $a_{2}$ thus evolving to $6_{\emph{\ssfe}}$. The bijection becomes $\{ (a_{3},
a_{3}), (a_{2}, a_{2}) \}$.

\item If $6_{\emph{\ssfe}}$ performs $b_{2}$ thus evolving to $13_{\emph{\ssfe}}$, then $6_{\emph{\ssff}}$
can only respond by performing $b_{3}$ thus evolving to $12_{\emph{\ssff}}$. The bijection becomes $\{
(a_{3}, a_{3}), (a_{2}, a_{2}), (b_{2}, b_{3}) \}$ but causality is not preserved because $a_{2}$ causes
$b_{2}$ in $\emph{\ssfe}$ whereas $a_{2}$ does not cause $b_{3}$ in $\emph{\ssff}$ as can be seen in the
originating event structures in \blue{Figure~\ref{fig:cex_frbbrm_not_hhpb_es}}.

	\end{itemize}

On the other hand, there is no way of telling $\emph{\ssfe}$ and $\emph{\ssff}$ apart based on $\sbis{\rm
FRB:brm}$, as all configurations reached after three transitions in the forward direction feature the
backward ready multiset $\lmp a, b \rmp$. \blue{Note that this would not happen if the six considered events
had different labels -- e.g., of the form $a_{i}$ and $b_{i}$ -- because in that case the forward
bisimulation game alone would distinguish the two configuration structures: the sequence of transitions
$1_{\emph{\ssfe}} \arrow{a_{3}}{\emph{\ssfe}} 4_{\emph{\ssfe}} \arrow{a_{2}}{\emph{\ssfe}} 6_{\emph{\ssfe}}
\arrow{b_{2}}{\emph{\ssfe}} 13_{\emph{\ssfe}}$ (resp.\ $1_{\emph{\ssfe}} \arrow{a_{2}}{\emph{\ssfe}}
3_{\emph{\ssfe}} \arrow{b_{2}}{\emph{\ssfe}} 8_{\emph{\ssfe}} \arrow{a_{3}}{\emph{\ssfe}}
13_{\emph{\ssfe}}$) would not be matched by $1_{\emph{\ssff}} \arrow{a_{3}}{\emph{\ssff}} 4_{\emph{\ssff}}
\arrow{a_{2}}{\emph{\ssff}} 6_{\emph{\ssff}} \arrow{b_{3}}{\emph{\ssfe}} 12_{\emph{\ssff}}$ (resp.\
$1_{\emph{\ssff}} \arrow{a_{2}}{\emph{\ssff}} 3_{\emph{\ssff}} \arrow{b_{2}}{\emph{\ssff}} 8_{\emph{\ssff}}
\arrow{a_{1}}{\emph{\ssfe}} 11_{\emph{\ssff}}$) as $b_{2} \neq b_{3}$ (resp.\ $a_{3} \neq a_{1}$).}


In process algebraic terms, following~\cite{GV03} we have that $\emph{\ssfe}$ and $\emph{\ssff}$ can be
respectively represented as:
\cws{0}{\begin{array}{l}
((a_{1} \, . \, b_{1}) \pco{\emptyset} (a_{2} \, . \, b_{2}) \pco{\emptyset} (a_{3} \, . \, b_{3}))
\; \pco{\{ a_{1}, a_{3}, b_{1}, b_{2}, b_{3} \}} \;
((a_{1} + a_{3}) \pco{\emptyset} (b_{1} + b_{2}) \pco{\{ b_{2} \}} (b_{2} + b_{3})) \\
\end{array}}
and:
\cws{0}{\begin{array}{l}
((a_{1} \, . \, b_{1}) \pco{\emptyset} (a_{2} \, . \, b_{2}) \pco{\emptyset} (a_{3} \, . \, b_{3}))
\; \pco{\{ a_{1}, a_{3}, b_{1}, b_{2} \}} \;
((a_{1} + a_{3}) \pco{\{ a_{3} \}} (a_{3} + b_{2}) \pco{\{ b_{2} \}} (b_{1} + b_{2})) \\
\end{array}}
where a relabeling mapping $a_{1}$, $a_{2}$, $a_{3}$ to $a$ and $b_{1}$, $b_{2}$, $b_{3}$ to $b$ has to be
finally applied. \blue{What can be observed is that there is autoconcurrency on $a$ -- on the lefthand side
of both processes the three $a$-labeled events are independent of each other -- and, at the same time,
autoconflict on $a$ and $b$ -- on the righthand side of both processes $a_{1}$ and $a_{3}$ as well as
$b_{2}$ and $b_{3}$ \linebreak are in conflict. It is such an interference between autoconcurrency and
autoconflict that causes $\sbis{\rm FRB:brm}$ not to achieve the same discriminating power as $\sbis{\rm
HHPB}$.

Our first characterization result holds under the stronger assumption that all autoconflicts are local,
where local means that neither conflicting event is concurrent with a third event.}


	\begin{defi}\label{def:no_local_autoconflicts}

\textcolor{blue}{
Let $\emph{\ssfc} = (\cale, \calc, \ell)$ be a stable configuration structure and $e_{1}, e_{2} \in \cale$
be such that $e_{1} \, \# \; e_{2}$ and $\ell(e_{1}) = \ell(e_{2})$. We say that the autoconflict between
$e_{1}$ and $e_{2}$ is \emph{local} iff there exist no $e \in \cale$ and $X_{1}, X_{2} \in \calc$ such that
$e \, \textit{co}_{X_{1}} \, e_{1}$ or $e \, \textit{co}_{X_{2}} \, e_{2}$.
}
\fullbox

	\end{defi}

	\begin{thm}\label{thm:hhpb_frb_brm_config_struct}

Let $\textsf{C}_{i} = (\cale_{i}, \calc_{i}, \ell_{i})$, $i \in \{ 1, 2 \}$, be two stable configuration
structures:

		\begin{itemize}

\item \blue{If $\textsf{C}_{1} \sbis{\rm HHPB} \textsf{C}_{2}$ then $\textsf{C}_{1} \sbis{\rm FRB:brm}
\textsf{C}_{2}$.}

\item \blue{If $\textsf{C}_{1} \sbis{\rm FRB:brm} \textsf{C}_{2}$ then $\textsf{C}_{1} \sbis{\rm HHPB}
\textsf{C}_{2}$ provided that all possible autoconflicts are local.}

		\end{itemize}

		\begin{proof}

\blue{We prove the two implications separately}:

			\begin{itemize}

\item Suppose that $\emph{\textsf{C}}_{1} \sbis{\rm HHPB} \emph{\textsf{C}}_{2}$ due to some hereditary
history-preserving bisimulation $\calb$. Then $\emph{\textsf{C}}_{1} \sbis{\rm FRB:brm}
\emph{\textsf{C}}_{2}$ follows by proving that $\calb' = \{ (X_{1}, X_{2}) \mid (X_{1}, X_{2}, f) \in \calb
\}$ is a brm-forward-reverse bisimulation. Observing that the starting clause and the clauses for outgoing
and incoming transitions matching of $\sbis{\rm FRB:brm}$ (see Definition~\ref{def:frb_brm_config_struct})
are a simplification of those of $\sbis{\rm HHPB}$ (see Definition~\ref{def:hhp_bisim}), given $(X_{1},
X_{2}) \in \calb'$, i.e., $(X_{1}, X_{2}, f) \in \calb$ for some labeling- and causality-preserving
bijection $f$ from $X_{1}$ to $X_{2}$, we just have to show that $\brm(X_{1}) = \brm(X_{2})$. \\
Suppose that this is not the case, say $X_{1}$ has fewer incoming $a$-transitions than $X_{2}$. Without loss
of generality, we can assume that $X_{1}$ has one incoming $a$-transition while $X_{2}$ has two. Then in
$\emph{\textsf{C}}_{2}$ there is a diamond closing into~$X_{2}$, i.e., there exist three configurations
$Y_{2}$, $X'_{2}$, and $X''_{2}$ and two $a$-labeled events $e'_{2}$ and $e''_{2}$ such that $Y_{2}
\xarrow{\ell_{2}(e'_{2})}{\emph{\textsf{C}}_{2}} X'_{2}$, $Y_{2}
\xarrow{\ell_{2}(e''_{2})}{\emph{\textsf{C}}_{2}} X''_{2}$, $X'_{2}
\xarrow{\ell_{2}(e''_{2})}{\emph{\textsf{C}}_{2}} X_{2}$, and $X''_{2}
\xarrow{\ell_{2}(e'_{2})}{\emph{\textsf{C}}_{2}} X_{2}$, with $e'_{2}$ and $e''_{2}$ concurrent in $X_{2}$.
\\
Due to $(X_{1}, X_{2}, f) \in \calb$, in $\emph{\textsf{C}}_{1}$ there must be two configurations $Y_{1}$
and $X'_{1}$ and two $a$-labeled events $e'_{1}$ and $e''_{1}$ such that $Y_{1}
\xarrow{\ell_{1}(e'_{1})}{\emph{\textsf{C}}_{1}} X'_{1} \xarrow{\ell_{1}(e''_{1})}{\emph{\textsf{C}}_{1}}
X_{1}$, with $e'_{1} \le_{X_{1}} e''_{1}$ because $X_{1}$ has a single incoming $a$-transition. Since
$\calb$ is a hereditary history-preserving bisimulation, $f$ should relate $e'_{1}, e''_{1} \in X_{1}$ with
$e'_{2}, e''_{2} \in X_{2}$ in a causality-preserving way, but this is not possible because $f(e'_{1})
\not\le_{X_{2}} f(e''_{1})$ where $f(e'_{1}) \in \{ e'_{2}, e''_{2} \}$ and $f(e''_{1}) \in \{ e'_{2},
e''_{2} \} \setminus \{ f(e'_{1}) \}$.

\item Suppose that $\emph{\textsf{C}}_{1} \sbis{\rm FRB:brm} \emph{\textsf{C}}_{2}$ due to some maximal
brm-forward-reverse bisimulation $\calb$. \linebreak Then, given $(X_{1}, X_{2}) \in \calb$, the existence
in $\emph{\textsf{C}}_{1}$ of a finite sequence of transitions \linebreak $X_{1, n} \xarrow{\ell_{1}(e_{1,
n})}{\emph{\textsf{C}}_{1}} X_{1, n - 1} \dots X_{1, 1} \xarrow{\ell_{1}(e_{1, 1})}{\emph{\textsf{C}}_{1}}
X_{1}$ implies the existence in $\emph{\textsf{C}}_{2}$ of a finite sequence of transitions $X_{2, n}
\xarrow{\ell_{2}(e_{2, n})}{\emph{\textsf{C}}_{2}} X_{2, n - 1} \dots X_{2, 1} \xarrow{\ell_{2}(e_{2,
1})}{\emph{\textsf{C}}_{2}} X_{2}$ such that $(X_{1, h}, X_{2, h}) \in \calb$ and $\ell_{1}(e_{1, h}) =
\ell_{2}(e_{2, h})$ for all $h = 1, \dots, n$, and vice versa. Note that $e_{1, h} \neq e_{1, k}$ and $e_{2,
h} \neq e_{2, k}$ for all $h \neq k$ because in each transition the source configuration and the target
configuration differ by one event, which is precisely the executed event (see Definition~\ref{def:trans}).
\\
Thus $\emph{\textsf{C}}_{1} \sbis{\rm HHPB} \emph{\textsf{C}}_{2}$ follows by proving that:
\blue{
\cws{0}{\hspace*{-0.4cm}\begin{array}{l}
\calb' \: = \: \{ (X_{1}, X_{2}, \{ (e_{1, h}, e_{2, h}) \mid h \in H = \{ 1, \dots, n \} \textrm{ for some
} n \in \natns_{\ge 1} \}) \mid \\
\hspace*{3.1cm} (X_{1}, X_{2}) \in \calb \, \land \\
\hspace*{3.1cm} \emptyset = X_{i, |H|} \xarrow{\ell_{i}(e_{i, |H|})}{\emph{\textsf{C}}_{i}} X_{i, |H| - 1}
\dots X_{i, 1} \xarrow{\ell_{i}(e_{i, 1})}{\emph{\textsf{C}}_{i}} X_{i} \textrm{ for all } i \in \{ 1, 2 \}
\, \land \\
\hspace*{3.1cm} (X_{1, h}, X_{2, h}) \in \calb \textrm{ for all } h \in H \, \land \\
\hspace*{3.1cm} \ell_{1}(e_{1, h}) = \ell_{2}(e_{2, h}) \textrm{ for all } h \in H \}
\end{array}}
}
is a hereditary history-preserving bisimulation. Observing that $(\emptyset, \emptyset) \in \calb$ implies
$(\emptyset, \emptyset, \emptyset) \in \calb'$, take $(X_{1}, X_{2}, \{ (e_{1, h}, e_{2, h}) \mid h \in H
\}) \in \calb'$, so that $(X_{1}, X_{2}) \in \calb$:

				\begin{itemize}

\item If $X_{1} \arrow{a}{\emph{\textsf{C}}_{1}} X'_{1}$ due to some $a$-labeled event $e_{1}$, then $X_{2}
\arrow{a}{\emph{\textsf{C}}_{2}} X'_{2}$ due to some \linebreak $a$-labeled event $e_{2}$ such that
$(X'_{1}, X'_{2}) \in \calb$. Since $e_{1} \notin X_{1}$ and $e_{2} \notin X_{2}$, it holds that $(X'_{1},
X'_{2}, \{ (e_{1, h}, e_{2, h}) \mid h \in H \} \cup \{ (e_{1}, e_{2}) \}) \in \calb'$. \\
If we start from $X_{2} \arrow{a}{\emph{\textsf{C}}_{2}} X'_{2}$, then we reason in the same way.

\item If $X'_{1} \arrow{a}{\emph{\textsf{C}}_{1}} X_{1}$ due to some $a$-labeled event $e_{1}$, then $X'_{2}
\arrow{a}{\emph{\textsf{C}}_{2}} X_{2}$ due to some $a$-labeled event $e_{2}$ such that $(X'_{1}, X'_{2})
\in \calb$. Since $e_{1} \notin X'_{1}$, $e_{2} \notin X'_{2}$, and $\brm(X_{1}) = \brm(X_{2})$, the latter
transition can be selected in such a way to satisfy $\{ (e_{1, h}, e_{2, h}) \mid h \in H \} \restr X'_{1} =
\{ (e_{1, h}, e_{2, h}) \mid h \in H \} \setminus \{ (e_{1}, e_{2}) \}$, hence $(X'_{1}, X'_{2}, \{ (e_{1,
h}, e_{2, h}) \mid h \in H \} \setminus \{ (e_{1}, e_{2}) \}) \in \calb'$. \\
If we start from $X'_{2} \arrow{a}{\emph{\textsf{C}}_{2}} X_{2}$, then we reason in the same way.

\item $f = \{ (e_{1, h}, e_{2, h}) \mid h \in H \}$ certainly is a bijection from $X_{1}$ to $X_{2}$ -- as
the events along either computation are $(i)$ different from each other, so the two reached configurations
$X_{1}$ and $X_{2}$ contain the same number of events, and $(ii)$ paired in a stepwise manner -- that
preserves labeling -- by definition of $\calb'$. If $|H| \le 1$ then causality is trivially preserved too,
\blue{so we assume that $|H| \ge 2$ and proceed by contradiction}. \\
Suppose that $X_{1}$ and $X_{2}$ break causality and, among all the pairs of $\calb$-related configurations
that break causality, they are the closest ones to $\emptyset$ and $\emptyset$ (in terms of number of
transitions to be executed from either empty configuration). Breaking causality means that there exists
$X_{1} \arrow{a}{\emph{\textsf{C}}_{1}} X'_{1}$, originated by $e'_{1} \notin X_{1}$ with $e_{1}
\le_{X'_{1}} e'_{1}$ for some $e_{1} \in X_{1}$, such that all the possible responses $X_{2}
\arrow{a}{\emph{\textsf{C}}_{2}} X'_{2}$ complying with $\calb$, which is maximal, satisfy $g(e_{1})
\not\le_{X'_{2}} g(e'_{1})$ where $g$ is $f$ extended with the new pair of events \blue{that originate the
two aforementioned transitions of $X_{1}$ and $X_{2}$ respectively}. \\
(In Figure~\ref{fig:cex_frbbrm_not_hhpb_cs}, $X_{1}$ and $X_{2}$ are states $6_{\emph{\ssfe}}$ and
$6_{\emph{\ssff}}$, $X'_{1}$ and $X'_{2}$ are states $13_{\emph{\ssfe}}$ and $12_{\emph{\ssff}}$, $e_{1} =
a_{2}$, $e'_{1} = b_{2}$, $g(e_{1}) = a_{2}$, $g(e'_{1}) = b_{3}$, $e_{1} \le_{X'_{1}} e'_{1}$, $g(e_{1})
\not\le_{X'_{2}} g(e'_{1})$.) \\
From $X_{1}$ and $X_{2}$ we go backward by following the respective computations undertaken in the forward
direction with respect to $\calb'$ -- without undoing $e_{1}$ -- until we reach $Y_{1}$ and~$Y_{2}$ with
$(Y_{1}, Y_{2}) \in \calb$ that have several incoming transitions; \blue{the existence of $Y_{1}$ and
$Y_{2}$ will be discussed later on}. Note that $Y_{1}$ and $Y_{2}$ must have the same number of incoming
transitions because $\calb$ is a brm-forward-reverse bisimulation. \\
(In Figure~\ref{fig:cex_frbbrm_not_hhpb_cs}, $Y_{1}$ and $Y_{2}$ are again states $6_{\emph{\ssfe}}$ and
$6_{\emph{\ssff}}$.) \\
In $Y_{1}$ and $Y_{2}$ we undo two identically labeled transitions matched by $\calb$, respectively
originated by two events $e''_{1}$ and $e''_{2}$ that are different from the two events $e_{1}$ and
$g(e_{1})$ \linebreak that originated the two transitions with which we arrived at $Y_{1}$ and $Y_{2}$ via
the computations undertaken in the forward direction with respect to $\calb'$. \blue{Note that $e''_{1} \,
\textit{co}_{Y_{1}} \, e_{1}$ and $e''_{2} \, \textit{co}_{Y_{2}} \, g(e_{1})$ because their labels
respectively belong to the backward ready multisets of $Y_{1}$ and $Y_{2}$.} Let $Y'_{1}$ and $Y'_{2}$ be
the two newly reached configurations \blue{in the backward direction} such that $(Y'_{1}, Y'_{2}) \in
\calb$. \\
(In Figure~\ref{fig:cex_frbbrm_not_hhpb_cs}, $Y'_{1}$ and $Y'_{2}$ are states $3_{\emph{\ssfe}}$ and
$3_{\emph{\ssff}}$, $e''_{1} = a_{3}$, $e''_{2} = a_{3}$, $e''_{1} \, \textit{co}_{Y_{1}} \, e_{1}$,
$e''_{2} \, \textit{co}_{Y_{2}} \, g(e_{1})$.) \\
There are two cases:

					\begin{itemize}

\item If in $Y'_{1}$ it is possible to perform a transition to some $Y''_{1}$ originated by $e'_{1}$, then
we execute it so as to recreate $e_{1} \le_{Y''_{1}} e'_{1}$ along a different computation of
$\emph{\textsf{C}}_{1}$ with respect to $\calb'$. Therefore $Y'_{2}$ responds by executing an identically
labeled transition to some $Y''_{2}$ originated by some event $e'_{2}$ such that $h(e_{1}) \le_{Y''_{2}}
h(e'_{1}) = e'_{2}$, where $h$ is the resulting bijection, as $Y'_{1}$ and $Y'_{2}$ are closer to
$\emptyset$ and $\emptyset$ than $X_{1}$ and $X_{2}$ -- remember the assumption that $X_{1}$ and $X_{2}$ are
the first $\calb$-related configurations that break causality. The two newly reached configurations \blue{in
the forward direction} satisfy $(Y''_{1}, Y''_{2}) \in \calb$. \\
(In Figure~\ref{fig:cex_frbbrm_not_hhpb_cs}, $Y''_{1}$ and $Y''_{2}$ are states $8_{\emph{\ssfe}}$ and
$8_{\emph{\ssff}}$, $e'_{2} = b_{2}$, $h(e_{1}) \le_{Y''_{2}} h(e'_{1}) = e'_{2}$.) \\
In $Y''_{1}$ it is possible to perform a transition to $X'_{1}$ originated by $e''_{1}$ otherwise we would
not have been able to go from $\emptyset$ to $X'_{1}$ via $Y'_{1}$ and $X_{1}$ by encountering $e''_{1}$ and
$e'_{1}$. \linebreak In other words, \blue{$e''_{1} \, \textit{co}_{X'_{1}} \, e'_{1}$}. \blue{Therefore
$Y''_{2}$ responds by performing an identically labeled transition to some $X'_{2}$ originated by $e''_{2}$.
But then $X_{2}$ responds to $X_{1}$ by executing a transition to $X'_{2}$ originated by some $e'_{2}$ with
the same label as $e'_{1}$ where $e''_{2} \, \textit{co}_{X'_{2}} \, e'_{2}$}, which contradicts the
assumption that $X_{2}$ cannot respond to $X_{1}$ in a causality-preserving way.
Consequently $e''_{2} \, \# \; e'_{2}$ (note that $e'_{2} \notin X_{2}$ and $e''_{2} \notin Y''_{2}$). \\
\blue{If $e''_{2}$ and $e'_{2}$ are identically labeled, then they form an autoconflict and this contradicts
in turn the assumption that all possible autoconflicts are local because $e''_{2} \, \textit{co}_{X_{2}} \,
g(e_{1})$. If instead $e''_{2}$ and $e'_{2}$ are not identically labeled, then we observe that $e'_{2}$ and
$g(e'_{1})$ must be identically labeled -- as $e'_{2}$ is one of the possible responses to $e'_{1}$ -- and
in conflict with each other -- otherwise $X_{2}$ would have an outgoing transition originated by $e'_{2}$
that preserves causality. Furthermore, $g(e_{1}) \, \textit{co}_{X'_{2}} \, g(e'_{1})$ because $g(e_{1})
\not\le_{X'_{2}} g(e'_{1})$ and they cannot be in conflict with each other as both events belong to
$X'_{2}$.  This again contradicts the assumption that all possible autoconflicts are local.}


\item If in $Y'_{1}$ it is not possible to perform a transition originated by $e'_{1}$, then we keep going
forward until we find it by executing transitions originated by all the events that have been undone by
going from $X_{1}$ and $X_{2}$ back to $Y_{1}$ and $Y_{2}$ respectively.

					\end{itemize}

\noindent
If the aforementioned $Y_{1}$ and $Y_{2}$ did not exist, then while going backward from $X_{1}$ and~$X_{2}$
by following the respective computations undertaken in the forward direction with respect to $\calb'$ --
without undoing $e_{1}$ -- we should reach $Z_{1}$ and $Z_{2}$ such that $(Z_{1}, Z_{2}) \in \calb$ with at
least one of them having several outgoing transitions. The reason is that if neither $Y_{1}$ and $Y_{2}$ nor
$Z_{1}$ and $Z_{2}$ existed, then there would be a single computation from $\emptyset$ to $X_{1}$ and a
single computation from $\emptyset$ to $X_{2}$ -- in which all events are causally related and hence there
is no concurrency -- thus contradicting $e_{1} \le_{X'_{1}} e'_{1}$ and $g(e_{1}) \not\le_{X'_{2}}
g(e'_{1})$ as it would hold that $g(e_{1}) \le_{X'_{2}} g(e'_{1})$. \\
However, since after leaving $Z_{1}$ and $Z_{2}$ towards $X_{1}$ and $X_{2}$ there would be neither
\linebreak $\calb$-related configurations with several incoming transitions, nor $\calb$-related
configurations at least one of which has several outgoing transitions, there would be a single computation
from $Z_{1}$ to $X_{1}$ and a single computation from $Z_{2}$ to $X_{2}$ -- in which all events are causally
related and hence there is no concurrency -- thus contradicting again $e_{1} \le_{X'_{1}} e'_{1}$ and
$g(e_{1}) \not\le_{X'_{2}} g(e'_{1})$ as it would hold that $g(e_{1}) \le_{X'_{2}} g(e'_{1})$.
\qedhere

				\end{itemize}
        
			\end{itemize}
    
		\end{proof}

	\end{thm}

%
%
\section{Operational Characterization}
\label{sec:alt_char_operational}
%
%

The second characterization that we provide for $\sbis{\rm HHPB}$ is operational. More precisely, we present
a variant of the syntax (Section~\ref{sec:rpc_syntax}) and the proved operational semantics
(Section~\ref{sec:rpc_proved_op_sem}) of PRPC, the reversible process calculus of~\cite{BEM24,Esp25},
followed by a redefinition of brm-forward-reverse bisimilarity on that variant (Section~\ref{sec:frb_brm}).
Then we develop a denotational semantics for the modified calculus based on stable configuration structures
(Section~\ref{sec:den_sem_scs}), so as to import the notion of HHPB. Finally, we prove that, in the absence
of non-local autoconflicts, the stable configuration structures associated with two processes are hereditary
history-preserving bisimilar iff the two processes are brm-forward-reverse bisimilar
(Section~\ref{sec:oper_char_res}).

%
\subsection{Syntax of Reversible Concurrent Processes}
\label{sec:rpc_syntax}
%

In the representation of a process, we are used to describe only its future behavior. However, in order to
support reversibility in the style of~\cite{PU07a}, we need to equip the syntax with information about the
past, in particular the actions that have already been executed, \blue{which we decorate with $\dag$ as
in~\cite{BR23}}. Taking inspiration from CCS~\cite{Mil89a} and CSP~\cite{BHR84}, given a countable set
$\cala$ of actions including an unobservable one denoted by~$\tau$, we extend the syntax of PRPC -- Proved
Reversible Process Calculus~\cite{BEM24,Esp25} as follows:
\cws{0}{\begin{array}{rcl}
P & \!\! ::= \!\! & \nil \mid a \, . \, P \mid a^{\dag \color{blue}{\xi}} . \, P \mid P + P \mid P \pco{L} P
\\
\xi & \!\! ::= \!\! & \varepsilon \mid \langle \theta, \theta \rangle_{L} \\
\end{array}}
where $a \in \cala$, $L \subseteq \cala \setminus \{ \tau \}$, $\varepsilon$ is the empty string
\blue{(which will be omitted next to $\dag$)}, $\theta$ is a proof term (which will be defined in
Section~\ref{sec:rpc_proved_op_sem}), and:

	\begin{itemize}

\item $\nil$ is the terminated process.

\item $a \, . \, P$ is a process that can execute action $a$ and whose forward continuation is $P$
(unexecuted action prefix).

\item $a^{\dag \xi} . \, P$ is a process that executed action $a$ and whose forward continuation is inside
$P$; \linebreak it can undo action $a$ after all executed actions within $P$ have been undone (executed
action prefix). \blue{The $\xi$ superscript, explained in Section~\ref{sec:rpc_proved_op_sem}, is the
novelty with respect to PRPC.}

\item $P_{1} + P_{2}$ expresses a nondeterministic choice between $P_{1}$ and $P_{2}$ as far as neither has
executed any action yet, otherwise only the one that was selected in the past can move (past-sensitive
alternative composition).

\item $P_{1} \pco{L} P_{2}$ expresses that $P_{1}$ and $P_{2}$ proceed independently of each other on
actions in $\overline{L} = \cala \setminus L$, while they have to synchronize on every action in $L$
(parallel composition).

	\end{itemize}

We now characterize two important classes of processes via as many predicates. Firstly, we define
\emph{initial} processes, in which all actions are unexecuted and hence no $\dag$-decoration appears,
through the following predicate:
\cws{0}{\begin{array}{rcl}
\initial(\nil) & & \\
\initial(a \, . \, P) & \!\! \textrm{iff} \!\! & \initial(P) \\
\initial(P_{1} + P_{2}) & \!\! \textrm{iff} \!\! & \initial(P_{1}) \land \initial(P_{2}) \\
\initial(P_{1} \pco{L} P_{2}) & \!\! \textrm{iff} \!\! & \initial(P_{1}) \land \initial(P_{2}) \\
\end{array}}

Secondly, we define \emph{well-formed} processes, whose set we denote by $\calp$, in which both unexecuted
and executed actions can occur in certain circumstances:
\cws{0}{\begin{array}{rcl}
\wf(\nil) & & \\
\wf(a \, . \, P) & \!\! \textrm{iff} \!\! & \initial(P) \\
\wf(a^{\dag \xi} . \, P) & \!\! \textrm{iff} \!\! & \wf(P) \\
\wf(P_{1} + P_{2}) & \!\! \textrm{iff} \!\! & (\wf(P_{1}) \land \initial(P_{2})) \lor (\initial(P_{1}) \land
\wf(P_{2})) \\
\wf(P_{1} \pco{L} P_{2}) & \!\! \textrm{iff} \!\! & \wf(P_{1}) \land \wf(P_{2}) \\
\end{array}}
Well formedness not only imposes that every unexecuted action is followed by an initial process, but also
that in every alternative composition at least one subprocess is initial. Multiple options may arise in the
presence of both alternative and parallel compositions. However, at each occurrence of the former, only the
subprocess chosen for execution can move. Although not selected, the other subprocess is kept as an initial
subprocess within the overall process, in the same way as executed actions are kept inside the
syntax~\cite{BC94,PU07a}, so as to support reversibility. As an example, in $a^{\dag} . \, b \, . \, \nil +
c \, . \, d \, . \, \nil$ the subprocess $c \, . \, d \, . \, \nil$ cannot move because $a$ was selected in
the choice between $a$ and $c$.

It is worth noting that:

	\begin{itemize}

\item $\nil$ is both initial and well-formed.

\item Any initial process is well-formed too.

\item $\calp$ also contains processes that are not initial like, e.g., $a^{\dag} . \, b \, . \, \nil$, which
can either do $b$ or undo $a$.

\item In $\calp$ the relative positions of already executed actions and actions to be executed matter.
Precisely, an action of the former kind can never occur after one of the latter kind. \linebreak For
instance, $a^{\dag} . \, b \, . \, \nil \in \calp$ whereas $b \, . \, a^{\dag} . \, \nil \notin \calp$.

\item In $\calp$ the subprocesses of an alternative composition can be both initial, but cannot be both
non-initial. For example, $a \, . \, \nil + b \, . \, \nil \in \calp$ while $a^{\dag} . \, \nil + b^{\dag} .
\, \nil \notin \calp$.

	\end{itemize}

\blue{In the following, we will sometimes need to remove all $\dag$-decorations from a process, which
corresponds to bring that process back to its initial version.} This is accomplished through function
$\toinitial : \calp \rightarrow \calp_{\rm init}$, with $\calp_{\rm init}$ being the set of initial
processes of $\calp$, which is defined as follows:
\cws{4}{\begin{array}{rcll}
\toinitial(P) & \!\! = \!\! & P & \quad \textrm{if $\initial(P)$} \\
\toinitial(a^{\dag \xi} . \, P') & \!\! = \!\! & a \, . \, \toinitial(P') & \\
\toinitial(P_{1} + P_{2}) & \!\! = \!\! & \toinitial(P_{1}) + \toinitial(P_{2}) & \quad \textrm{if
$\lnot\initial(P_{1}) \lor \lnot\initial(P_{2})$} \\
\toinitial(P_{1} \pco{L} P_{2}) & \!\! = \!\! & \toinitial(P_{1}) \pco{L} \toinitial(P_{2}) & \quad
\textrm{if $\lnot\initial(P_{1}) \lor \lnot\initial(P_{2})$} \\
\end{array}}

%
\subsection{Proved Operational Semantics}
\label{sec:rpc_proved_op_sem}
%

According to~\cite{PU07a}, dynamic operators such as action prefix and alternative composition have to be
made static in the operational semantic rules, so as to retain within the syntax all the information needed
to enable reversibility. Unlike~\cite{PU07a}, we do not generate a forward transition relation and a
backward one, but a single transition relation that we deem to be symmetric in order to enforce the
\emph{loop property}~\cite{DK04}: every executed action can be undone and every undone action can be redone.
A backward transition from $P'$ to~$P$ is subsumed by the corresponding forward transition $t$ from $P$ to
$P'$. As already done in Sections~\ref{sec:config_struct_hhp_bisim} and~\ref{sec:alt_char_config_struct} as
well as in~\cite{DMV90,Bed91}, we will view $t$ as an \emph{outgoing} transition of $P$ when going forward,
while we will view $t$ as an \emph{incoming} transition of $P'$ when going backward.

Following~\cite{BEM24,Esp25}, for the considered extension of PRPC we provide an operational semantics based
on~\cite{DP92}, which is very concrete as every transition is labeled with a \emph{proof
term}~\cite{BC88b,BC94}. This is an action preceded by the sequence of operator symbols in the scope of
which the action occurs inside the source process of the transition. In the case of a binary operator, the
corresponding symbol also specifies whether the action occurs to the left or to the right. In other words, a
proof term describes a path inside a process that uniquely identifies an action. The syntax for the set
$\Theta$ of proof terms is the following:
\cws{0}{\theta \: ::= \: a \mid ._{a} \theta \mid \lplu \theta \mid \rplu \theta \mid \,\, \lpar_{L} \theta
\mid \rpar_{L} \theta \mid \langle \theta, \theta \rangle_{L}}
where $a \in \cala$, $L \subseteq \cala \setminus \{ \tau \}$, and:

\blue{
	\begin{itemize}

\item $._{a} \theta$ pinpoints an action occurring inside the scope of the subprocess of an $a$-action
prefix.

\item $\lplu \theta$ pinpoints an action occurring inside the scope of the left subprocess of an alternative
composition.

\item $\rplu \theta$ pinpoints an action occurring inside the scope of the right subprocess of an
alternative composition.

\item $\lpar_{L}$ pinpoints an action not in $L$ occurring inside the scope of the left subprocess of a
parallel composition.

\item $\rpar_{L}$ pinpoints an action not in $L$ occurring inside the scope of the right subprocess of a
parallel composition.

\item $\langle \theta_{1}, \theta_{2} \rangle_{L}$ pinpoints two equal actions in $L$ that respectively
occur inside the left subprocess and the right subprocess of a parallel composition.

	\end{itemize}
}

The proved operational semantic rules are in Table~\ref{tab:proved_semantics} and generate the proved
labeled transition system $(\calp, \Theta, \! \arrow{}{} \!)$ where $\! \arrow{}{} \! \subseteq \calp \times
\Theta \times \calp$ is the proved transition relation. We denote by $\procs \subsetneq \calp$ the set of
processes that are \emph{reachable} from an initial one via $\! \arrow{}{} \!$. Not all well-formed
processes are reachable; for example, $a^{\dag} . \, \nil \pco{\{ a \}} \nil$ is not reachable from $a \, .
\, \nil \pco{\{ a \}} \nil$ as action $a$ on the left cannot synchronize with any action on the right. From
now on, we consider only~$\procs$ and denote by $\procs_{\rm init}$ the subset of its initial processes.
Every process in~$\procs$ may have several outgoing transitions and, if it is not initial, has at least one
incoming transition.

	\begin{table}[t]

\[
\begin{array}{|ccc|}
\hline
\inferrule*[left=(Act$_{\rm f}$)]{\initial(P)}{a \, . \, P \arrow{a}{} a^{\dag} . \, P} & \qquad &
\inferrule*[left=(Act$_{\rm p}$)]{P \arrow{\theta}{} P'}{a^{\dag \xi} . \, P
\xarrow{{\color{magenta}{._{a}}} \theta}{} a^{\dag \xi} . \, P'} \\[0.2cm]
\inferrule*[left=(Cho$_{\rm l}$)]{P_{1} \arrow{\theta}{} P'_{1} \quad \initial(P_{2})}{P_{1} + P_{2}
\arrow{{\color{magenta}{\lplu}} \: \theta}{} P'_{1} + P_{2}} & &
\inferrule*[left=(Cho$_{\rm r}$)]{P_{2} \arrow{\theta}{} P'_{2} \quad \initial(P_{1})}{P_{1} + P_{2}
\arrow{{\color{magenta}{\rplu}} \: \theta}{} P_{1} + P'_{2}} \\[0.2cm]
\inferrule*[left=(Par$_{\rm l}$)]{P_{1} \arrow{\theta}{} P'_{1} \quad \act(\theta) \notin L}{P_{1} \pco{L}
P_{2} \xarrow{{\color{magenta}{\lpar_{L}}} \theta}{} P'_{1} \pco{L} P_{2}} & &
\inferrule*[left=(Par$_{\rm r}$)]{P_{2} \arrow{\theta}{} P'_{2} \quad \act(\theta) \notin L}{P_{1} \pco{L}
P_{2} \xarrow{{\color{magenta}{\rpar_{L}}} \theta}{} P_{1} \pco{L} P'_{2}} \\[0.2cm]
& & \hspace*{-6.4cm} \inferrule*[left=(Syn)]{P_{1} \arrow{\theta_{1}}{} P'_{1} \quad P_{2}
\arrow{\theta_{2}}{} P'_{2} \quad \act(\theta_{1}) = \act(\theta_{2}) \in L}{P_{1} \pco{L} P_{2}
\xarrow{{\color{magenta}{\langle}} \theta_{1} \! {\color{magenta}{,}} \theta_{2}
{\color{magenta}{\rangle_{L}}}}{} \blue{\enr({\color{black}{P'_{1} \pco{L} P'_{2}}}, \langle \theta_{1},
\theta_{2} \rangle_{L})}} \\
\hline
\end{array}
\]

\caption{Proved operational semantic rules for extended PRPC}
\label{tab:proved_semantics}

	\end{table}

The first rule for action prefix (\textsc{Act}$_{\rm f}$ where f stands for forward), \blue{which} applies
only if $P$ is initial, retains the executed action in the target process of the generated forward
transition by decorating the action itself with $\dag$. The second rule (\textsc{Act}$_{\rm p}$ where p
stands for propagation) propagates actions of inner initial subprocesses by putting an $a$-dot before them
in the label for each outer executed $a$-action prefix that is encountered.

In both rules for alternative composition (\textsc{Cho}$_{\rm l}$ and \textsc{Cho}$_{\rm r}$ where l stands
for left and r~stands for right), the subprocess that has not been selected for execution is retained as an
initial subprocess in the target process of the generated transition. When both subprocesses are initial,
both rules for alternative composition are applicable, otherwise only one of them can be applied and in that
case it is the non-initial subprocess that can move, because the other one has been discarded at the moment
of the selection. The symbol $\lplu$ or $\rplu$ is added at the beginning of the proof term.

\blue{
Using action prefix and alternative composition only, we can build the set $\procs_{\rm seq}$ of
\emph{sequential} processes, i.e., processes in which there are no occurrences of parallel composition. Due
to the $\dag$-decorations of executed actions inside the process syntax, over $\procs_{\rm seq}$ every
non-initial process has exactly one incoming transition, proved labeled transition systems turn out to be
trees, and well formedness coincides with reachability.
}

	\begin{exa}\label{ex:single_incom_trans}

The proved labeled transition system underlying the initial sequential process $a \, . \, \nil$ has a single
transition $a \, . \, \nil \arrow{a}{} a^{\dag} . \, \nil$. In contrast, the proved labeled transition
system underlying the initial sequential process $a \, . \, \nil + a \, . \, \nil$ has the two transitions
$a \, . \, \nil + a \, . \, \nil \arrow{\lplu a}{} a^{\dag} . \, \nil + a \, . \, \nil$ and $a \, . \, \nil
+ a \, . \, \nil \arrow{\rplu a}{} a \, . \, \nil + a^{\dag} . \, \nil$. Note that the two target processes
are different from each other due to the presence of action decorations, whereas a single $a$-transition
from $a \, . \, \nil + a \, . \, \nil$ to $\nil$ would be generated in the setting of a forward-only process
calculus.
\fullbox

	\end{exa}

The three rules for parallel composition use partial function $\act : \Theta \parfun \cala$ to extract an
action from a proof term $\theta$. This function, which will be used throughout the paper, \linebreak is
defined by induction on the syntactical structure of $\theta$ as follows:
\cws{0}{\begin{array}{rcl}
\act(a) & \!\! = \!\! & a \\
\act(._{a} \theta') & \!\! = \!\! & \act(\theta') \\
\act(\lplu \theta') \: = \: \act(\hspace{-0.04cm} \rplu \theta') & \!\! = \!\! & \act(\theta') \\
\act(\lpar_{L} \theta') \: = \: \act(\rpar_{L} \theta') & \!\! = \!\! & \act(\theta') \\
\act(\langle \theta_{1}, \theta_{2} \rangle_{L}) & \!\! = \!\! & \left\{ \begin{array}{ll}
\act(\theta_{1}) & \quad \textrm{if $\act(\theta_{1}) = \act(\theta_{2})$} \\
\textrm{undefined} & \quad \textrm{otherwise} \\
\end{array} \right. \\
\end{array}}
In the first two rules (\textsc{Par}$_{\rm l}$ and \textsc{Par}$_{\rm r}$), a single subprocess proceeds by
performing an action not belonging to~$L$, with $\lpar_{L}$ or $\rpar_{L}$ being placed at the beginning of
the proof term. In the third rule (\textsc{Syn}), both subprocesses synchronize on an action in $L$ and the
resulting proof term contains both individual proof terms. If $L = \emptyset$ or $L = \cala \setminus \{
\tau \}$, then the two subprocesses are fully independent or fully synchronized, respectively, on observable
actions. \blue{When writing $\act(\theta) \notin L$ or $\act(\theta_{1}) = \act(\theta_{2}) \in L$, it is
understood that $\act$ is defined over those proof terms.}

\blue{The novelty with respect to PRPC is that} the natural target process $P'_{1} \pco{L} P'_{2}$ of a
synchronization has to be suitably manipulated in rule \textsc{Syn} to correctly reflect causality and
concurrency. More precisely, the $\dag$-decoration of every executed action participating in the
synchronization has to be enriched with a proof term of the form $\langle \theta_{1}, \theta_{2}
\rangle_{L}$. This is accomplished by taking $\enr(P'_{1} \pco{L} P'_{2}, \langle \theta_{1}, \theta_{2}
\rangle_{L}) = \enr'(P'_{1} \pco{L} P'_{2}, \langle \theta_{1}, \theta_{2} \rangle_{L}, \langle \theta_{1},
\theta_{2} \rangle_{L})$ as target process of a synchronization. The partial function $\enr' : \procs \times
\Theta \times \Theta \parfun \procs$ is defined by induction on the syntactical structure of its first
argument $P \in \procs$ as follows, \blue{where its second argument $\theta \in \Theta$ is used to pinpoint
the synchronizing actions while its third argument $\bar{\theta} \in \Theta$, which does not change
throughout the induction, is used to further decorate those actions}:
\cws{6}{\begin{array}{rcl}
\enr'(\nil, \theta, \bar{\theta}) & \!\! = \!\! & \nil \\
\enr'(a \, . \, P', \theta, \bar{\theta}) & \!\! = \!\! & \textrm{undefined} \\
\enr'(a^{\dag \xi} . \, P', \theta, \bar{\theta}) & \!\! = \!\! & \left\{ \begin{array}{ll}
a^{\dag {\color{blue}{\bar{\theta}}}} . \, P' & \quad \textrm{if $\theta = a$} \\
a^{\dag \xi} . \, \enr'(P', \theta', \bar{\theta}) & \quad \textrm{if $\theta = ._{a} \theta'$} \\
\textrm{undefined} & \quad \textrm{otherwise} \\
\end{array} \right. \\
\enr'(P_{1} + P_{2}, \theta, \bar{\theta}) & \!\! = \!\! & \left\{ \begin{array}{ll}
\enr'(P_{1}, \theta', \bar{\theta}) + P_{2} & \quad \textrm{if $\theta = \lplu \theta'$} \\
P_{1} + \enr'(P_{2}, \theta', \bar{\theta}) & \quad \textrm{if $\theta = \rplu \theta'$} \\
\textrm{undefined} & \quad \textrm{otherwise} \\
\end{array} \right. \\
\enr'(P_{1} \pco{L} P_{2}, \theta, \bar{\theta}) & \!\! = \!\! & \left\{ \begin{array}{ll}
\enr'(P_{1}, \theta', \bar{\theta}) \pco{L} P_{2} & \quad \textrm{if $\theta = \, \lpar_{L} \theta'$} \\
P_{1} \pco{L} \enr'(P_{2}, \theta', \bar{\theta}) & \quad \textrm{if $\theta = \rpar_{L} \theta'$} \\
\enr'(P_{1}, \theta_{1}, \bar{\theta}) \pco{L} \enr'(P_{2}, \theta_{2}, \bar{\theta}) & \quad \textrm{if
$\theta = \langle \theta_{1}, \theta_{2} \rangle_{L}$} \\
\textrm{undefined} & \quad \textrm{otherwise} \\
\end{array} \right. \\
\end{array}}

	\begin{exa}\label{ex:autoconc_autocaus}

The proved labeled transition system of the initial process $(a \, . \, \nil \pco{\emptyset} a \, . \, \nil)
\pco{\{ a \}} \linebreak a \, . \, a \, . \, \nil$, which is the synchronization of autoconcurrency with
autocausation, has the following two maximal transition sequences, \blue{where maximality refers to sequence
length}:

		\begin{itemize}

\item $(a \, . \, \nil \pco{\emptyset} a \, . \, \nil) \pco{\{ a \}} a \, . \, a \, . \, \nil \\
\hspace*{-0.1cm} \xarrow{{\color{violet}{\langle \lpar_{\emptyset} a, a \rangle_{\{ a \}}}}}{} \! (a^{\dag
\color{violet}{\langle \lpar_{\emptyset} a, a \rangle_{\{ a \}}}} . \, \nil \pco{\emptyset} a \, . \, \nil)
\pco{\{ a \}} a^{\dag \color{violet}{\langle \lpar_{\emptyset} a, a \rangle_{\{ a \}}}} . \, a \, . \, \nil
\\
\hspace*{-0.1cm} \xarrow{{\color{cyan}{\langle \rpar_{\emptyset} a, ._{a} a \rangle_{\{ a \}}}}}{} \!
(a^{\dag \color{violet}{\langle \lpar_{\emptyset} a, a \rangle_{\{ a \}}}} . \, \nil \pco{\emptyset} a^{\dag
\color{cyan}{\langle \rpar_{\emptyset} a, ._{a} a \rangle_{\{ a \}}}} . \, \nil) \pco{\{ a \}} a^{\dag
\color{violet}{\langle \lpar_{\emptyset} a, a \rangle_{\{ a \}}}} . \, a^{\dag \color{cyan}{\langle
\rpar_{\emptyset} a, ._{a} a \rangle_{\{ a \}}}} . \, \nil$

\item $(a \, . \, \nil \pco{\emptyset} a \, . \, \nil) \pco{\{ a \}} a \, . \, a \, . \, \nil \\
\hspace*{-0.1cm} \xarrow{{\color{brown}{\langle \rpar_{\emptyset} a, a \rangle_{\{ a \}}}}}{} \! (a \, . \,
\nil \pco{\emptyset} a^{\dag \color{brown}{\langle \rpar_{\emptyset} a, a \rangle_{\{ a \}}}} . \, \nil)
\pco{\{ a \}} a^{\dag \color{brown}{\langle \rpar_{\emptyset} a, a \rangle_{\{ a \}}}} . \, a \, . \, \nil
\\
\hspace*{-0.1cm} \xarrow{{\color{orange}{\langle \lpar_{\emptyset} a, ._{a} a \rangle_{\{ a \}}}}}{} \!
(a^{\dag {\color{orange}{\langle \lpar_{\emptyset} a, ._{a} a \rangle_{\{ a \}}}}} . \, \nil \pco{\emptyset}
a^{\dag \color{brown}{\langle \rpar_{\emptyset} a, a \rangle_{\{ a \}}}} . \, \nil) \pco{\{ a \}} a^{\dag
\color{brown}{\langle \rpar_{\emptyset} a, a \rangle_{\{ a \}}}} . \, a^{\dag {\color{orange}{\langle
\lpar_{\emptyset} a, ._{a} a \rangle_{\{ a \}}}}} . \, \nil$

		\end{itemize}

\noindent
Note that the target processes of the two sequences are different thanks to the different additional
decorations of the pairs of synchronizing executed actions. Without those decorations, the two sequences
would end up in the same process $(a^{\dag} . \, \nil \pco{\emptyset} a^{\dag} . \, \nil) \pco{\{ a \}}
a^{\dag} . \, a^{\dag} . \, \nil$ -- thus yielding a diamond-shaped transition system -- which would not
reflect the fact that the two executed $a$-actions in $a^{\dag} . \, a^{\dag} . \, \nil$ cannot be undone in
any order as the first one causes the second one.
\fullbox

	\end{exa}

	\begin{exa}\label{ex:autoconc_autoconc}

The proved labeled transition system of the initial process $(a \, . \, \nil \pco{\emptyset} a \, . \, \nil)
\pco{\{ a \}} \linebreak (a \, . \, \nil \pco{\emptyset} a \, . \, \nil)$, which is the synchronization of
autoconcurrency with itself, has the following four maximal transition sequences:

		\begin{itemize}

\item $(a \, . \, \nil \pco{\emptyset} a \, . \, \nil) \pco{\{ a \}} (a \, . \, \nil \pco{\emptyset} a \, .
\, \nil) \\
\xarrow{{\color{violet}{\langle \lpar_{\emptyset} a, \lpar_{\emptyset} a \rangle_{\{ a \}}}}}{} (a^{\dag
\color{violet}{\langle \lpar_{\emptyset} a, \lpar_{\emptyset} a \rangle_{\{ a \}}}} . \, \nil
\pco{\emptyset} a \, . \, \nil) \pco{\{ a \}} (a^{\dag \color{violet}{\langle \lpar_{\emptyset} a,
\lpar_{\emptyset} a \rangle_{\{ a \}}}} . \, \nil \pco{\emptyset} a \, . \, \nil) \\
\xarrow{{\color{cyan}{\langle \rpar_{\emptyset} a, \rpar_{\emptyset} a \rangle_{\{ a \}}}}}{} (a^{\dag
\color{violet}{\langle \lpar_{\emptyset} a, \lpar_{\emptyset} a \rangle_{\{ a \}}}} . \, \nil
\pco{\emptyset} a^{\dag \color{cyan}{\langle \rpar_{\emptyset} a, \rpar_{\emptyset} a \rangle_{\{ a \}}}} .
\, \nil) \pco{\{ a \}}
(a^{\dag \color{violet}{\langle \lpar_{\emptyset} a, \lpar_{\emptyset} a \rangle_{\{ a \}}}} . \, \nil
\pco{\emptyset} a^{\dag \color{cyan}{\langle \rpar_{\emptyset} a, \rpar_{\emptyset} a \rangle_{\{ a \}}}} .
\, \nil)$

\item $(a \, . \, \nil \pco{\emptyset} a \, . \, \nil) \pco{\{ a \}} (a \, . \, \nil \pco{\emptyset} a \, .
\, \nil) \\
\xarrow{{\color{cyan}{\langle \rpar_{\emptyset} a, \rpar_{\emptyset} a \rangle_{\{ a \}}}}}{} (a \, . \,
\nil \pco{\emptyset} a^{\dag \color{cyan}{\langle \rpar_{\emptyset} a, \rpar_{\emptyset} a \rangle_{\{ a
\}}}} . \, \nil) \pco{\{ a \}} (a \, . \, \nil \pco{\emptyset} a^{\dag \color{cyan}{\langle
\rpar_{\emptyset} a, \rpar_{\emptyset} a \rangle_{\{ a \}}}} . \, \nil) \\
\xarrow{{\color{violet}{\langle \lpar_{\emptyset} a, \lpar_{\emptyset} a \rangle_{\{ a \}}}}}{} (a^{\dag
\color{violet}{\langle \lpar_{\emptyset} a, \lpar_{\emptyset} a \rangle_{\{ a \}}}} . \, \nil
\pco{\emptyset} a^{\dag \color{cyan}{\langle \rpar_{\emptyset} a, \rpar_{\emptyset} a \rangle_{\{ a \}}}} .
\, \nil) \pco{\{ a \}}
(a^{\dag \color{violet}{\langle \lpar_{\emptyset} a, \lpar_{\emptyset} a \rangle_{\{ a \}}}} . \, \nil
\pco{\emptyset} a^{\dag \color{cyan}{\langle \rpar_{\emptyset} a, \rpar_{\emptyset} a \rangle_{\{ a \}}}} .
\, \nil)$

\item $(a \, . \, \nil \pco{\emptyset} a \, . \, \nil) \pco{\{ a \}} (a \, . \, \nil \pco{\emptyset} a \, .
\, \nil) \\
\xarrow{{\color{brown}{\langle \lpar_{\emptyset} a, \rpar_{\emptyset} a \rangle_{\{ a \}}}}}{} (a^{\dag
\color{brown}{\langle \lpar_{\emptyset} a, \rpar_{\emptyset} a \rangle_{\{ a \}}}} . \, \nil \pco{\emptyset}
a \, . \, \nil) \pco{\{ a \}} (a \, . \, \nil \pco{\emptyset} a^{\dag \color{brown}{\langle
\lpar_{\emptyset} a, \rpar_{\emptyset} a \rangle_{\{ a \}}}} . \, \nil) \\
\xarrow{{\color{orange}{\langle \rpar_{\emptyset} a, \lpar_{\emptyset} a \rangle_{\{ a \}}}}}{} (a^{\dag
\color{brown}{\langle \lpar_{\emptyset} a, \rpar_{\emptyset} a \rangle_{\{ a \}}}} . \, \nil \pco{\emptyset}
a^{\dag \color{orange}{\langle \rpar_{\emptyset} a, \lpar_{\emptyset} a \rangle_{\{ a \}}}} . \, \nil)
\pco{\{ a \}}
(a^{\dag \color{orange}{\langle \rpar_{\emptyset} a, \lpar_{\emptyset} a \rangle_{\{ a \}}}} . \, \nil
\pco{\emptyset} a^{\dag \color{brown}{\langle \lpar_{\emptyset} a, \rpar_{\emptyset} a \rangle_{\{ a \}}}} .
\, \nil)$

\item $(a \, . \, \nil \pco{\emptyset} a \, . \, \nil) \pco{\{ a \}} (a \, . \, \nil \pco{\emptyset} a \, .
\, \nil) \\
\xarrow{{\color{orange}{\langle \rpar_{\emptyset} a, \lpar_{\emptyset} a \rangle_{\{ a \}}}}}{} (a \, . \,
\nil \pco{\emptyset} a^{\dag \color{orange}{\langle \rpar_{\emptyset} a, \lpar_{\emptyset} a \rangle_{\{ a
\}}}} . \, \nil) \pco{\{ a \}} (a^{\dag \color{orange}{\langle \rpar_{\emptyset} a, \lpar_{\emptyset} a
\rangle_{\{ a \}}}} . \, \nil \pco{\emptyset} a \, . \, \nil) \\
\xarrow{{\color{brown}{\langle \lpar_{\emptyset} a, \rpar_{\emptyset} a \rangle_{\{ a \}}}}}{} (a^{\dag
\color{brown}{\langle \lpar_{\emptyset} a, \rpar_{\emptyset} a \rangle_{\{ a \}}}} . \, \nil \pco{\emptyset}
a^{\dag \color{orange}{\langle \rpar_{\emptyset} a, \lpar_{\emptyset} a \rangle_{\{ a \}}}} . \, \nil)
\pco{\{ a \}}
(a^{\dag \color{orange}{\langle \rpar_{\emptyset} a, \lpar_{\emptyset} a \rangle_{\{ a \}}}} . \, \nil
\pco{\emptyset} a^{\dag \color{brown}{\langle \lpar_{\emptyset} a, \rpar_{\emptyset} a \rangle_{\{ a \}}}} .
\, \nil)$

		\end{itemize}

\noindent
While the target processes of the first (resp.\ last) two sequences are equal, the target process of the
first two sequences is different from the one of the last two sequences due to the different additional
decorations of the pairs of synchronizing executed actions. This results in a transition system \blue{whose
shape consists of two alternative diamonds} like the one of $(a \, . \, \nil \pco{\emptyset} a \, . \, \nil)
+ (a \, . \, \nil \pco{\emptyset} a \, . \, \nil)$. Without those decorations, the four sequences would end
up in the same process $(a^{\dag} . \, \nil \pco{\emptyset} a^{\dag} . \, \nil) \pco{\{ a \}} (a^{\dag} . \,
\nil \pco{\emptyset} a^{\dag} . \, \nil)$, thus yielding a transition system \blue{whose shape is a single
diamond}.
\fullbox

	\end{exa}

%
\subsection{Forward-Reverse Bisimilarity and Backward Ready Multisets}
\label{sec:frb_brm}
%

We now redefine brm-forward-reverse bisimilarity over $\procs$. Unlike stable configuration structures, in
the case of processes we can syntactically construct their (finite) backward ready multisets, intended as
the multisets of actions occurring in the labels of their incoming transitions. In the following, we use
$\oplus$ for multiset union, which adds multiplicities of identical elements, and $\otimes$ for multiset
intersection, which multiplies the multiplicities of those elements.

	\begin{defi}\label{def:brm_proc}

\blue{
The \emph{backward ready multiset} of $P \in \procs$ is inductively defined as follows:
\cws{0}{\begin{array}{rcl}
\brm(\nil) & \!\! = \!\! & \emptyset \\
\brm(a \, . \, P') & \!\! = \!\! & \emptyset \\
\brm(a^{\dag \xi} . \, P') & \!\! = \!\! & \left\{ \begin{array}{ll}
\lmp a \rmp & \quad \textrm{if $\initial(P')$} \\
\brm(P') & \quad \textrm{if $\lnot\initial(P')$} \\
\end{array} \right. \\
\brm(P_{1} + P_{2}) & \!\! = \!\! & \left\{ \begin{array}{ll}
\emptyset & \quad \textrm{if $\initial(P_{1}) \land \initial(P_{2})$} \\
\brm(P_{1}) & \quad \textrm{if $\lnot\initial(P_{1}) \land \initial(P_{2})$} \\
\brm(P_{2}) & \quad \textrm{if $\initial(P_{1}) \land \lnot\initial(P_{2})$} \\
\end{array} \right. \\
\brm(P_{1} \pco{L} P_{2}) & \!\! = \!\! & (\brm(P_{1}) \otimes \overline{L}) \oplus (\brm(P_{2}) \otimes
\overline{L}) \oplus (\brm(P_{1}) \otimes \brm(P_{2}) \otimes L) \\
\end{array}}
where $\overline{L} = \cala \setminus L$.
\fullbox
}

	\end{defi}

\blue{
For example, $\brm(a^{\dag} . \, \nil \pco{\emptyset} a^{\dag} . \, \nil) = \lmp a, a \rmp$ whereas
$\brm(a^{\dag} . \, a^{\dag} . \, \nil + a \, . \, a \, . \, \nil) = \lmp a \rmp$.
}

The first two clauses stated below are analogous to the ones of forward-reverse bisimilarity $\sbis{\rm
FRB}$ over a single transition relation defined in~\cite{BEM24,Esp25}. \blue{The difference is given by} the
use of function $\act$ to abstract from operator symbols inside transition labels.

	\begin{defi}\label{def:frb_brm}

We say that two processes $P_{1}, P_{2} \in \procs$ are \emph{brm-forward-reverse bisimilar}, written $P_{1}
\sbis{\rm FRB:brm} P_{2}$, iff $P_{1}$ and $P_{2}$ are related by a brm-forward-reverse bisimulation, i.e.,
a symmetric relation $\calb$ over $\procs$ such that, whenever $(Q_{1}, Q_{2}) \in \calb$, then:

		\begin{itemize}

\item For each $Q_{1} \arrow{\theta_{1}}{} Q'_{1}$ there exists $Q_{2} \arrow{\theta_{2}}{} Q'_{2}$ such
that $\act(\theta_{1}) = \act(\theta_{2})$ and $(Q'_{1}, Q'_{2}) \in \calb$.

\item For each $Q'_{1} \arrow{\theta_{1}}{} Q_{1}$ there exists $Q'_{2} \arrow{\theta_{2}}{} Q_{2}$ such
that $\act(\theta_{1}) = \act(\theta_{2})$ and $(Q'_{1}, Q'_{2}) \in \calb$.

\item $\brm(Q_{1}) = \brm(Q_{2})$.
\fullbox

		\end{itemize}

	\end{defi}

	\begin{exa}\label{ex:auto_frb_brm}

$a \, . \, \nil \pco{\emptyset} a \, . \, \nil \not\sbis{\rm FRB:brm} a \, . \, a \, . \, \nil$ because in
the forward bisimulation game they respectively reach $a^{\dag} . \, \nil \pco{\emptyset} a^{\dag} \, . \,
\nil$ and $a^{\dag} . \, a^{\dag} . \, \nil$ after performing two $a$-transitions, where $\brm(a^{\dag} .
\, \nil \pco{\emptyset} a^{\dag} \, . \, \nil) = \lmp a, a \rmp \neq \lmp a \rmp = \brm(a^{\dag} . \,
a^{\dag} . \, \nil)$. Likewise, $(a \, . \, \nil \pco{\emptyset} a \, . \, \nil) \pco{\{ a \}} a \, . \, a
\, . \, \nil \linebreak \not\sbis{\rm FRB:brm} (a \, . \, \nil \pco{\emptyset} a \, . \, \nil)$. In
contrast, $(a \, . \, \nil \pco{\emptyset} a \, . \, \nil) \pco{\{ a \}} (a \, . \, \nil \pco{\emptyset} a
\, . \, \nil) \sbis{\rm FRB:brm} (a \, . \, \nil \pco{\emptyset} a \, . \, \nil) + (a \, . \, \nil
\pco{\emptyset} a \, . \, \nil) \sbis{\rm FRB:brm} a \, . \, \nil \pco{\emptyset} a \, . \, \nil$.
\fullbox

	\end{exa}

An axiomatization of $\sbis{\rm FRB:brm}$ over $\procs$ can be derived from the one of $\sbis{\rm FRB}$
in~\cite{BEM24}. The idea is to use backward ready multisets instead of backward ready sets \blue{during the
procedure in which all action prefixes inside a process are extended with suitable additional information,
so as to distinguish, e.g., $a \, . \, \nil \pco{\emptyset} b \, . \, \nil$ from $a \, . \, b \, . \, \nil +
b \, . \, a \, . \, \nil$}.

%
\subsection{Denotational Semantics on Stable Configuration Structures}
\label{sec:den_sem_scs}
%

To enable a comparison between hereditary history-preserving bisimilarity over stable configuration
structures and brm-forward-reverse bisimilarity over processes, we proceed with the introduction of a
denotational semantics for $\procs$ based on stable configuration structures. The first step consists
\blue{in} redefining the process operators of Section~\ref{sec:rpc_syntax} over stable configuration
structures. Taking inspiration from~\cite{BC94}, we do this by using proof terms in $\Theta$ to formalize
events:

	\begin{itemize}

\item The terminated stable configuration structure $\emph{\ssfn}$ is defined as $(\emptyset, \{ \emptyset
\}, \emptyset)$.

\item Let $a \in \cala$ and $\emph{\textsf{C}} = (\cale, \calc, \ell)$ be a stable configuration structure
such that $\cale \subseteq \Theta$. \linebreak The action prefix $a \, . \, \emph{\textsf{C}}$ is defined as
$(\cale', \calc', \ell')$ where:

		\begin{itemize}

\item $\cale' \, = \, \{ a \} \cup \{ ._{a} \theta \mid \theta \in \cale \}$.

\item $\calc' \, = \, \{ \emptyset \} \cup \{ X' \in \powerset_{\rm fin}(\cale') \mid \exists X \in \calc
\ldotp X' = \{ a \} \cup \{ ._{a} \theta \mid \theta \in X \} \}$.

\item $\ell' \, = \, \{ (a, a) \} \cup \{ (._{a} \theta, \act(._{a} \theta)) \mid \exists X \in \calc \ldotp
\theta \in X \}$.

		\end{itemize}

\item Let $\emph{\textsf{C}}_{i} = (\cale_{i}, \calc_{i}, \ell_{i})$ be a stable configuration structure
such that $\cale_{i} \subseteq \Theta$ for $i \in \{ 1, 2 \}$. \linebreak The alternative composition
$\emph{\textsf{C}}_{1} + \emph{\textsf{C}}_{2}$ is defined as $(\cale, \calc, \ell)$ where:

		\begin{itemize}

\item $\cale \, = \, \{ \lplu \theta \mid \theta \in \cale_{1} \} \cup \{ \hspace{-0.02cm} \rplu \theta \mid
\theta \in \cale_{2} \}$.

\item $\calc \, = \, \{ X \in \powerset_{\rm fin}(\cale) \mid \exists X_{1} \in \calc_{1} \ldotp X = \{
\lplu \theta \mid \theta \in X_{1} \} \} \cup \\
\hspace*{0.86cm} \{ X \in \powerset_{\rm fin}(\cale) \mid \exists X_{2} \in \calc_{2} \ldotp X = \{
\hspace{-0.02cm} \rplu \theta \mid \theta \in X_{2} \} \}$.

\item $\ell \, = \, \{ (\hspace{0.01cm} \lplu \theta, \act(\hspace{0.01cm} \lplu \theta)) \mid \exists X_{1}
\in \calc_{1} \ldotp \theta \in X_{1} \} \cup \\
\hspace*{0.8cm} \{ (\hspace{-0.01cm} \rplu \theta, \hspace{-0.02cm} \act(\rplu \theta)) \mid \exists X_{2}
\in \calc_{2} \ldotp \theta \in X_{2} \}$.

		\end{itemize}

\item Let $\emph{\textsf{C}}_{i} = (\cale_{i}, \calc_{i}, \ell_{i})$ be a stable configuration structure
such that $\cale_{i} \subseteq \Theta$ for $i \in \{ 1, 2 \}$ and $L \subseteq \cala \setminus \{ \tau \}$.
The parallel composition $\emph{\textsf{C}}_{1} \pco{L} \emph{\textsf{C}}_{2}$ is defined as $(\cale, \calc,
\ell)$ where:

		\begin{itemize}

\item $\cale \, = \, \{ \lpar_{L} \theta \mid \theta \in \cale_{1} \land \act(\theta) \notin L \} \cup \{
\hspace{-0.02cm} \rpar_{L} \theta \mid \theta \in \cale_{2} \land \act(\theta) \notin L \} \cup \\
\hspace*{0.87cm} \{ \langle \theta_{1}, \theta_{2} \rangle_{L} \mid \theta_{1} \in \cale_{1} \land
\theta_{2} \in \cale_{2} \land \act(\theta_{1}) = \act(\theta_{2}) \in L \}$.

\item $\calc \, = \, \{ X \in \powerset_{\rm fin}(\cale) \mid \proj_{1}(X) \in \calc_{1} \land \proj_{2}(X)
\in \calc_{2} \land \forall e, e' \in X \ldotp \\
\hspace*{1.0cm} ((\proj_{1}(\{ e \}) = \proj_{1}(\{ e' \}) \neq \emptyset \lor \proj_{2}(\{ e \}) =
\proj_{2}(\{ e' \}) \neq \emptyset) \Longrightarrow \\
\hspace*{1.2cm} e = e') \, \land \hspace{0.37cm} \textrm{[local injectivity of projections]} \\
\hspace*{1.0cm} (e \neq e' \Longrightarrow \hspace{0.23cm} \textrm{[coincidence freeness (a single event per
transition)]} \\
\hspace*{1.2cm} \exists Y \subseteq X \ldotp (\proj_{1}(Y) \in \calc_{1} \land \proj_{2}(Y) \in \calc_{2}
\land (e \in Y \Longleftrightarrow e' \notin Y))) \}$ \\
with projections being defined as follows:

			\begin{itemize}

\item $\proj_{1}(X) \, = \, \{ \theta_{1} \in \cale_{1} \mid \, \lpar_{L} \theta_{1} \in X \lor \exists
\theta_{2} \in \cale_{2} \ldotp \langle \theta_{1}, \theta_{2} \rangle_{L} \in X \}$.

\item $\proj_{2}(X) \, = \, \{ \theta_{2} \in \cale_{2} \mid \hspace{-0.02cm} \rpar_{L} \theta_{2} \in X
\lor \exists \theta_{1} \in \cale_{1} \ldotp \langle \theta_{1}, \theta_{2} \rangle_{L} \in X \}$.

			\end{itemize}

\item $\ell \, = \, \{ (\lpar_{L} \theta, \act(\lpar_{L} \theta)) \mid \exists X_{1} \in \calc_{1} \ldotp
\theta \in X_{1} \land \act(\theta) \notin L \} \cup \\
\hspace*{0.81cm} \{ (\hspace{-0.02cm} \rpar_{L} \theta, \act(\rpar_{L} \theta)) \mid \exists X_{2} \in
\calc_{2} \ldotp \theta \in X_{2} \land \act(\theta) \notin L \} \cup \\
\hspace*{0.81cm} \{ (\langle \theta_{1}, \theta_{2} \rangle_{L}, \act(\langle \theta_{1}, \theta_{2}
\rangle_{L})) \mid \exists X_{1} \in \calc_{1} \ldotp \exists X_{2} \in \calc_{2} \ldotp \\
\hspace*{6.0cm} \theta_{1} \in X_{1} \land \theta_{2} \in X_{2} \land \act(\theta_{1}) = \act(\theta_{2})
\in L \}$.

		\end{itemize}

	\end{itemize}

Our construction reflects a deliberate attempt to balance the fine-grained causal information needed by
hereditary history-preserving bisimilarity with the syntactic discipline imposed by the reversible calculus.
This requires treating occurrences of executed actions not simply as nodes in a partial order, but as
evidence of how alternative or parallel subprocesses have been selected. The key design choice is to let
proof terms determine the identity and causal placement of events. Each transition of the process calculus
introduces exactly one event in the configuration structure, with the shape of its proof term determining
how the event embeds into the partial order: prefix constructors contribute subject causality~\cite{BS98},
choice constructors determine branching structure, and the synchronization constructor $(\langle \theta_{1},
\theta_{2} \rangle_{L}$) inserts the necessary causal links between the two participating sides, i.e.,
object causality~\cite{BS98}. In this way, the construction makes concurrency and causality explicit,
without resorting to artificial identifiers: the structure of the proof term is the identifier.

With each process $P \in \procs$ we then denotationally associate a stable configuration structure semantics
in a way similar to~\cite{Win86,AC20}, with the notable difference that we represent events via proof terms.
More precisely, each process is given a pair formed by a stable configuration structure, built by using the
operators above, and a configuration of that structure. The idea is that all processes reachable from the
same initial process share the same configuration structure. In contrast, the designated configuration
uniquely identifies the specific process through the proof terms labeling a sequence of proved transitions
by means of which the considered process is reached from the initial one.

Note that such a sequence is empty if $P$ is initial -- which corresponds to the empty configuration -- and
unique if $P$ is sequential. In the case that $P$ is neither initial nor sequential, if there are several
transition sequences reaching it -- meaning that non-synchronizing actions of different parallel
subprocesses have been executed -- then they result in the same configuration~\cite{BC94}, because
independent actions can be executed in any order and the order of the elements within a configuration --
which is a set -- does not matter.

	\begin{defi}\label{def:scs_sem}

The \emph{stable configuration structure semantics} of $P \in \procs$ is the pair $\lsp P \rsp =
(\emph{\ssfc}_{P}, X_{P})$ where:

		\begin{itemize}

\item $\emph{\ssfc}_{P} = \scs(\toinitial(P))$, with the stable configuration structure $\scs(Q)$ associated
with an initial process $Q \in \procs$ being defined by induction on the syntactical structure of $Q$
\linebreak as follows:

			\begin{itemize}

\item $\scs(\nil) = \emph{\textsf{N}}$.

\item $\scs(a \, . \, Q') = a \, . \, \scs(Q')$.

\item $\scs(Q_{1} + Q_{2}) = \scs(Q_{1}) + \scs(Q_{2})$.

\item $\scs(Q_{1} \pco{L} Q_{2}) = \scs(Q_{1}) \pco{L} \scs(Q_{2})$.

			\end{itemize}

\item $X_{P} = \emptyset$ if $P$ is initial, otherwise $X_{P} = \{ \theta_{i} \mid 1 \le i \le n \}$ for
some $n \in \natns_{\ge 1}$ such that there exists $P_{i - 1} \arrow{\theta_{i}}{} P_{i}$ for all $1 \le i
\le n$ with $P_{0} = \toinitial(P)$ and $P_{n} = P$.
\fullbox

		\end{itemize}

	\end{defi}

\blue{
We conclude by proving that $\emph{\ssfc}_{P}$ is indeed a stable configuration structure.

	\begin{thm}\label{thm:cp_stable_config_struct}

Let $P \in \procs$. Then $\emph{\ssfc}_{P}$ is a stable configuration structure.

		\begin{proof}

Since $\emph{\ssfc}_{P} = \scs(\toinitial(P))$, it suffices to prove by structural induction on the initial
process $Q = \toinitial(P)$ that $\scs(Q)$ is stable.
 
			\begin{itemize}

\item If $Q$ is $\nil$ then $\scs(\nil) = (\emptyset, \{ \emptyset \}, \emptyset)$ is trivially rooted,
connected, and closed under bounded unions and intersections.

\item Let $Q$ be $a \, . \, Q'$ and suppose that $\scs(Q') = (\cale, \calc, \ell)$ is a stable configuration
structure. Then the configurations of $\scs(Q)$ are $\emptyset$ and those of the form $X_{a} = \{ a \} \cup
\{ ._{a} \theta \mid \theta \in X \}$ for some $X \in \calc$, hence $\scs(Q)$ is rooted and connected.
Observing that $\emptyset \cup X_{a} = X_{a}$ and $\emptyset \cap X_{a} = \emptyset$ with $\emptyset \cup
X_{a} \subseteq X_{a}$, if $X_{a} \cup Y_{a} \subseteq Z_{a}$ then $X_{a} \cup Y_{a} = (\{ a \} \cup \{
._{a} \theta \mid \theta \in X \}) \cup \linebreak (\{ a \} \cup \{ ._{a} \theta \mid \theta \in Y \}) = \{
a \} \cup \{ ._{a} \theta \mid \theta \in X \cup Y \}$ and $X_{a} \cap Y_{a} = (\{ a \} \cup \{ ._{a} \theta
\mid \theta \in X \}) \cap \linebreak (\{ a \} \cup \{ ._{a} \theta \mid \theta \in Y \}) = \{ a \} \cup \{
._{a} \theta \mid \theta \in X \cap Y \}$ are configurations of $\scs(Q)$ because $X \cup Y$ and $X \cap Y$
are configurations of $\scs(Q')$ by the induction hypothesis (note that $X \cup Y \subseteq Z$).

\item Let $Q$ be $Q_{1} + Q_{2}$ and suppose that $\scs(Q_{i}) = (\cale_{i}, \calc_{i}, \ell_{i})$ is a
stable configuration structure for $i \in \{ 1, 2 \}$. Then the configurations of $\scs(Q)$ are of the form
$X_{\rm l} = \{ \lplu \theta \mid \theta \in X_{1} \}$ for some $X_{1} \in \calc_{1}$ or $X_{\rm r} = \{
\rplu \theta \mid \theta \in X_{2} \}$ for some $X_{2} \in \calc_{2}$, which comprise $\emptyset$, hence
$\scs(Q)$ is rooted and connected. Observing that in order for the union of two configurations to be
included in a configuration it must be the case that the two configurations are either of the form $X_{\rm
l}$ and $Y_{\rm l}$ or of the form $X_{\rm r}$ and $Y_{\rm r}$, if $X_{\rm l} \cup Y_{\rm l} \subseteq
Z_{\rm l}$ then $X_{\rm l} \cup Y_{\rm l} = \{ \lplu \theta \mid \theta \in X_{1} \} \linebreak \cup \{
\lplu \theta \mid \theta \in Y_{1} \} = \{ \lplu \theta \mid \theta \in X_{1} \cup Y_{1} \}$ and $X_{\rm l}
\cap Y_{\rm l} = \{ \lplu \theta \mid \theta \in X_{1} \} \cap \{ \lplu \theta \mid \theta \in Y_{1} \} = \{
\lplu \theta \mid \theta \in X_{1} \cap Y_{1} \}$ are configurations of $\scs(Q)$ because $X_{1} \cup Y_{1}$
and $X_{1} \cap Y_{1}$ are configurations of $\scs(Q_{1})$ by the induction hypothesis (note that $X_{1}
\cup Y_{1} \subseteq Z_{1}$).

\item Let $Q$ be $Q_{1} \pco{L} Q_{2}$ and suppose that $\scs(Q_{i}) = (\cale_{i}, \calc_{i}, \ell_{i})$ is
a stable configuration structure for $i \in \{ 1, 2 \}$. Then every configuration $X$ of $\scs(Q)$ is such
that $(i)$ its two projections respectively belong to $\calc_{1}$ and $\calc_{2}$ and $(ii)$ any two events
in $X$ satisfy local injectivity of projections and coincidence freeness, hence $\scs(Q)$ is rooted. \\
As for connectedness, consider a non-empty configuration $X$ of $\scs(Q)$ and an event $e \in X$ that is
maximal with respect to the causality relation. If $e$ is of the form $\lpar_{L} \theta$ where $\theta \in
\cale_{1}$ with $\act(\theta) \notin L$, then $\theta$ is maximal in $\proj_{1}(X) \in \calc_{1}$, hence
$\proj_{1}(X) \setminus \{ \theta \} \in \calc_{1}$ by the induction hypothesis with $\proj_{1}(X \setminus
\{ e \}) = \proj_{1}(X) \setminus \{ \theta \}$ and $\proj_{2}(X \setminus \{ e \}) = \proj_{2}(X)$, i.e.,
$X \setminus \{ e \}$ is a configuration of $\scs(Q)$. If $e$ is of the form $\rpar_{L} \theta$ where
$\theta \in \cale_{2}$ with $\act(\theta) \notin L$, then we reason in a similar way. If $e$ is of the form
$\langle \theta_1, \theta_2 \rangle_{L}$ where $\theta_{1} \in \cale_{1}$, $\theta_{2} \in \cale_{2}$, and
$\act(\theta_{1}) = \act(\theta_{2}) \in L$, then $\theta_{1}$ and $\theta_{2}$ are maximal in $\proj_{1}(X)
\in \calc_{1}$ and $\proj_{2}(X) \in \calc_{2}$ respectively, hence $\proj_{1}(X) \setminus \{ \theta_{1} \}
\in \calc_{1}$ and $\proj_{2}(X) \setminus \{ \theta_{2} \} \in \calc_{2}$ by the induction hypothesis with
$\proj_{1}(X \setminus \{ e \}) = \proj_{1}(X) \setminus \{ \theta_{1} \}$ and $\proj_{2}(X \setminus \{ e
\}) = \proj_{2}(X) \setminus \{ \theta_{2} \}$, i.e., $X \setminus \{ e \}$ is a configuration of $\scs(Q)$.
\\
As for closure under bounded unions and intersections, consider any configurations $X, Y, Z$ of $\scs(Q)$
such that $X \cup Y \subseteq Z$. Then $X \cup Y$ and $X \cap Y$ are configurations of $\scs(Q)$ because,
for $i \in \{ 1, 2 \}$, $\proj_{i}(X \cup Y) = \proj_{i}(X) \cup \proj_{i}(Y)$ and $\proj_{i}(X \cap Y) =
\proj_{i}(X) \cap \proj_{i}(Y)$ where $\proj_{i}(X) \cup \proj_{i}(Y)$ and $\proj_{i}(X) \cap \proj_{i}(Y)$
are configurations of $\scs(Q_{i})$ by the induction hypothesis. Note that local injectivity of projections
and coincidence freeness are inherited from $Z$ because $X \cup Y, X \cap Y \subseteq Z$.
\qedhere

			\end{itemize}

		\end{proof}

	\end{thm}
}

	\begin{exa}\label{ex:auto_hhpb}

$\lsp a \, . \, \nil \pco{\emptyset} a \, . \, \nil \rsp$ comprises (see
Figure~\ref{fig:auto_conc_caus_confl}(a)):

		\begin{itemize}

\item The two events $\lpar_{\emptyset} a$ and $\rpar_{\emptyset} a$.

\item The four configurations $\emptyset$, $\{ \lpar_{\emptyset} a \}$, $\{ \rpar_{\emptyset} a \}$, $\{
\lpar_{\emptyset} a, \rpar_{\emptyset} a \}$.

\item The two maximal transition sequences $\emptyset \arrow{a}{\emph{\textsf{C}}_{a \, . \, \nil
\pco{\emptyset} a \, . \, \nil}} \{ \lpar_{\emptyset} a \} \arrow{a}{\emph{\textsf{C}}_{a \, . \, \nil
\pco{\emptyset} a \, . \, \nil}} \{ \lpar_{\emptyset} a, \rpar_{\emptyset} a \}$ and \linebreak $\emptyset
\arrow{a}{\emph{\textsf{C}}_{a \, . \, \nil \pco{\emptyset} a \, . \, \nil}} \{ \rpar_{\emptyset} a \}
\arrow{a}{\emph{\textsf{C}}_{a \, . \, \nil \pco{\emptyset} a \, . \, \nil}} \{ \lpar_{\emptyset} a,
\rpar_{\emptyset} a \}$.

		\end{itemize}

\noindent
In contrast, $\lsp a \, . \, a \, . \, \nil \rsp$ comprises (see Figure~\ref{fig:auto_conc_caus_confl}(b)):

		\begin{itemize}

\item The two events $a$ and $._{a} a$.

\item The three configurations $\emptyset$, $\{ a \}$, $\{ a, ._{a} a \}$.

\item The only maximal transition sequence $\emptyset \arrow{a}{\emph{\textsf{C}}_{a \, . \, a \, . \,
\nil}} \{ a \} \arrow{a}{\emph{\textsf{C}}_{a \, . \, a \, . \, \nil}} \{ a, ._{a} a \}$.

		\end{itemize}

\noindent
Therefore, $\lsp a \, . \, \nil \pco{\emptyset} a \, . \, \nil \rsp \not\sbis{\rm HHPB} \lsp a \, . \, a \,
. \, \nil \rsp$ because $a$ causally precedes $._{a} a$ while $\lpar_{\emptyset} a$ and $\rpar_{\emptyset}
a$ are independent of each other and hence no (labeling- and) causality-preserving bijection would relate
the former two events to the latter two.
\fullbox

	\end{exa}

%
\subsection{Operational Characterization Result}
\label{sec:oper_char_res}
%

We start by establishing a connection between proved transitions of processes and transitions of stable
configuration structures associated with processes.

	\begin{lem}\label{lem:proved_trans_scs_trans}

Let $P, P' \in \procs$ and $\theta \in \Theta$. Then $P \arrow{\theta}{} P'$ iff $X_{P}
\xarrow{\act(\theta)}{\textsf{C}_{P}} X_{P'}$.

		\begin{proof}

The proof is divided into two parts:

			\begin{itemize}

\item Suppose that $P \arrow{\theta}{} P'$. We show that $X_{P} \xarrow{\act(\theta)}{\emph{\textsf{C}}_{P}}
X_{P'}$ by proceeding by induction on the number $n \in \natns_{\ge 1}$ of applications of operational
semantic rules in Table~\ref{tab:proved_semantics} that are necessary to derive the transition $P
\arrow{\theta}{} P'$:

				\begin{itemize}

\item If $n = 1$ then $P$ is $a \, . \, Q$ with $\initial(Q)$, $a \, . \, Q \arrow{a}{} a^{\dag} . \, Q$ by
rule \textsc{Act}$_{\rm f}$, and $P'$ is $a^{\dag} . \, Q$. From $\initial(P)$ it follows that $\lsp P \rsp
= (\emph{\textsf{C}}_{P}, \emptyset)$ and $\lsp P' \rsp = (\emph{\textsf{C}}_{P}, \{ a \})$, hence $X_{P}
\arrow{a}{\emph{\textsf{C}}_{P}} X_{P'}$.

\item If $n > 1$ then there are three cases:

					\begin{itemize}

\item Let $P$ be $a^{\dag} . \, Q$ so that $P \xarrow{._{a} \theta'}{} P'$ with $P'$ being $a^{\dag} . \,
Q'$. Then $Q \arrow{\theta'}{} Q'$ by rule \textsc{Act}$_{\rm p}$, hence $X_{Q}
\xarrow{\act(\theta')}{\emph{\textsf{C}}_{Q}} X_{Q'}$ by the induction hypothesis with $\lsp Q \rsp =
(\emph{\textsf{C}}_{Q}, X_{Q})$ and $\lsp Q' \rsp = (\emph{\textsf{C}}_{Q}, X_{Q'})$. \\
Let $\toinitial(Q) \arrow{\theta_{1}}{} \dots \arrow{\theta_{n}}{} Q \arrow{\theta'}{} Q'$ with $n \in
\natns$, so $X_{Q} = \{\theta_{i} \mid 1 \le i \le n \}$ and $X_{Q'} = X_{Q} \cup \{ \theta' \}$. Then
$\toinitial(P) = a \, . \, \toinitial(Q) \arrow{a}{} \xarrow{._{a} \theta_{1}}{} \dots \xarrow{._{a}
\theta_{n}}{} P \xarrow{._{a} \theta'}{} P'$, $\lsp P \rsp = (a \, . \, \emph{\textsf{C}}_{Q}, X_{P})$ with
$X_{P} = \{ a \} \cup \{ ._{a} \theta_{i} \mid \theta_{i} \in X_{Q} \}$, and $\lsp P' \rsp = (a \, . \,
\emph{\textsf{C}}_{Q}, X_{P'})$ with $X_{P'} = X_{P} \cup \{ ._{a} \theta' \}$ and $._{a} \theta' \notin
X_{P}$. Thus $X_{P} \xarrow{\act(._{a} \theta')}{\emph{\textsf{C}}_{P}} X_{P'}$ with $\emph{\textsf{C}}_{P}
= a \, . \, \emph{\textsf{C}}_{Q}$.

\item Let $P$ be $P_{1} + P_{2}$. There are two subcases:

						\begin{itemize}

\item If $P_{1}$ moves, i.e., $P \xarrow{\lplu \theta'}{} P'$ with $P'$ being $P'_{1} + P_{2}$ and
$\initial(P_{2})$, then $P_{1} \arrow{\theta'}{} P'_{1}$ by rule \textsc{Cho}$_{\rm l}$, hence $X_{P_{1}}
\xarrow{\act(\theta')}{\emph{\textsf{C}}_{P_{1}}} X_{P'_{1}}$ by the induction hypothesis with $\lsp P_{1}
\rsp = (\emph{\textsf{C}}_{P_{1}}, X_{P_{1}})$ and $\lsp P'_{1} \rsp = (\emph{\textsf{C}}_{P_{1}},
X_{P'_{1}})$. \\
Let $\toinitial(P_{1}) \arrow{\theta_{1}}{} \dots \arrow{\theta_{n}}{} P_{1} \arrow{\theta'}{} P'_{1}$ with
$n \in \natns$, so $X_{P_{1}} = \{\theta_{i} \mid 1 \le i \le n \}$ and $X_{P'_{1}} = X_{P_{1}} \cup \{
\theta' \}$. Then $\toinitial(P) = \toinitial(P_{1}) + P_{2} \xarrow{\lplu \theta_{1}}{} \dots \xarrow{\lplu
\theta_{n}}{} P_{1} + P_{2} \xarrow{\lplu \theta'}{} P'$, $\lsp P \rsp = (\emph{\textsf{C}}_{P_{1}} +
\emph{\textsf{C}}_{P_{2}}, X_{P})$ with $X_{P} = \{ \lplu \theta_{i} \mid \theta_{i} \in X_{P_{1}} \}$, and
$\lsp P' \rsp = (\emph{\textsf{C}}_{P_{1}} + \emph{\textsf{C}}_{P_{2}}, X_{P'})$ with $X_{P'} = X_{P} \cup
\{ \lplu \theta' \}$ and $\lplu \theta' \notin X_{P}$. Thus $X_{P} \xarrow{\act(\lplu
\theta')}{\emph{\textsf{C}}_{P}} X_{P'}$ with $\emph{\textsf{C}}_{P} = \emph{\textsf{C}}_{P_{1}} +
\emph{\textsf{C}}_{P_{2}}$.

\item The subcase in which $P_{2}$ moves and $P_{1}$ is initial is like the previous one.

						\end{itemize}

\item Let $P$ be $P_{1} \pco{L} P_{2}$. Given two sequences $\overrightarrow{\theta_{1}}$ and
$\overrightarrow{\theta_{2}}$ of proof terms labeling two sequences of proved transitions respectively
departing from $P_{1}$ and $P_{2}$, we characterize their interleaving and synchronization through the
function $\zip : \Theta^{*} \times \Theta^{*} \times 2^{\cala \setminus \{ \tau \}} \rightarrow \Theta^{*}$
defined by induction on the sum of the lengths of its first two arguments $\overrightarrow{\theta_{1}},
\overrightarrow{\theta_{2}} \in \Theta^{*}$ as follows:
\cws{0}{\hspace*{-0.9cm} \zip(\overrightarrow{\theta_{1}}, \overrightarrow{\theta_{2}}, L) \, = \left\{
\begin{array}{ll}
\lpar_{L} \theta' \zip(\overrightarrow{\theta'_1}, \overrightarrow{\theta_{2}}) & \textrm{if }
\overrightarrow{\theta_{1}} = \theta' \overrightarrow{\theta'_{1}} \land \act(\theta') \notin L \land
(\overrightarrow{\theta_{2}} = \varepsilon \, \lor \\
& \hspace{0.5cm} (\overrightarrow{\theta_{2}} = \theta'' \overrightarrow{\theta''_{2}} \land (\act(\theta'')
\in L \lor |\overrightarrow{\theta_{1}}| \ge |\overrightarrow{\theta_{2}}|))) \\
\rpar_{L} \theta'' \zip(\overrightarrow{\theta_{1}}, \overrightarrow{\theta''_{2}}) & \textrm{if }
\overrightarrow{\theta_{2}} = \theta'' \overrightarrow{\theta''_{2}} \land \act(\theta'') \notin L \land
(\overrightarrow{\theta_{1}} = \varepsilon \, \lor \\
& \hspace{0.5cm} (\overrightarrow{\theta_{1}} = \theta' \overrightarrow{\theta'_{1}} \land (\act(\theta')
\in L \lor |\overrightarrow{\theta_{1}}| < |\overrightarrow{\theta_{2}}|))) \\
\langle \theta', \theta'' \rangle_{L} \zip(\overrightarrow{\theta_{1}}, \overrightarrow{\theta_{2}}) &
\textrm{if } \overrightarrow{\theta_{1}} = \theta' \overrightarrow{\theta'_{1}} \land
\overrightarrow{\theta_{2}} = \theta'' \overrightarrow{\theta''_{2}} \, \land \\
& \hspace{0.5cm} \act(\theta') = \act(\theta'') \in L \\
\varepsilon & \textrm{otherwise} \\
\end{array}\right.}
\blue{
The first two clauses of the $\zip$ function respectively represent the cases in which $P_{1}$ or $P_{2}$
moves by itself, i.e., without synchronizing with the other one. This is possible if the first proof term in
the sequence, i.e., $\theta'$, embodies an action not in~$L$ and $(i)$ the sequence of proof terms of the
other process is empty or $(ii)$ its first element embodies an action in $L$ that thus requires
synchronization or $(iii)$ it is not longer than the sequence of proof terms of the former process. In the
event that both processes can move alone, the choice between the two aforementioned clauses is thus made by
comparing $|\overrightarrow{\theta_{1}}|$ and $|\overrightarrow{\theta_{2}}|$. The third clause applies when
the two sequences of proof terms are not empty and their first elements embody the same action in $L$.
} \\
There are three subcases:

						\begin{itemize}

\item If $\act(\theta) \notin L$ and $P_{1}$ moves, i.e., $P \xarrow{\lpar_{L} \theta'}{} P'$ with $P'$
being $P'_{1} \pco{L} P_{2}$, then $P_{1} \arrow{\theta'}{} P'_{1}$ by rule \textsc{Par}$_{\rm l}$, hence
$X_{P_{1}} \xarrow{\act(\theta')}{\emph{\textsf{C}}_{P_{1}}} X_{P'_{1}}$ by the induction hypothesis with
$\lsp P_{1} \rsp = (\emph{\textsf{C}}_{P_{1}}, X_{P_{1}})$ and $\lsp P'_{1} \rsp =
(\emph{\textsf{C}}_{P_{1}}, X_{P'_{1}})$. \\
Let $\toinitial(P_{1}) \xarrow{\theta_{1, 1}}{} \dots \xarrow{\theta_{1, n_{1}}}{} P_{1} \arrow{\theta'}{}
P'_{1}$ with $n_{1} \in \natns$, so $X_{P_{1}} = \{ \theta_{1, i} \mid 1 \le i \le n_{1} \}$ and $X_{P'_{1}}
= X_{P_{1}} \cup \{ \theta' \}$. Also let $\toinitial(P_{2}) \xarrow{\theta_{2, 1}}{} \dots
\xarrow{\theta_{2, n_{2}}}{} P_{2}$ with $n_{2} \in \natns$, so $X_{P_{2}} = \{\theta_{2, i} \mid 1 \le i
\le n_{2} \}$. We denote by $\overrightarrow{\theta_{1}}$ and $\overrightarrow{\theta_{2}}$ the two
sequences of proof terms. Then $\toinitial(P) = \toinitial(P_{1}) \pco{L} \toinitial(P_{2})$ reaches $P$ via
a sequence of proved transitions labeled with $\zip(\overrightarrow{\theta_{1}},
\overrightarrow{\theta_{2}}, L)$ and afterwards $P'$ via $P \xarrow{\lpar_{L} \theta'}{} P'$, $\lsp P \rsp =
(\emph{\textsf{C}}_{P_{1}} \pco{L} \emph{\textsf{C}}_{P_{2}}, X_{P})$ with $X_{P} = \{ \bar{\theta} \mid
\bar{\theta} \textrm{ in } \zip(\overrightarrow{\theta_{1}}, \overrightarrow{\theta_{2}}, L) \}$, and $\lsp
P' \rsp = (\emph{\textsf{C}}_{P_{1}} \pco{L} \emph{\textsf{C}}_{P_{2}}, X_{P'})$ with $X_{P'} = X_{P} \cup
\{ \lpar_{L} \theta' \}$ and $\lpar_{L} \theta' \notin X_{P}$. Thus $X_{P} \xarrow{\act(\lpar_{L}
\theta')}{\emph{\textsf{C}}_{P}} X_{P'}$ with $\emph{\textsf{C}}_{P} = \emph{\textsf{C}}_{P_{1}} \pco{L}
\emph{\textsf{C}}_{P_{2}}$.

\item The subcase in which $\act(\theta) \notin L$ and $P_{2}$ moves is like the previous one.

\item If $\act(\theta) \in L$, i.e., $P \xarrow{\langle \theta_{1}, \theta_{2} \rangle_{L}}{} P'$ with $P'$
being $\enr(P'_{1} \pco{L} P'_{2}, \langle \theta_{1}, \theta_{2} \rangle_{L})$, then $P_{k}
\arrow{\theta_{k}}{} P'_{k}$ for $k \in \{ 1, 2 \}$ by rule \textsc{Syn}, hence $X_{P_{k}}
\xarrow{\act(\theta_{k})}{\emph{\textsf{C}}_{P_{k}}} X_{P'_{k}}$ by the induction hypothesis with $\lsp
P_{k} \rsp = (\emph{\textsf{C}}_{P_{k}}, X_{P_{k}})$ and $\lsp P'_{k} \rsp = (\emph{\textsf{C}}_{P_{k}},
X_{P'_{k}})$. \\
For $k \in \{ 1, 2 \}$ let $\toinitial(P_{k}) \xarrow{\theta_{k, 1}}{} \dots \xarrow{\theta_{k, n_{k}}}{}
P_{k} \arrow{\theta_{k}}{} P'_{k}$ with $n_{k} \in \natns$, so $X_{P_{k}} = \{ \theta_{k, i} \mid 1 \le i
\le n_{k} \}$ and $X_{P'_{k}} = X_{P_{k}} \cup \{ \theta_{k} \}$; we denote by $\overrightarrow{\theta_{k}}$
the sequence of proof terms. Then $\toinitial(P) = \toinitial(P_{1}) \pco{L} \toinitial(P_{2})$ reaches $P$
via a sequence of proved transitions labeled with $\zip(\overrightarrow{\theta_{1}},
\overrightarrow{\theta_{2}}, L)$ and afterwards $P'$ via $P \xarrow{\langle \theta_{1}, \theta_{2}
\rangle_{L}}{} P'$, $\lsp P \rsp = (\emph{\textsf{C}}_{P_{1}} \pco{L} \emph{\textsf{C}}_{P_{2}}, X_{P})$
with $X_{P} = \{ \bar{\theta} \mid \bar{\theta} \textrm{ in } \zip(\overrightarrow{\theta_{1}},
\overrightarrow{\theta_{2}}, L) \}$, and $\lsp P' \rsp = (\emph{\textsf{C}}_{P_{1}} \pco{L}
\emph{\textsf{C}}_{P_{2}}, X_{P'})$ with $X_{P'} = X_{P} \cup \{ \langle \theta_{1}, \theta_{2} \rangle_{L}
\}$ and $\langle \theta_{1}, \theta_{2} \rangle_{L} \notin X_{P}$. Thus $X_{P} \xarrow{\act(\langle
\theta_{1}, \theta_{2} \rangle_{L})}{\emph{\textsf{C}}_{P}} X_{P'}$ with $\emph{\textsf{C}}_{P} =
\emph{\textsf{C}}_{P_{1}} \pco{L} \emph{\textsf{C}}_{P_{2}}$.

						\end{itemize}

					\end{itemize}

				\end{itemize}

\item Suppose that $X_{P} \xarrow{\act(\theta)}{\emph{\textsf{C}}_{P}} X_{P'}$. Consider a variant $\Theta'$
of $\Theta$ in which $\lplu$, $\rplu$, $\lpar_{L}$, and $\rpar_{L}$ are given a subscript equal to the
subprocess that does not move \blue{-- so that they respectively become $\lplu_{Q}$, $\rplu_{Q}$, $\lpar_{L,
Q}$, and $\rpar_{L, Q}$ --} and this is applied to the corresponding rules in
Table~\ref{tab:proved_semantics} as well as the corresponding operations on stable configuration structures
in Section~\ref{sec:den_sem_scs}. \blue{For instance, the proof term generated when moving from $a \, . \,
\nil + b \, . \, \nil$ to $a^{\dag} . \, \nil + b \, . \, \nil$ becomes $\lplu_{b \, . \, \nil} \, a$.}
Then, although the configuration structure transition $X_{P} \xarrow{\act(\theta)}{\emph{\textsf{C}}_{P}}
X_{P'}$ is not generated inductively, from the only event $\theta$ -- enriched as described above -- in
$X_{P'} \setminus X_{P}$ we can pinpoint $\act(\theta)$ within $P$ and $P'$ as we now show. \\
Assuming that $\theta$ does not have subterms of the form $\langle \theta_{1}, \! \theta_{2} \rangle_{L}$,
meaning that $X_{P} \xarrow{\act(\theta)}{\emph{\textsf{C}}_{P}} \! X_{P'}$ originated from a single action,
we define the notion of process context $\textrm{C}[\bullet] = \ctx(\theta)$ by induction on the syntactical
structure of $\theta \in \Theta'$ as follows:
\cws{0}{\hspace*{-0.9cm} \ctx(\theta) \: = \: \left\{ \begin{array}{ll}
\bullet & \quad \textrm{if $\theta \in \cala$} \\
a^{\dag} . \, \ctx(\theta') & \quad \textrm{if $\theta = ._{a} \theta'$} \\
\ctx(\theta') + Q & \quad \textrm{if $\theta = \hspace{0.03cm} \lplu_{Q} \theta'$ where $\initial(Q)$} \\
Q + \ctx(\theta') & \quad \textrm{if $\theta = \rplu_{Q} \theta'$ where $\initial(Q)$} \\
\ctx(\theta') \pco{L} Q & \quad \textrm{if $\theta = \,\, \lpar_{L, Q} \theta'$} \\
Q \pco{L} \ctx(\theta') & \quad \textrm{if $\theta = \rpar_{L, Q} \theta'$} \\
\end{array}\right.}
\blue{which becomes a process when the only $\bullet$-hole is replaced with a process.} Since $X_{P}$ is the
set of proof terms labeling a sequence of proved transitions from $\toinitial(P)$ to $P$, from $X_{P}
\xarrow{\act(\theta)}{\emph{\textsf{C}}_{P}} X_{P'}$ it follows that $P$ must contain an occurrence of
$\act(\theta)$ in an initial subprocess of the form $\act(\theta) \, . \, \bar{P}$. Then $P =
\textrm{C}[\act(\theta) \, . \, \bar{P}]$ with $\act(\theta) \, . \, \bar{P} \xarrow{\act(\theta)}{}
\act(\theta)^{\dag} . \, \bar{P}$ and hence $P = \textrm{C}[\act(\theta) \, . \, \bar{P}] \arrow{\theta}{}
\textrm{C}[\act(\theta)^{\dag} . \, \bar{P}] = P'$. \\
If $\theta$ has subterms of the form $\langle \theta_{1}, \theta_{2} \rangle_{L}$, then we proceed in a
similar way by constructing a context with as many $\bullet$-holes as there are synchronizing subprocesses.
The following clause has to be added to the inductive definition of process context: $\ctx(\theta) =
\ctx(\theta_{1}) \pco{L} \ctx(\theta_{2})$ if $\theta = \langle \theta_{1}, \theta_{2} \rangle_{L}$.
\qedhere

			\end{itemize}

		\end{proof}

	\end{lem}

We are now in a position of proving our operational characterization result under the same assumption as the
denotational characterization result of Theorem~\ref{thm:hhpb_frb_brm_config_struct}.

%
%
%

	\begin{thm}\label{thm:hhpb_frb_brm_proc}

Let $P_{1}, P_{2} \in \procs$:

		\begin{itemize}

\item \blue{If $\lsp P_{1} \rsp \sbis{\rm HHPB} \lsp P_{2} \rsp$ then $P_{1} \sbis{\rm FRB:brm} P_{2}$.}

\item \blue{If $P_{1} \sbis{\rm FRB:brm} P_{2}$ then $\lsp P_{1} \rsp \sbis{\rm HHPB} \lsp P_{2} \rsp$
provided that all possible autoconflicts are local.}

		\end{itemize}

		\begin{proof}

\blue{We prove the two implications separately}:

			\begin{itemize}

\item Suppose that $\lsp P_{1} \rsp \sbis{\rm HHPB} \lsp P_{2} \rsp$. Then $P_{1} \sbis{\rm FRB:brm} P_{2}$
follows by proving that the symmetric relation $\calb = \{ (Q_{1}, Q_{2}) \mid \lsp Q_{1} \rsp \sbis{\rm
HHPB} \lsp Q_{2} \rsp \}$ is a brm-forward-reverse bisimulation. Let $(Q_{1}, Q_{2}) \in \calb$, so that
$\lsp Q_{1} \rsp \sbis{\rm HHPB} \lsp Q_{2} \rsp$:

				\begin{itemize}

\item If $Q_{1} \arrow{\theta_{1}}{} Q'_{1}$ then $X_{Q_{1}}
\xarrow{\act(\theta_{1})}{\emph{\textsf{C}}_{Q_{1}}} X_{Q'_{1}}$ due to
Lemma~\ref{lem:proved_trans_scs_trans}, hence $X_{Q_{2}}
\xarrow{\act(\theta_{2})}{\emph{\textsf{C}}_{Q_{2}}} X_{Q'_{2}}$ with $\act(\theta_{1}) = \act(\theta_{2})$
and $\lsp Q'_{1} \rsp \sbis{\rm HHPB} \lsp Q'_{2} \rsp$ because $\lsp Q_{1} \rsp \sbis{\rm HHPB} \lsp Q_{2}
\rsp$, from which \linebreak it follows that $Q_{2} \arrow{\theta_{2}}{} Q'_{2}$ due to
Lemma~\ref{lem:proved_trans_scs_trans} with $(Q'_{1}, Q'_{2}) \in \calb$.

\item If $Q'_{1} \arrow{\theta_{1}}{} Q_{1}$ then $X_{Q'_{1}}
\xarrow{\act(\theta_{1})}{\emph{\textsf{C}}_{Q'_{1}}} X_{Q_{1}}$ due to
Lemma~\ref{lem:proved_trans_scs_trans}, hence $X_{Q'_{2}}
\xarrow{\act(\theta_{2})}{\emph{\textsf{C}}_{Q'_{2}}} X_{Q_{2}}$ with $\act(\theta_{1}) = \act(\theta_{2})$
and $\lsp Q'_{1} \rsp \sbis{\rm HHPB} \lsp Q'_{2} \rsp$ because $\lsp Q_{1} \rsp \sbis{\rm HHPB} \lsp Q_{2}
\rsp$, from which \linebreak it follows that $Q'_{2} \arrow{\theta_{2}}{} Q_{2}$ due to
Lemma~\ref{lem:proved_trans_scs_trans} with $(Q'_{1}, Q'_{2}) \in \calb$.

\item The stable configuration structure semantics preserves actions inside transition labels
(Lemma~\ref{lem:proved_trans_scs_trans}), uniquely identifies different occurrences of the same action in a
process via as many different proof terms in the set of events, and is able to distinguish between causality
and concurrency like the proved operational semantics of Table~\ref{tab:proved_semantics}. Therefore, from
$\lsp Q_{1} \rsp \sbis{\rm HHPB} \lsp Q_{2} \rsp$ -- in particular the incoming transition matching between
$X_{Q_{1}}$ and $X_{Q_{2}}$ and the labeling- and causality-preserving bijection from $X_{Q_{1}}$ to
$X_{Q_{2}}$ -- it follows that there must be a one-to-one correspondence between the incoming proved
transitions of $Q_{1}$ and the incoming proved transitions of $Q_{2}$, hence $\brm(Q_{1}) = \brm(Q_{2})$.

				\end{itemize}

\item Suppose that $P_{1} \sbis{\rm FRB:brm} P_{2}$. Then the existence of a finite sequence of proved
transitions $\toinitial(P_{1}) = P_{1, 1} \xarrow{\theta_{P_{1, 1}}}{} P_{1, 2} \dots P_{1, n}
\xarrow{\theta_{P_{1, n}}}{} P_{1}$ implies the existence of a finite sequence of proved transitions
$\toinitial(P_{2}) = P_{2, 1} \xarrow{\theta_{P_{2, 1}}}{} P_{2, 2} \dots P_{2, n} \xarrow{\theta_{P_{2,
n}}}{} P_{2}$ such that $P_{1, h} \sbis{\rm FRB:brm} P_{2, h}$ and $\act(\theta_{P_{1, h}}) =
\act(\theta_{P_{2, h}})$ for all $h = 1, \dots, n$, and vice versa. Note that $\theta_{P_{1, h}} \neq
\theta_{P_{1, k}}$ and $\theta_{P_{2, h}} \neq \theta_{P_{2, k}}$ for all $h \neq k$ because different
occurrences of the same action in a process are identified by different proof terms. \\
Thus $\lsp P_{1} \rsp \sbis{\rm HHPB} \lsp P_{2} \rsp$ follows by proving that:
\blue{
\cws{0}{\hspace*{-0.4cm}\begin{array}{l}
\calb \, = \, \{ (X_{Q_{1}}, X_{Q_{2}}, \{ (\theta_{Q_{1, h}}, \theta_{Q_{2, h}}) \mid h \in H = \{ 1,
\dots, n \} \textrm{ for some } n \in \natns_{\ge 1} \}) \mid \\
\hspace*{3.5cm} Q_{1} \sbis{\rm FRB:brm} Q_{2} \, \land \\
\hspace*{3.5cm} \toinitial(Q_{i}) = Q_{i, 1} \xarrow{\theta_{Q_{i, 1}}}{} Q_{i, 2} \dots Q_{i, |H|}
\xarrow{\theta_{Q_{i, |H|}}}{} Q_{i} \textrm{ for all } i \in \{ 1, 2 \} \, \land \\
\hspace*{3.5cm} Q_{1, h} \sbis{\rm FRB:brm} Q_{2, h} \textrm{ for all } h \in H \, \land \\
\hspace*{3.5cm} \act(\theta_{Q_{1, h}}) = \act(\theta_{Q_{2, h}}) \textrm{ for all } h \in H \}
\end{array}}
}
is a hereditary-history preserving bisimulation. Observing that $Q_{1} \sbis{\rm FRB:brm} Q_{2}$ implies
$(\emptyset, \emptyset, \emptyset) \in \calb$ when $Q_{1}$ and $Q_{2}$ are both initial, take $(X_{Q_{1}},
X_{Q_{2}}, \{ (\theta_{Q_{1, h}}, \theta_{Q_{2, h}}) \mid h \in H \}) \in \calb$, so that $Q_{1} \sbis{\rm
FRB:brm} Q_{2}$:

				\begin{itemize}

\item If $X_{Q_{1}} \xarrow{\act(\theta_{1})}{\emph{\textsf{C}}_{Q_{1}}} X_{Q'_{1}}$ then $Q_{1}
\arrow{\theta_{1}}{} Q'_{1}$ due to Lemma~\ref{lem:proved_trans_scs_trans}, hence $Q_{2}
\arrow{\theta_{2}}{} Q'_{2}$ with $\act(\theta_{1}) = \act(\theta_{2})$ and $Q'_{1} \sbis{\rm FRB:brm}
Q'_{2}$ because $Q_{1} \sbis{\rm FRB:brm} Q_{2}$, from which it follows that $X_{Q_{2}}
\xarrow{\act(\theta_{2})}{\emph{\textsf{C}}_{Q_{2}}} X_{Q'_{2}}$ due to
Lemma~\ref{lem:proved_trans_scs_trans}. Since $\theta_{1} \notin X_{Q_{1}}$ and $\theta_{2} \notin
X_{Q_{2}}$, \linebreak it holds that $(X_{Q'_{1}}, X_{Q'_{2}}, \{ (\theta_{Q_{1, h}}, \theta_{Q_{2, h}})
\mid h \in H \} \cup \{ (\theta_{1}, \theta_{2}) \}) \in \calb$. \\
If we start from $X_{Q_{2}} \xarrow{\act(\theta_{2})}{\emph{\textsf{C}}_{Q_{2}}} X_{Q'_{2}}$, then we reason
in the same way.

\item If $X_{Q'_{1}} \xarrow{\act(\theta_{1})}{\emph{\textsf{C}}_{Q'_{1}}} X_{Q_{1}}$ then $Q'_{1}
\arrow{\theta_{1}}{} Q_{1}$ due to Lemma~\ref{lem:proved_trans_scs_trans}, hence $Q'_{2}
\arrow{\theta_{2}}{} Q_{2}$ with $\act(\theta_{1}) \! = \act(\theta_{2})$ and $Q'_{1} \sbis{\rm FRB:brm}
Q'_{2}$ because $Q_{1} \sbis{\rm FRB:brm} Q_{2}$, from which it follows that $X_{Q'_{2}}
\xarrow{\act(\theta_{2})}{\emph{\textsf{C}}_{Q'_{2}}} X_{Q_{2}}$ due to
Lemma~\ref{lem:proved_trans_scs_trans}. Since $\theta_{1} \notin X_{Q'_{1}}$, $\theta_{2} \notin
X_{Q'_{2}}$, and $\brm(Q_{1}) = \brm(Q_{2})$, the latter transition can be selected in such a way to satisfy
$\{ (\theta_{Q_{1, h}}, \theta_{Q_{2, h}}) \mid h \in H \} \upharpoonright X_{Q'_{1}} = \{ (\theta_{Q_{1,
h}}, \theta_{Q_{2, h}}) \mid h \in H \} \setminus \{ (\theta_{1}, \theta_{2}) \}$, hence $(X_{Q'_{1}},
X_{Q'_{2}}, \{ (\theta_{Q_{1, h}}, \theta_{Q_{2, h}}) \mid h \in H \} \setminus \{ (\theta_{1}, \theta_{2})
\}) \in \calb$. \\
If we start from $X_{Q'_{2}} \xarrow{\act(\theta_{2})}{\emph{\textsf{C}}_{Q'_{2}}} X_{Q_{2}}$, then we
reason in the same way.

\item $f = \{ (\theta_{Q_{1, h}}, \theta_{Q_{2, h}}) \mid h \in H \}$ certainly is a bijection from
$X_{Q_{1}}$ to $X_{Q_{2}}$ -- as the events along either computation are $(i)$ different from each other, so
the two reached configurations $X_{Q_{1}}$ and $X_{Q_{2}}$ contain the same number of events, and $(ii)$
paired in a stepwise manner -- that preserves labeling -- by definition of $\calb$. If $|H| \le 1$ then
causality is trivially preserved too, \blue{so we assume that $|H| \ge 2$ and proceed by contradiction}. \\
Suppose that $X_{Q_{1}}$ and $X_{Q_{2}}$ break causality \blue{-- in the same sense as the corresponding
part of the proof of Theorem~\ref{thm:hhpb_frb_brm_config_struct} --} and, among all the pairs of
configurations associated with $\sbis{\rm FRB:brm}$-equivalent processes that break causality, they are the
closest ones to $\emptyset$ and $\emptyset$ (in terms of number of transitions to be executed from either
empty configuration). The rest of the proof is like the one of the corresponding part of the proof of
Theorem~\ref{thm:hhpb_frb_brm_config_struct}, with Lemma~\ref{lem:proved_trans_scs_trans} being exploited as
well.
\qedhere

				\end{itemize}

			\end{itemize}

		\end{proof}

	\end{thm}

%
%
\section{Relationships between $\call_{\rm EI}$ and $\call_{\rm BRM}$}
\label{sec:rel_eil_brml}
%
%

\blue{After recalling $\call_{\rm EI}$, the \emph{event identifier logic}~\cite{PU14} that characterizes
hereditary history-preserving bisimilarity over stable event structures (Section~\ref{sec:eil}), and
introducing $\call_{\rm BRM}$, a modal logic that characterizes brm-forward-reverse bisimilarity over
processes (Section~\ref{sec:brml}), in this section we investigate the relationships between the two logics
because, in the absence of non-local autoconflicts, due to Theorem~\ref{thm:hhpb_frb_brm_proc} two processes
satisfy the same formulas of $\call_{\rm BRM}$ iff their associated stable configuration structures satisfy
the same formulas of $\call_{\rm EI}$.} On the one hand, we show how to reinterpret $\call_{\rm EI}$ over
processes (Section~\ref{sec:eil_over_procs}) and $\call_{\rm BRM}$ over stable configuration structures
(Section~\ref{sec:brml_over_scs}). On the other hand, we discuss how to translate $\call_{\rm BRM}$ into
$\call_{\rm EI}$ (Section~\ref{sec:brml_into_eil}) and then $\call_{\rm EI}$ into $\call_{\rm BRM}$
(Section~\ref{sec:eil_into_brml}).

%
\subsection{\blue{Event Identifier Logic}}
\label{sec:eil}
%

The set $\call_{\rm EI}$ of its formulas is generated by the following syntax~\cite{PU14}:
\cws{0}{\phi \: ::= \: \true \mid \lnot\phi \mid \phi \land \phi \mid \eildiam{x : a}{} \phi \mid (x : a)
\phi \mid \eilbdiam{x}{} \phi}
where $a \in \cala$ and $x \in \id$, with $\id$ being a countable set of identifiers. The unary operators
$\eildiam{x : a}{} \_$ and $(x : a) \_$ act as binders for the identifiers inside them. Therefore, the set
of identifiers that occur \emph{free} in $\phi \in \call_{\rm EI}$ is defined by induction on the
syntactical structure of~$\phi$ as follows:
\cws{0}{\begin{array}{rcl}
\fid(\true) & \!\! = \!\! & \emptyset \\
\fid(\lnot\phi) & \!\! = \!\! & \fid(\phi) \\
\fid(\phi_{1} \land \phi_{2}) & \!\! = \!\! & \fid(\phi_{1}) \cup \fid(\phi_{2}) \\
\fid(\eildiam{x:a}{} \phi) & \!\! = \!\! & \fid(\phi) \setminus \{ x \} \\
\fid((x:a) \phi) & \!\! = \!\! & \fid(\phi) \setminus \{ x \} \\
\fid(\eilbdiam{x}{} \phi) & \!\! = \!\! & \fid(\phi) \cup \{ x \} \\
\end{array}}
where we say that $\phi$ is \emph{closed} if $\fid(\phi) = \emptyset$, \emph{open} otherwise.

In order to assign meaning to open formulas, environments are employed to indicate what events the free
identifiers are bound to. Given a configuration structure $\emph{\ssfc} = (\cale, \calc, \ell)$, an
\emph{environment} is a partial function $\rho : \id \parfun \cale$. Given $X \in \calc$ and $\phi \in
\call_{\rm EI}$, we say that $\rho$ is a \emph{permissible} environment for $X$ and $\phi$, \blue{written
$\rho \in \perm(X, \phi)$}, iff $\rho$ maps every free identifier in $\phi$ to an event in~$X$. Denoting
with $\dom(\rho)$ the domain of $\rho$, $\rge(\rho)$ the codomain of $\rho$, and $\rho_{\phi}$ the
restriction $\rho \restr \fid(\phi)$, permissibility is formalized as $\fid(\phi) \subseteq \dom(\rho)$ --
\blue{every free identifier in $\phi$ is mapped by $\rho$} -- and $\rge(\rho_{\phi}) \subseteq X$ --
\blue{each such mapping is to an event in $X$}.

The satisfaction relation $\models \: \subseteq (\calc \times \cale^{\id}) \times \call_{\rm EI}$, with
$\cale^{\id}$ being the set of functions from $\id$ to $\cale$, i.e., the set of environments, is defined by
induction on the syntactical structure of $\phi \in \call_{\rm EI}$ as follows~\cite{PU14}:
\cws{0}{\begin{array}{rclcl}
X & \!\! \models_{\rho} \!\! & \true & \\
X & \!\! \models_{\rho} \!\! & \lnot\phi' & \textrm{iff} & X \not\models_{\rho} \phi' \\
X & \!\! \models_{\rho} \!\! & \phi_{1} \land \phi_{2} & \textrm{iff} & X \models_{\rho} \phi_{1} \textrm{
and } X \models_{\rho} \phi_{2} \\
X & \!\! \models_{\rho} \!\! & \eildiam{x : a}{} \phi' & \textrm{iff} & \textrm{there exists } X
\arrow{\ell(e)}{\emph{\textsf{C}}} X' \textrm{ such that } \ell(e) = a \textrm{ and } X' \models_{\rho [x
\mapsto e]} \phi' \\
X & \!\! \models_{\rho} \!\! & (x : a) \phi' & \textrm{iff} & \textrm{there exists } e \in X \textrm{ such
that } \ell(e) = a \textrm{ and } X \models_{\rho [x \mapsto e]} \phi' \\
X & \!\! \models_{\rho} \!\! & \eilbdiam{x}{} \phi' & \textrm{iff} & \textrm{there exists } X'
\arrow{\ell(e)}{\emph{\textsf{C}}} X \textrm{ such that } \rho(x) = e \textrm{ and } X' \models_{\rho} \phi'
\\
\end{array}}
where it is understood that the environment in the subscript of every occurrence of $\models$ is permissible
for the configuration on the left and the formula on the right. Moreover, $\rho [x \mapsto e]$ is equal to
$\rho \setminus \{ (x, \rho(x)) \} \cup \{ (x, e) \}$ if $x \in \dom(\rho)$, $\rho \cup \{ (x, e) \}$
otherwise.

Let $\call_{\rm EI}^{\rm c}$ be the set of closed formulas of $\call_{\rm EI}$. Given $\phi \in \call_{\rm
EI}^{\rm c}$, we write $X \models \phi$ as a shorthand for $X \models_{\emptyset} \phi$ and
$\emph{\textsf{C}} \models \phi$ as a shorthand for $\emptyset \models \phi$. Image finiteness means that no
configuration has infinitely many outgoing transitions with the same label.

	\begin{thm}[\cite{PU14}]\label{thm:hhpb_log_char}

Let $\textsf{C}_{i} = (\cale_{i}, \calc_{i}, \ell_{i})$, $i \in \{ 1, 2 \}$, be two image-finite stable
configuration structures. Then $\textsf{C}_{1} \sbis{\rm HHPB} \textsf{C}_{2}$ iff $\forall \phi \in
\call_{\rm EI}^{\rm c} \ldotp \textsf{C}_{1} \models \phi \Longleftrightarrow \textsf{C}_{2} \models \phi$.
\fullbox

	\end{thm}

%
\subsection{\blue{Backward Ready Multiset Logic}}
\label{sec:brml}
%

We now develop a modal logic characterization for $\sbis{\rm FRB:brm}$ inspired by the one of $\sbis{\rm
FRB}$ in~\cite{BE23b}. The set $\call_{\rm BRM}$ of formulas of the \emph{backward ready multiset logic} is
generated by the following syntax:
\cws{0}{\phi \: ::= \: \true \mid M \mid \lnot\phi \mid \phi \land \phi \mid \diam{a}{} \phi \mid
\diam{a^{\dag}}{} \phi}
where $M : \cala \rightarrow \natns$ and $a \in \cala$. The satisfaction relation $\models \: \subseteq
\procs \times \call_{\rm BRM}$ is defined by induction on the syntactical structure of $\phi \in \call_{\rm
BRM}$ as follows:
\cws{0}{\begin{array}{rclcl}
P & \!\! \models \!\! & \true & \\
P & \!\! \models \!\! & M & \textrm{iff} & \brm(P) = M \\
P & \!\! \models \!\! & \lnot\phi' & \textrm{iff} & P \not\models \phi' \\
P & \!\! \models \!\! & \phi_{1} \land \phi_{2} & \textrm{iff} & P \models \phi_{1} \textrm{ and } P \models
\phi_{2} \\
P & \!\! \models \!\! & \diam{a}{} \phi' & \textrm{iff} & \textrm{there exists } P \arrow{\theta}{} P'
\textrm{ such that } \act(\theta) = a \textrm{ and } P' \models \phi' \\
P & \!\! \models \!\! & \diam{a^\dag}{} \phi' & \textrm{iff} & \textrm{there exists } P' \arrow{\theta}{} P
\textrm{ such that } \act(\theta) = a \textrm{ and } P' \models \phi' \\
\end{array}}
\indent
Note that every $P \in \procs$ is image finite, i.e., it has finitely many outgoing (and incoming)
transitions labeled with proof terms containing the same action. Since it is instrumental to prove the next
theorem, we define the depth of $\phi \in \call_{\rm BRM}$ -- intended as an upper bound \blue{on} the depth
of the syntax tree of the considered formula -- by induction on the syntactical structure of~$\phi$ as
follows:
\cws{4}{\begin{array}{rcl}
\depth(\true) & \!\! = \!\! & 0 \\
\depth(M) & \!\! = \!\! & 0 \\
\depth(\lnot\phi') & \!\! = \!\! & 1 + \depth(\phi') \\
\depth(\phi_{1} \land \phi_{2}) & \!\! = \!\! & 1 + \textrm{max}(\depth(\phi_{1}), \depth(\phi_{2})) \\
\depth(\diam{a}{} \phi') & \!\! = \!\! & 1 + \depth(\phi') \\
\depth(\diam{a^\dag}{} \phi') & \!\! = \!\! & 1 + \depth(\phi') \\
\end{array}}

	\begin{thm}\label{thm:frb_brm_log_char}

Let $P_{1}, P_{2} \in \procs$. Then $P_{1} \sbis{\rm FRB:brm} P_{2}$ iff $\forall \phi \in \call_{\rm BRM}
\ldotp P_{1} \models \phi \Longleftrightarrow P_{2} \models \phi$.

		\begin{proof}

The proof is divided into two parts:

			\begin{itemize}

\item Assuming that $P_{1} \sbis{\rm FRB:brm} P_{2}$ and $P_{1} \models \phi$ for an arbitrary formula $\phi
\in \call_{\rm FRB:brm}$, \linebreak we prove that $P_{2} \models \phi$ too by proceeding by induction on $k
= \depth(\phi)$:

				\begin{itemize}

\item If $k = 0$ then $\phi$ is either $\true$ or $M$. In the former case, it is trivially satisfied by
$P_{2}$ too. In the latter case, it is satisfied by $P_{2}$ too because $P_{1} \sbis{\rm FRB:brm} P_{2}$ and
hence $\brm(P_{1}) = \brm(P_{2})$.

\item If $k \ge 1$ then there are four cases:

					\begin{itemize}

\item If $\phi$ is $\lnot\phi'$ then from $P_{1} \models \lnot\phi'$ we derive that $P_{1} \not\models
\phi'$. If it were $P_{2} \models \phi'$ then by the induction hypothesis it would hold that $P_{1} \models
\phi'$, which is not the case. Therefore $P_{2} \not\models \phi'$ and hence $P_{2} \models \lnot\phi'$ too.

\item If $\phi$ is $\phi_{1} \land \phi_{2}$ then from $P_{1} \models \phi_{1} \land \phi_{2}$ we derive
that $P_{1} \models \phi_{1}$ and $P_{1} \models \phi_{2}$. From the induction hypothesis it follows that
$P_{2} \models \phi_{1}$ and $P_{2} \models \phi_{2}$ and hence $P_{2} \models \phi_{1} \land \phi_{2}$ too.

\item If $\phi$ is $\diam{a}{} \phi'$ then from $P_{1} \models \diam{a}{} \phi'$ we derive that there exists
$P_{1} \arrow{\theta_{1}}{} P'_{1}$ such that $\act(\theta_{1}) = a$ and $P'_{1} \models \phi'$. From $P_{1}
\sbis{\rm FRB:brm} P_{2}$ it then follows that there exists $P_{2} \arrow{\theta_{2}}{} P'_{2}$ such that
$\act(\theta_{2}) = a$ and $P'_{1} \sbis{\rm FRB:brm} P'_{2}$. By applying the induction hypothesis we
derive that $P'_{2} \models \phi'$ and hence $P_{2} \models \diam{a}{} \phi'$ too.

\item If $\phi$ is $\diam{a^{\dag}}{} \phi'$ then from $P_{1} \models \diam{a^{\dag}}{} \phi'$ we derive
that there exists $P'_{1} \arrow{\theta_{1}}{} P_{1}$ such that $\act(\theta_{1}) = a$ and $P'_{1} \models
\phi'$. From $P_{1} \sbis{\rm FRB:brm} P_{2}$ it then follows that there exists $P'_{2} \arrow{\theta_{2}}{}
P_{2}$ such that $\act(\theta_{2}) = a$ and $P'_{1} \sbis{\rm FRB:brm} P'_{2}$. By applying the induction
hypothesis we derive that $P'_{2} \models \phi'$ and hence $P_{2} \models \diam{a^{\dag}}{} \phi'$ too.

					\end{itemize}

				\end{itemize}

\item Assuming that $P_{1}$ and $P_{2}$ satisfy the same formulas in $\call_{\rm BRM}$, we prove that the
symmetric relation $\calb = \{ (Q_{1}, Q_{2}) \in \procs \times \procs \mid Q_{1} \textrm{ and } Q_{2}
\textrm{ satisfy the same formulas in } \call_{\rm BRM} \}$ is a brm-forward-reverse bisimulation. \\
Given $(Q_{1}, Q_{2}) \in \calb$:

				\begin{itemize}

\item If $Q_{1} \arrow{\theta_{1}}{} Q'_{1}$ suppose by contradiction that there is no $Q'_{2}$ satisfying
the same formulas as $Q'_{1}$ such that $Q_{2} \arrow{\theta_{2}}{} Q'_{2}$ and $\act(\theta_{1}) =
\act(\theta_{2})$, i.e., $(Q'_{1}, Q'_{2}) \in \calb$ for no $Q'_{2}$ $\act(\theta_{1})$-reachable from
$Q_{2}$. Since $Q_{2}$ has finitely many outgoing transitions, the set of processes that $Q_{2}$ can reach
by performing an $\act(\theta_{1})$-transition is finite, say $\{ Q'_{2, 1}, \dots, Q'_{2, n} \}$ with $n
\ge 0$. Since none of the processes in the set satisfies the same formulas as $Q'_{1}$, for each $1 \le i
\le n$ there exists $\phi_{i} \in \call_{\rm BRM}$ such that $Q'_{1} \models \phi_{i}$ but $Q'_{2, i}
\not\models \phi_{i}$. \\
We can then construct the formula $\diam{\act(\theta_{1})}{} \bigwedge\limits_{i = 1}^{n} \phi_{i}$ that is
satisfied by~$Q_{1}$ but not by $Q_{2}$; if $n = 0$ then it is sufficient to take $\diam{\act(\theta_{1})}{}
\true$. This formula violates $(Q_{1}, Q_{2}) \in \calb$, hence there must exist at least one $Q'_{2}$
satisfying the same formulas as $Q'_{1}$ such that $Q_{2} \arrow{\theta_{2}}{} Q'_{2}$ and $\act(\theta_{1})
= \act(\theta_{2})$, so that $(Q'_{1}, Q'_{2}) \in \calb$.

\item If $Q'_{1} \arrow{\theta_{1}}{} Q_{1}$ suppose by contradiction that there is no $Q'_{2}$ satisfying
the same formulas as $Q'_{1}$ such that $Q'_{2} \arrow{\theta_{2}}{} Q_{2}$ and $\act(\theta_{1}) =
\act(\theta_{2})$, i.e., $(Q'_{1}, Q'_{2}) \in \calb$ for no $Q'_{2}$ $\act(\theta_{1})$-reaching $Q_{2}$.
Since $Q_{2}$ has finitely many incoming transitions, the set of processes that can reach $Q_{2}$ by
performing an $\act(\theta_{1})$-transition is finite, say $\{ Q'_{2, 1}, \dots, Q'_{2, n} \}$ with $n \ge
0$. Since none of the processes in the set satisfies the same formulas as $Q'_{1}$, for each $1 \le i \le n$
there exists $\phi_{i} \in \call_{\rm BRM}$ such that $Q'_{1} \models \phi_{i}$ but $Q'_{2, i} \not\models
\phi_{i}$. \\
We can then construct the formula $\diam{\act(\theta_{1})^{\dag}}{} \bigwedge\limits_{i = 1}^{n} \phi_{i}$
that is satisfied by~$Q_{1}$ but not by $Q_{2}$; if $n = 0$ then it is sufficient to take
$\diam{\act(\theta_{1})^{\dag}}{} \true$.  This formula violates $(Q_{1}, Q_{2}) \in \calb$, hence there
must exist at least one $Q'_{2}$ satisfying the same formulas as $Q'_{1}$ such that $Q'_{2}
\arrow{\theta_{2}}{} Q_{2}$ and $\act(\theta_{1}) = \act(\theta_{2})$, so that $(Q'_{1}, Q'_{2}) \in \calb$.

\item The fact that $\brm(Q_{1}) = \brm(Q_{2})$ follows from the fact that $Q_{1}$ and $Q_{2}$ satisfy, in
particular, the same formulas of the form $M$.
\qedhere

				\end{itemize}

			\end{itemize}

		\end{proof}

	\end{thm}

%
\subsection{Reinterpreting $\call_{\rm EI}$ over $\procs$}
\label{sec:eil_over_procs}
%

The only non-trivial case is the one of the binder $(x : a)$. The process analogous of an event, which is
labeled with a certain action, in a configuration \linebreak is a subprocess starting with an executed
occurrence of that action. Indicating with $\sbpr(P)$ \linebreak the set of all subprocesses of $P$, let
$\apt(a^{\dag} . \, P', P)$ be the proof term associated with the execution of action $a$ of the subprocess
$a^{\dag} . \, P'$ of $P$. Formally, $\apt(a^{\dag} . \, P', P) = \theta$ iff $a^{\dag} . \, P' \in
\sbpr(P)$ and there exist $P'', P''' \in \procs$ such that $a \, . \, \toinitial(P') \in \sbpr(P'')$, $P''
\arrow{\theta}{} P'''$, $\act(\theta) = a$, and $a^{\dag} . \, \toinitial(P') \in \sbpr(P''')$. \blue{In the
case of subprocesses composed in parallel each of which starts with the same executed non-synchronizing
action, the result of $\apt$ is made unique by considering the leftmost subprocess: for instance,
$\apt(a^{\dag} . \, \nil, a^{\dag} . \, \nil \pco{\emptyset} a^{\dag} . \, \nil) = \,\, \lpar_{\emptyset} \,
a$. \linebreak The reason is that those actions correspond to concurrent events, all of which belong to the
same configuration and can be undone in any order.}

The satisfaction relation $\models \: \subseteq (\procs \times \Theta^{\id}) \times \call_{\rm EI}$ is
defined by induction on the syntactical structure of $\phi \in \call_{\rm EI}$ as follows: 
\cws{0}{\begin{array}{rclcl}
P & \!\! \models_{\rho} \!\! & \true & \\
P & \!\! \models_{\rho} \!\! & \lnot\phi' & \textrm{iff} & P \not\models_{\rho} \phi' \\
P & \!\! \models_{\rho} \!\! & \phi_{1} \land \phi_{2} & \textrm{iff} & P \models_{\rho} \phi_{1} \textrm{
and } P \models_{\rho} \phi_{2} \\
P & \!\! \models_{\rho} \!\! & \eildiam{x : a}{} \phi' & \textrm{iff} & \textrm{there exists } P
\arrow{\theta}{} P' \textrm{ such that } \act(\theta) = a \textrm{ and } P' \models_{\rho[x \mapsto \theta]}
\phi' \\
P & \!\! \models_{\rho} \!\! & (x : a) \phi' & \textrm{iff} & \textrm{there exists } a^{\dag} . \, P' \in
\sbpr(P) \textrm{ such that } P \models_{\rho[x \mapsto \apt(a^{\dag} . \, P', P)]} \phi' \\
P & \!\! \models_{\rho} \!\! & \eilbdiam{x}{} \phi' & \textrm{iff} & \textrm{there exists } P'
\arrow{\theta}{} P \textrm{ such that } \rho(x) = \theta \textrm{ and } P' \models_{\rho} \phi' \\
\end{array}}
where it is understood that the environment in the subscript of every occurrence of $\models$ is permissible
for the configuration identifying (in the associated denotational semantics) the process on the left --
e.g., $X_{P}$ in the case of process $P$ -- and the formula on the right.

Since it is instrumental to prove the next results, we define the depth of $\phi \in \call_{\rm EI}$
\linebreak by induction on the syntactical structure of $\phi$ as follows:
\cws{4}{\begin{array}{rcl}
\depth(\true) & \!\! = \!\! & 0 \\
\depth(\lnot\phi') & \!\! = \!\! & 1 + \depth(\phi') \\
\depth(\phi_{1} \land \phi_{2}) & \!\! = \!\! & 1 + \textrm{max}(\depth(\phi_{1}), \depth(\phi_{2})) \\
\depth(\eildiam{x : a}{} \phi') & \!\! = \!\! & 1 + \depth(\phi') \\
\depth((x : a) \phi') & \!\! = \!\! & 1 + \depth(\phi') \\
\depth(\eilbdiam{x}{} \phi') & \!\! = \!\! & 1 + \depth(\phi') \\
\end{array}}

	\begin{thm}\label{thm:proc_scs_sat_eil}

Let $P \in \procs$. Then $\forall \phi \in \call_{\rm EI} \ldotp \forall \rho \in \perm(X_{P}, \phi) \ldotp
P \models_{\rho} \phi \Longleftrightarrow \lsp P \rsp \models_{\rho} \phi$.

		\begin{proof}

We proceed by induction on $k = \depth(\phi)$:

    			\begin{itemize}

\item If $k = 0$ then $\phi$ is $\true$ and both $P$ and $\lsp P \rsp$ satisfy it.

\item If $k \ge 1$ then there are five cases:

				\begin{itemize}

\item If $\phi$ is $\lnot\phi'$ then by the induction hypothesis:
\cws{4}{\hspace*{-1.0cm}\begin{array}{rcl}
P \models_{\rho} \lnot\phi' & \textrm{iff} & P \not\models_{\rho} \phi' \\
& \textrm{iff} & \lsp P \rsp \not\models_{\rho} \phi' \\
& \textrm{iff} & \lsp P \rsp \models_{\rho} \lnot\phi' \\
\end{array}}

\item If $\phi$ is $\phi_{1} \land \phi_{2}$ then by the induction hypothesis:
\cws{4}{\hspace*{-1.0cm}\begin{array}{rcl}
P \models_{\rho} \phi_{1} \land \phi_{2} & \textrm{iff} & P \models_{\rho} \phi_{1} \textrm{ and } P
\models_{\rho} \phi_{2} \\
& \textrm{iff} & \lsp P \rsp \models_{\rho} \phi_{1} \textrm{ and } \lsp P \rsp \models_{\rho} \phi_{2} \\
& \textrm{iff} & \lsp P \rsp \models_{\rho} \phi_{1} \land \phi_{2} \\
\end{array}}

\item If $\phi$ is $\eildiam{x : a}{} \phi'$ then by Lemma~\ref{lem:proved_trans_scs_trans} and the
induction hypothesis:
\cws{4}{\hspace*{-1.0cm}\begin{array}{rcl}
P \models_{\rho} \eildiam{x : a}{} \phi' & \textrm{iff} & \textrm{there is } P \arrow{\theta}{} P' \textrm{
s.t. } \act(\theta) = a \textrm{ and } P' \models_{\rho [x \mapsto \theta]} \phi' \\
& \textrm{iff} & \textrm{there is } X_{P} \xarrow{\act(\theta)}{\emph{\textsf{C}}_{P}} X_{P'} \textrm{ s.t.
} \act(\theta) = a \textrm{ and } \lsp P' \rsp \models_{\rho [x \mapsto \theta]} \phi' \\
& \textrm{iff} & \lsp P \rsp \models_{\rho} \eildiam{x : a}{} \phi' \\
\end{array}}

\item If $\phi$ is $(x : a) \phi'$ then by the induction hypothesis:
\cws{4}{\hspace*{-1.0cm}\begin{array}{rcl}
P \models_{\rho} (x : a) \phi' & \textrm{iff} & \textrm{there is } a^{\dag} . \, P' \in \sbpr(P) \textrm{
s.t. } P \models_{\rho [x \mapsto \apt(a^{\dag} . \, P', P)]} \phi' \\
& \textrm{iff} & \textrm{there is } \theta \in X_{P} \textrm{ s.t. } \theta = \apt(a^{\dag} . \, P', P)
\textrm{ and } \\
& & \hspace*{3.4cm} \lsp P \rsp \models_{\rho[x \mapsto \apt(a^{\dag} . \, P', P)]} \phi' \\
& \textrm{iff} & \textrm{there is } \theta \in X_{P} \textrm{ s.t. } \act(\theta) = a \textrm{ and } \lsp P
\rsp \models_{\rho [x \mapsto \theta]} \phi' \\
& \textrm{iff} & \lsp P \rsp \models_{\rho} (x : a) \phi' \\
\end{array}}

\item If $\phi$ is $\eilbdiam{x}{} \phi'$ then by Lemma~\ref{lem:proved_trans_scs_trans} and the induction
hypothesis:
\cws{8}{\hspace*{-1.0cm}\begin{array}{rcl}
P \models_{\rho} \eilbdiam{x}{} \phi' & \textrm{iff} & \textrm{there is } P' \arrow{\theta}{} P \textrm{
s.t. } \rho(x) = \theta \textrm{ and } P' \models_{\rho} \phi' \\
& \textrm{iff} & \textrm{there is } X_{P'} \xarrow{\act(\theta)}{\emph{\textsf{C}}_{P'}} X_{P} \textrm{ s.t.
} \rho(x) = \theta \textrm{ and } \lsp P' \rsp \models_{\rho} \phi' \\
& \textrm{iff} & \lsp P \rsp \models_{\rho} \eilbdiam{x}{} \phi' \\
\end{array}}
\qedhere

				\end{itemize}

			\end{itemize}

		\end{proof}

	\end{thm}

To prove that, consequently, $\call_{\rm EI}$ reinterpreted over $\procs$ characterizes $\sbis{\rm
FRB:brm}$, we follow~\cite{PU14} and hence first show that any substitution of the variables freely
occurring in a formula requires a modification of the permissible environment.

	\begin{lem}\label{lem:free_var_subst_env_mod}

Let $P \in \procs$, $\phi \in \call_{\rm EI}$, and $\rho \in \perm(X_{P}, \phi)$. Given a substitution
$\sigma$ that -- not necessarily injectively -- maps $\fid(\phi)$ to a set of fresh identifiers that do not
occur either free or bound in $\phi$, let $\sigma(\phi)$ be the formula obtained from $\phi$ by replacing
each occurrence of $x \in \fid(\phi)$ with $\sigma(x)$ and let $\rho^{\sigma} \in \perm(X_{P},
\sigma(\phi))$ be the environment obtained from $\rho$ by replacing each $x \in \fid(\phi)$ with
$\sigma(x)$. Then $P \models_{\rho} \phi$ iff $P \models_{\rho^{\sigma}} \sigma(\phi)$.

		\begin{proof}

We proceed by induction on $k = \depth(\phi)$:

			\begin{itemize}

\item If $k = 0$ then $\phi$ must be $\true$. Since $\true$ is closed, we have that $\sigma(\true) = \true$
and $\rho^{\sigma} = \rho$, hence the result trivially follows.

\item If $k \ge 1$ then there are five cases: 

				\begin{itemize}

\item If $\phi$ is $\lnot\phi'$ then by the induction hypothesis:
\cws{4}{\hspace*{-1.0cm}\begin{array}{rcl}
P \models_{\rho} \lnot\phi' & \textrm{iff} & P \not\models_{\rho} \phi' \\
& \textrm{iff} & P \not\models_{\rho^{\sigma}} \sigma(\phi') \\
& \textrm{iff} & P \models_{\rho^{\sigma}} \sigma(\lnot\phi') \\
\end{array}}

\item If $\phi$ is $\phi_{1} \land \phi_{2}$ then by the induction hypothesis:
\cws{0}{\hspace*{-1.0cm}\begin{array}{rcl}
P \models_{\rho} \phi_{1} \land \phi_{2} & \textrm{iff} & P \models_{\rho} \phi_{1} \textrm{ and } P
\models_{\rho} \phi_{2} \\
& \textrm{iff} & P \models_{\rho^{\sigma_{1}}} \sigma_{1}(\phi_{1}) \textrm{ and } P
\models_{\rho^{\sigma_{2}}} \sigma_{2}(\phi_{2}) \\
& \textrm{iff} & P \models_{\rho^{\sigma}} \sigma(\phi_{1}) \textrm{ and } P \models_{\rho^{\sigma}}
\sigma(\phi_{2}) \\
& \textrm{iff} & P \models_{\rho^{\sigma}} \sigma (\phi_{1} \land \phi_{2}) \\
\end{array}}
provided that $\sigma_{1}(x) = \sigma_{2}(x)$ for all $x \in \fid(\phi_{1}) \cap \fid(\phi_{2})$ and
$\sigma \restr \fid(\phi_{k}) = \sigma_{k}$ for $k \in \{ 1, 2 \}$.
\blue{
These conditions are satisfied by choosing $\sigma_{k} = \sigma \restr \fid(\phi_{k})$. Indeed, these
restrictions satisfy the freshness requirement of the induction hypothesis and, for every $x \in
\fid(\phi_{1}) \cap \fid(\phi_{2})$, we have that $\sigma_{1}(x) = \sigma(x) = \sigma_{2}(x)$. Moreover,
$\rho^{\sigma_{k}}$ and $\rho^{\sigma}$ agree on all free identifiers of $\sigma_{k}(\phi_{k})$, with
satisfaction depending only on the values assigned to free identifiers.
}

\item If $\phi$ is $\eildiam{x : a}{} \phi'$ then by the induction hypothesis:
\cws{0}{\hspace*{-1.0cm}\begin{array}{rcl}
P \models_{\rho} \eildiam{x : a}{} \phi' & \textrm{iff} & \textrm{there is } P \arrow{\theta}{} P' \textrm{
s.t. } \act(\theta) = a \textrm { and } P' \models_{\rho [x \mapsto \theta]} \phi' \\
& \textrm{iff} & \textrm{there is } P \arrow{\theta}{} P' \textrm{ s.t. } \act(\theta) = a \textrm { and }
P' \models_{(\rho [x \mapsto \theta])^{\sigma'}} \sigma'(\phi') \\
& \textrm{iff} & \textrm{there is } P \arrow{\theta}{} P' \textrm{ s.t. } \act(\theta) = a \textrm { and }
P' \models_{\rho^{\sigma} [x \mapsto \theta]} \sigma(\phi') \\
& \textrm{iff} & P \models_{\rho^{\sigma}} \sigma(\eildiam{x : a}{} \phi') \\
\end{array}}
provided that $\sigma = \sigma'$ if $x \notin \fid(\phi')$ while $\sigma[x \mapsto \sigma'(x)] = \sigma'$ if
$x \in \fid(\phi')$.
\blue{
These conditions are satisfied by choosing $\sigma'$ to agree with $\sigma$ on all identifiers and, whenever
$x \in \fid(\phi')$, \linebreak to impose $\sigma'(x) = x$. The equivalence between the first and the second
line above follows by applying the induction hypothesis to $\phi'$ and $\sigma'(\phi')$ after renaming their
corresponding free identifiers to common fresh identifiers. Both formulas have smaller depth than $\phi$ and
the resulting environments agree on all free identifiers.
}

\item If $\phi$ is $(x : a) \phi'$ then by the induction hypothesis:
\cws{0}{\hspace*{-1.0cm}\begin{array}{rcl}
P \models_{\rho} (x : a) \phi' & \textrm{iff} & \textrm{there is } a^{\dag} . \, P' \in \sbpr(P) \textrm{
s.t. } P \models_{\rho [x \mapsto \apt(a^{\dag} . \, P', P)]} \phi' \\
& \textrm{iff} & \textrm{there is } a^{\dag} . \, P' \in \sbpr(P) \textrm{ s.t. } P \models_{(\rho [x
\mapsto \apt(a^{\dag} . \, P', P)])^{\sigma'}} \sigma'(\phi') \\
& \textrm{iff} & \textrm{there is } a^{\dag} . \, P' \in \sbpr(P) \textrm{ s.t. } P \models_{\rho^{\sigma}
[x \mapsto \apt(a^{\dag} . \, P', P)]} \sigma(\phi') \\
& \textrm{iff} & P \models_{\rho^{\sigma}} \sigma((x : a) \phi') \\
\end{array}}
provided that $\sigma = \sigma'$ if $x \notin \fid(\phi')$ while $\sigma[x \mapsto \sigma'(x)] = \sigma'$ if
$x \in \fid(\phi')$.
\blue{These conditions are satisfied by choosing $\sigma$ and $\sigma'$ as in the previous case.}

\item If $\phi$ is $\eilbdiam{x}{} \phi'$ then by the induction hypothesis:
\cws{0}{\hspace*{-1.0cm}\begin{array}{rcl}
P \models_{\rho} \eilbdiam{x}{} \phi' & \textrm{iff} & \textrm{there is } P' \arrow{\theta}{} P \textrm{
s.t. } \rho(x) = \theta \textrm { and } P' \models_{\rho} \phi' \\
& \textrm{iff} & \textrm{there is } P' \arrow{\theta}{} P \textrm{ s.t. } \rho^{\sigma'}(\sigma'(x)) =
\theta \textrm { and } P' \models_{\rho^{\sigma'}} \sigma'(\phi') \\
& \textrm{iff} & \textrm{there is } P' \arrow{\theta}{} P \textrm{ s.t. } \rho^{\sigma}(\sigma(x)) = \theta
\textrm { and } P' \models_{\rho^{\sigma}} \sigma(\phi') \\
& \textrm{iff} & P \models_{\rho^{\sigma}} \sigma(\eilbdiam{x}{} \phi') \\
\end{array}}
as $\rho^{\sigma'}(\sigma'(x)) = \rho(x)$, provided that $\sigma \setminus \{ (x, \sigma(x)) \} = \sigma'$
if $x \notin \fid(\phi')$ while $\sigma = \sigma'$ if $x \in \fid(\phi')$.
\blue{
These conditions are satisfied by choosing $\sigma' = \sigma \restr \fid(\phi')$. This restriction satisfies
the freshness requirement of the induction hypothesis. Moreover, $\rho^{\sigma'}$ and $\rho^{\sigma}$ agree
on all free identifiers of $\sigma'(\phi')$, with satisfaction depending only on the values assigned to free
identifiers.
}
\qedhere

				\end{itemize}  

			\end{itemize}
    
		\end{proof}

	\end{lem}

	\begin{cor}\label{cor:eil_char_frb_brm}

Let $P_{1}, P_{2} \in \procs$. Then $P_{1} \sbis{\rm FRB:brm} P_{2}$ iff $\exists f_{1, 2} \ldotp \forall
\phi \in \call_{\rm EI} \ldotp \forall \rho \in \perm(X_{P_{1}}, \phi) \ldotp \linebreak P_{1}
\models_{\rho} \phi \Longleftrightarrow P_{2} \models_{f_{1, 2} \circ \rho} \phi$ where $f_{1, 2}$ is a
labeling-preserving bijection from $X_{P_{1}}$ to $X_{P_{2}}$.

		\begin{proof}

The proof is divided into two parts:

			\begin{itemize}

\item Assuming that $P_{1} \sbis{\rm FRB:brm} P_{2}$, we observe that the existence of a sequence of proved
transitions $\toinitial(P_{1}) \xarrow{\theta_{P_{1}, 1}}{} \dots \xarrow{\theta_{P_{1}, n}}{} P_{1}$
implies the existence of a sequence of proved transitions $\toinitial(P_{2}) \xarrow{\theta_{P_{2}, 1}}{}
\dots \xarrow{\theta_{P_{2}, n}}{} P_{2}$ such that $\act(\theta_{P_{1}, h}) = \act(\theta_{P_{2}, h})$ for
all $h = 1, \dots, n$, and vice versa, where $\{ \theta_{P_{1}, h} \mid 1 \le h \le n \} = X_{P_{1}}$ and
$\{ \theta_{P_{2}, h} \mid 1 \le h \le n \} = X_{P_{2}}$. Note that $n = 0$ when $P_{1}$ and $P_{2}$ are
both initial; moreover $\theta_{P_{1}, h} \neq \theta_{P_{1}, k}$ and $\theta_{P_{2}, h} \neq \theta_{P_{2},
k}$ for all $h \neq k$. Let $f_{1, 2} = \{ (\theta_{P_{1}, h}, \theta_{P_{2}, h}) \mid 1 \le h \le n \}$,
which clearly is a labeling-preserving bijection from $X_{P_{1}}$ to $X_{P_{2}}$. \\
We proceed by induction on $k = \depth(\phi)$:

				\begin{itemize}

\item If $k = 0$ then $\phi$ must be $\true$, which is trivially satisfied by $P_{1}$ and $P_{2}$ regardless
of their respective permissible environments.

\item If $k \ge 1$ then there are five cases:

					\begin{itemize}

\item If $\phi$ is $\lnot\phi'$ then by the induction hypothesis:
\cws{4}{\hspace*{-1.4cm}\begin{array}{rcl}
P_{1} \models_{\rho} \lnot\phi' & \textrm{iff} & P_{1} \not\models_{\rho} \phi' \\
& \textrm{iff} & P_{2} \not\models_{f_{1, 2} \circ \rho} \phi' \\
& \textrm{iff} & P_{2} \models_{f_{1, 2} \circ \rho} \lnot\phi' \\
\end{array}}

\item If $\phi$ is $\phi_{1} \land \phi_{2}$ then by the induction hypothesis:
\cws{4}{\hspace*{-1.4cm}\begin{array}{rcl}
P_{1} \models_{\rho} \phi_{1} \land \phi_{2} & \textrm{iff} & P_{1} \models_{\rho} \phi_{1} \textrm{ and }
P_{1} \models_{\rho} \phi_{2} \\
& \textrm{iff} & P_{2} \models_{f_{1, 2} \circ \rho} \phi_{1} \textrm{ and } P_{2} \models_{f_{1, 2} \circ
\rho} \phi_{2} \\
& \textrm{iff} & P_{2} \models_{f_{1, 2} \circ \rho} \phi_{1} \land \phi_{2} \\
\end{array}}

\item If $\phi$ is $\eildiam{x : a}{} \phi'$ then by $P_{1} \sbis{\rm FRB:brm} P_{2}$ and the induction
hypothesis:
\cws{0}{\hspace*{-1.4cm}\begin{array}{rcl}
P_{1} \models_{\rho} \eildiam{x : a}{} \phi' & \textrm{iff} & \textrm{there is } P_{1} \arrow{\theta_{1}}{}
P'_{1} \textrm{ s.t. } \act(\theta_{1}) = a \textrm { and } P'_{1} \models_{\rho [x \mapsto \theta_{1}]}
\phi' \\
& \textrm{iff} & \textrm{there is } P_{2} \arrow{\theta_{2}}{} P'_{2} \textrm{ s.t. } \act(\theta_{2}) = a
\textrm { and } P'_{2} \models_{f'_{1, 2} \circ \rho [x \mapsto \theta_{1}]} \phi' \\
& \textrm{iff} & P_{2} \models_{f_{1, 2} \circ \rho} \eildiam{x : a}{} \phi' \\
\end{array}}
provided that $f_{1, 2} \cup \{ (\theta_{1}, \theta_{2}) \} = f'_{1, 2}$.

\item If $\phi$ is $(x : a) \phi'$ then by $P_{1} \sbis{\rm FRB:brm} P_{2}$ and the induction hypothesis:
\cws{4}{\hspace*{-1.4cm}\begin{array}{rcl}
P_{1} \models_{\rho} (x : a) \phi' & \textrm{iff} & \textrm{there is } a^{\dag} . \, P'_{1} \in \sbpr(P_{1})
\textrm{ s.t. } P_{1} \models_{\rho [x \mapsto \apt(a^{\dag} . \, P'_{1}, P_{1})]} \phi' \\
& \textrm{iff} & \textrm{there is } a^{\dag} . \, P'_{2} \in \sbpr(P_{2}) \textrm{ s.t. } P_{2}
\models_{f_{1, 2} \circ \rho [x \mapsto \apt(a^{\dag} . \, P'_{1}, P_{1})]} \phi' \\
& \textrm{iff} & P_{2} \models_{f_{1, 2} \circ \rho} (x : a) \phi' \\
\end{array}}

\item If $\phi$ is $\eilbdiam{x}{} \phi'$ then by $P_{1} \sbis{\rm FRB:brm} P_{2}$ and the induction
hypothesis:
\cws{0}{\hspace*{-1.4cm}\begin{array}{rcl}
P_{1} \models_{\rho} \eilbdiam{x}{} \phi' & \textrm{iff} & \textrm{there is } P'_{1} \arrow{\theta_{1}}{}
P_{1} \textrm{ s.t. } \rho(x) = \theta_{1} \textrm { and } P'_{1} \models_{\rho} \phi' \\
& \textrm{iff} & \textrm{there is } P'_{2} \arrow{\theta_{2}}{} P_{2} \textrm{ s.t. } \rho(x) = \theta_{2}
\textrm { and } P'_{2} \models_{f'_{1, 2} \circ \rho} \phi' \\
& \textrm{iff} & P_{2} \models_{f_{1, 2} \circ \rho} \eilbdiam{x}{} \phi' \\
\end{array}}
provided that $f_{1, 2} = f'_{1, 2} \cup \{ (\theta_{1}, \theta_{2}) \}$.

					\end{itemize}

				\end{itemize}

\item Assuming that there exists a labeling-preserving bijection from $X_{P_{1}}$ to $X_{P_{2}}$ such that
$P_{1}$ and $P_{2}$ satisfy the same formulas of $\call_{\rm EI}$ under suitable permissible environments
related by the aforementioned bijection, the result follows by proving that the symmetric relation $\calb =
\{ (Q_{1}, Q_{2}) \mid \exists f_{1, 2} \ldotp \forall \phi \in \call_{\rm EI} \ldotp \forall \rho \in
\perm(Q_{1}, \phi) \ldotp Q_{1} \models_{\rho} \phi \Longleftrightarrow Q_{2} \models_{f_{1, 2} \circ \rho}
\phi \linebreak \textrm{where } f_{1, 2} \textrm{ is a labeling-preserving} \textrm{ bijection from }
X_{Q_{1}} \textrm{ to } X_{Q_{2}} \}$ is a brm-forward-reverse bisimulation. \\
Given $(Q_{1}, Q_{2}) \in \calb$:

				\begin{itemize}

\item If $Q_{1} \arrow{\theta_{1}}{} Q'_{1}$ suppose by contradiction that there is no $Q'_{2}$ satisfying
the same formulas as $Q'_{1}$ for some labeling-preserving bijection $f'_{1, 2}$ from $X_{Q'_{1}}$ to
$X_{Q'_{2}}$ such that $Q_{2} \arrow{\theta_{2}}{} Q'_{2}$ and $\act(\theta_{1}) = \act(\theta_{2})$, i.e.,
$(Q'_{1}, Q'_{2}) \in \calb$ for no $Q'_{2}$ $\act(\theta_{1})$-reachable from $Q_{2}$. Since $Q_{2}$ has
finitely many outgoing transitions, the set of processes that $Q_{2}$ can reach by performing an
$\act(\theta_{1})$-transition is finite, say $\{ Q'_{2, 1}, \dots, Q'_{2, n} \}$ with $n \ge 0$. Since none
of the processes in the set satisfies the same formulas as $Q'_{1}$, for each $1 \le i \le n$ there exists
$\phi_{i} \in \call_{\rm EI}$ such that $Q'_{1} \models_{\rho_{i}} \phi_{i}$ but $Q'_{2, i}
\not\models_{f'_{1, 2, i} \circ \rho_{i}} \phi_{i}$ for all labeling-preserving bijections $f'_{1, 2, i}$
from $X_{Q'_{1}}$ to $X_{Q'_{2, i}}$. \\
Since the formulas $\phi_{1}, \dots, \phi_{n}$ may contain different identifiers, let $\{ z_{\theta} \mid
\theta \in X_{Q'_{1}} \}$ be a set of fresh identifiers different from each other and the related
environment $\rho'$ be defined by $\rho'(z_{\theta}) = \theta$ for all $\theta \in X_{Q'_{1}}$. Also let
every substitution $\sigma_{i}$ be defined by $\sigma_{i}(x) = z_{\rho'(x)}$ for all $x \in \fid(\phi_{i})$,
so that $\rho_{i}(x) = \rho'(\sigma_{i}(x))$. It holds that $Q'_{1} \models_{\rho'} \sigma_{i}(\phi_{i})$ by
Lemma~\ref{lem:free_var_subst_env_mod} and $Q'_{2, i} \not\models_{f'_{1, 2, i} \circ \rho'}
\sigma_{i}(\phi_{i})$ for all labeling-preserving bijections $f'_{1, 2, i}$ from $X_{Q'_{1}}$ to $X_{Q'_{2,
i}}$. \\
We can then construct the formula $\eildiam{z_{\theta_{1}} : \act(\theta_{1})}{} \bigwedge\limits_{i =
1}^{n} \sigma_{i}(\phi_{i})$ that is satisfied by $Q_{1}$ under $\rho$ such that $\rho [z_{\theta_{1}}
\mapsto \theta_{1}] = \rho'$ but not by $Q_{2}$ under $f_{1, 2} \circ \rho$ for all labeling-preserving
bijections $f_{1, 2}$ from $X_{Q_{1}}$ to $X_{Q_{2}}$; if $n = 0$ then it is sufficient to take
$\eildiam{z_{\theta_{1}} : \act(\theta_{1})}{} \true$. This formula violates $(Q_{1}, Q_{2}) \in \calb$,
hence there must exist at least one $Q'_{2}$ satisfying the same formulas as $Q'_{1}$ for some
labeling-preserving bijection $f'_{1, 2}$ from $X_{Q'_{1}}$ to $X_{Q'_{2}}$ such that $Q_{2}
\arrow{\theta_{2}}{} Q'_{2}$ and $\act(\theta_{1}) = \act(\theta_{2})$, so that $(Q'_{1}, Q'_{2}) \in
\calb$.

\item If $Q'_{1} \arrow{\theta_{1}}{} Q_{1}$ suppose by contradiction that there is no $Q'_{2}$ satisfying
the same formulas as~$Q'_{1}$ for some labeling-preserving bijection $f'_{1, 2}$ from $X_{Q'_{1}}$ to
$X_{Q'_{2}}$ such that $Q'_{2} \arrow{\theta_{2}}{} Q_{2}$ and $\act(\theta_{1}) = \act(\theta_{2})$, i.e.,
$(Q'_{1}, Q'_{2}) \in \calb$ for no $Q'_{2}$ $\act(\theta_{1})$-reaching $Q_{2}$. Since $Q_{2}$ has finitely
many incoming transitions, the set of processes that can reach $Q_{2}$ by performing an
$\act(\theta_{1})$-transition is finite, say $\{ Q'_{2, 1}, \dots, Q'_{2, n} \}$ with $n \ge 0$. Since none
of the processes in the set satisfies the same formulas as $Q'_{1}$, for each $1 \le i \le n$ there exists
$\phi_{i} \in \call_{\rm EI}$ such that $Q'_{1} \models_{\rho_{i}} \phi_{i}$ but $Q'_{2, i}
\not\models_{f'_{1, 2, i} \circ \rho_{i}} \phi_{i}$. \\
Since the formulas $\phi_{1}, \dots, \phi_{n}$ may contain different identifiers, let $\{ z_{\theta} \mid
\theta \in X_{Q'_{1}} \}$ be a set of fresh identifiers different from each other and the related
environment $\rho'$ be defined by $\rho'(z_{\theta}) = \theta$ for all $\theta \in X_{Q'_{1}}$. Also let
every substitution $\sigma_{i}$ be defined by $\sigma_{i}(x) = z_{\rho'(x)}$ for all $x \in \fid(\phi_{i})$,
so that $\rho_{i}(x) = \rho'(\sigma_{i}(x))$. It holds that $Q'_{1} \models_{\rho'} \sigma_{i}(\phi_{i})$ by
Lemma~\ref{lem:free_var_subst_env_mod} and $Q'_{2, i} \not\models_{f'_{1, 2, i} \circ \rho'}
\sigma_{i}(\phi_{i})$ for all labeling-preserving bijections $f'_{1, 2, i}$ from $X_{Q'_{1}}$ to $X_{Q'_{2,
i}}$. \\
We can then construct the formula $\eilbdiam{z_{\theta_{1}}}{} \bigwedge\limits_{i = 1}^{n}
\sigma_{i}(\phi_{i})$ that is satisfied by $Q_{1}$ under $\rho = \rho'$ but not by $Q_{2}$ under $f_{1, 2}
\circ \rho$ for all labeling-preserving bijections $f_{1, 2}$ from $X_{Q_{1}}$ to $X_{Q_{2}}$; if $n = 0$
then it is sufficient to take $\eildiam{z_{\theta_{1}}}{} \true$. This formula violates $(Q_{1}, Q_{2}) \in
\calb$, hence there must exist at least one $Q'_{2}$ satisfying the same formulas as $Q'_{1}$ for some
labeling-preserving bijection $f'_{1, 2}$ from $X_{Q'_{1}}$ to $X_{Q'_{2}}$ such that $Q'_{2}
\arrow{\theta_{2}}{} Q_{2}$ and $\act(\theta_{1}) = \act(\theta_{2})$, so that $(Q'_{1}, Q'_{2}) \in \calb$.

\item Suppose by contradiction that $\brm(Q_{1}) \neq \brm(Q_{2})$. In order not to fall back into one of
the previous cases, we assume that there is an action $a$ with different nonzero multiplicities in
$\brm(Q_{1})$ and $\brm(Q_{2})$. Without loss of generality, we further assume that $a$ occurs with
multiplicity $2$ in $\brm(Q_{1})$ and $1$ in $\brm(Q_{2})$. \\
We can then construct the formula $\eilbdiam{x}{} \true \land \eilbdiam{y}{} \true$ that is satisfied by
$Q_{1}$ under $\rho$ such that $\act(\rho(x)) = \act(\rho(y)) = a$ and $\rho(x) \neq \rho(y)$ but not by
$Q_{2}$ under $f_{1, 2} \circ \rho$ for all labeling-preserving bijections $f_{1, 2}$ from $X_{Q_{1}}$
to~$X_{Q_{2}}$. This formula violates $(Q_{1}, Q_{2}) \in \calb$, hence it must be the case that
$\brm(Q_{1}) = \brm(Q_{2})$.
\qedhere

				\end{itemize}
				
			\end{itemize}

		\end{proof}

	\end{cor}

%
\subsection{Reinterpreting $\call_{\rm BRM}$ over Stable Configuration Structures}
\label{sec:brml_over_scs}
%

Let us denote by $\lsp \procs \rsp$ the set of all stable configuration structures -- whose events are proof
terms -- that turn out to be the denotational semantics of some $P \in \procs$. Recalling that $\lsp P \rsp
= (\emph{\textsf{C}}_{P}, X_{P})$, when writing $\lsp P \rsp \models \phi$ we mean $X_{P} \models \phi$.

The satisfaction relation $\models \: \subseteq \lsp \procs \rsp \times \call_{\rm BRM}$ is defined by
induction on the syntactical structure of $\phi \in \call_{\rm BRM}$ as follows:
\cws{0}{\begin{array}{rclcl}
\lsp P \rsp & \!\! \models \!\! & \true & \\
\lsp P \rsp & \!\! \models \!\! & M & \textrm{iff} & \lmp a \in \cala \mid X_{P'}
\arrow{a}{\emph{\textsf{C}}_{P'}} X_{P} \rmp = M \\
\lsp P \rsp & \!\! \models \!\! & \lnot\phi' & \textrm{iff} & \lsp P \rsp \not\models \phi' \\
\lsp P \rsp & \!\! \models \!\! & \phi_{1} \land \phi_{2} & \textrm{iff} & \lsp P \rsp \models \phi_{1}
\textrm{ and } \lsp P \rsp \models \phi_{2} \\
\lsp P \rsp & \!\! \models \!\! & \diam{a}{} \phi' & \textrm{iff} & \textrm{there exists } X_{P}
\arrow{a}{\emph{\textsf{C}}_{P}} X_{P'} \textrm{ such that } \lsp P' \rsp \models \phi' \\
\lsp P \rsp & \!\! \models \!\! & \diam{a^\dag}{} \phi' & \textrm{iff} & \textrm{there exists } X_{P'}
\arrow{a}{\emph{\textsf{C}}_{P'}} X_{P} \textrm{ such that } \lsp P' \rsp \models \phi' \\
\end{array}}
\indent
Every process and its associated stable configuration structure satisfy the same formulas of $\call_{\rm
BRM}$. As a consequence, $\call_{\rm BRM}$ reinterpreted over stable configuration structures characterizes
$\sbis{\rm HHPB}$ in the absence of non-local autoconflicts.

	\begin{thm}\label{thm:proc_scs_sat_brml}

Let $P \in \procs$. Then $\forall \phi \in \call_{\rm BRM} \ldotp P \models \phi \Longleftrightarrow \lsp P 
\rsp \models \phi$.

		\begin{proof}

We proceed by induction on $k = \depth(\phi)$:
	
			\begin{itemize}

\item If $k = 0$ then there are two cases:
			
				\begin{itemize}

\item If $\phi$ is $\true$ then both $P$ and $\lsp P \rsp$ satisfy it.

\item If $\phi$ is $M$ then by Lemma~\ref{lem:proved_trans_scs_trans}:
\cws{4}{\hspace*{-1.0cm}\begin{array}{rcl}
P \models M & \textrm{iff} & \brm(P) = M \\
& \textrm{iff} & \lmp \act(\theta) \mid P' \arrow{\theta}{} P \rmp = M \\
& \textrm{iff} & \lmp \act(\theta) \mid X_{P'} \xarrow{\act(\theta)}{\emph{\textsf{C}}_{P'}} X_{P} \rmp = M
\\
& \textrm{iff} & \lsp P \rsp \models M \\
\end{array}}

				\end{itemize}

\item If $k \ge 1$ then there are four cases:

				\begin{itemize}

\item If $\phi$ is $\lnot\phi'$ then by the induction hypothesis:
\cws{4}{\hspace*{-1.0cm}\begin{array}{rcl}
P \models \lnot\phi' & \textrm{iff} & P \not\models \phi' \\
& \textrm{iff} & \lsp P \rsp \not\models \phi' \\
& \textrm{iff} & \lsp P \rsp \models \lnot\phi' \\
\end{array}}

\item If $\phi$ is $\phi_{1} \land \phi_{2}$ then by the induction hypothesis:
\cws{4}{\hspace*{-1.0cm}\begin{array}{rcl}
P \models \phi_{1} \land \phi_{2} & \textrm{iff} & P \models \phi_{1} \textrm{ and } P \models \phi_{2} \\
& \textrm{iff} & \lsp P \rsp \models \phi_{1} \textrm{ and } \lsp P \rsp \models \phi_{2} \\
& \textrm{iff} & \lsp P \rsp \models \phi_{1} \land \phi_{2} \\
\end{array}}

\item If $\phi$ is $\diam{a}{} \phi'$ then by Lemma~\ref{lem:proved_trans_scs_trans} and the induction
hypothesis:
\cws{4}{\hspace*{-1.0cm}\begin{array}{rcl}
P \models \diam{a}{} \phi' & \textrm{iff} & \textrm{there is } P \arrow{\theta}{} P' \textrm{ s.t. }
\act(\theta) = a \textrm{ and } P' \models \phi' \\
& \textrm{iff} & \textrm{there is } X_{P} \xarrow{\act(\theta)}{\emph{\textsf{C}}_{P}} X_{P'} \textrm{ s.t.
} \act(\theta) = a \textrm{ and } \lsp P' \rsp \models \phi' \\
& \textrm{iff} & \lsp P \rsp \models \diam{a}{} \phi' \\
\end{array}}

\item If $\phi$ is $\diam{a^{\dag}}{} \phi'$ then by Lemma~\ref{lem:proved_trans_scs_trans} and the
induction hypothesis:
\cws{8}{\hspace*{-1.0cm}\begin{array}{rcl}
P \models \diam{a^{\dag}}{} \phi' & \textrm{iff} & \textrm{there is } P' \arrow{\theta}{} P \textrm{ s.t. }
\act(\theta) = a \textrm{ and } P' \models \phi' \\
& \textrm{iff} & \textrm{there is } X_{P'} \xarrow{\act(\theta)}{\emph{\textsf{C}}_{P'}} X_{P} \textrm{ s.t.
} \act(\theta) = a \textrm{ and } \lsp P' \rsp \models \phi' \\
& \textrm{iff} & \lsp P \rsp \models \diam{a^{\dag}}{} \phi' \\
\end{array}}
\qedhere

				\end{itemize}

			\end{itemize}
    
		\end{proof}

	\end{thm}

	\begin{cor}\label{cor:brml_char_hhpb}

Let $P_{1}, P_{2} \in \procs$. Then $\lsp P_{1} \rsp \sbis{\rm HHPB} \lsp P_{2} \rsp$ iff $\forall \phi \in
\call_{\rm BRM} \ldotp \lsp P_{1} \rsp \models \phi \Longleftrightarrow \lsp P_{2} \rsp \models \phi$
provided that all possible autoconflicts are local.

		\begin{proof}

From Theorems~\ref{thm:hhpb_frb_brm_proc}, \ref{thm:frb_brm_log_char}, and~\ref{thm:proc_scs_sat_brml} it
follows that:
\cws{8}{\begin{array}{rcl}
\lsp P_{1} \rsp \sbis{\rm HHPB} \lsp P_{2} \rsp & \textrm{iff} & P_{1} \sbis{\rm FRB:brm} P_{2} \\
& \textrm{iff} & \forall \phi \in \call_{\rm BRM} \ldotp P_{1} \models \phi \Longleftrightarrow P_{2}
\models \phi \\
& \textrm{iff} & \forall \phi \in \call_{\rm BRM} \ldotp \lsp P_{1} \rsp \models \phi \Longleftrightarrow
\lsp P_{2} \rsp \models \phi \\ 
\end{array}}
\qedhere

		\end{proof}

	\end{cor}

%
\subsection{Translating $\call_{\rm BRM}$ into $\call_{\rm EI}$}
\label{sec:brml_into_eil}
%

The main difficulty with this translation is the encoding of multisets, as they are not present in
$\call_{\rm EI}$. In the translation function we thus introduce two further parameters \blue{in addition to
a formula of $\call_{\rm BRM}$}:

	\begin{itemize}

\item The first one is a finite set $A$ of actions, e.g., those occurring in a process $P$. Since $P \models
M$ iff $\brm(P) = M$, the $\call_{\rm EI}$ formula corresponding to $M$ has to express the fact that every
action in the support of $M$, i.e., $\supp(M) = \{ a \in \cala \mid M(a) > 0 \}$, can be undone a number of
times equal to its multiplicity, while any action in $A \setminus \supp(M)$ cannot be undone at all. We
assume that $\supp(M)$ is finite to avoid infinite conjunctions in the translation \blue{of the $\call_{\rm
BRM}$ formula $M$}.

\item The second one is a finite sequence $\varrho_{n}: \{ 1, \dots, n \} \rightarrow \cali \times \cala$ of
pairs each formed by an identifier and an action. It acts like a stack-based memory that keeps track of
executed actions, bound to variables via $\eildiam{x : a}{}$ and $(x : a)$. \blue{This sequence, which is
empty in the case that the considered process $P$ is initial, is extended each time that an action
\linebreak is performed.}

	\end{itemize}

The translation function $\calt_{\rm BE} : \call_{\rm BRM} \times \powerset_{\rm fin}(\cala) \times \{
\varrho_{n} \mid n \in \natns_{\ge 1} \} \rightarrow \call_{\rm EI}$ is defined by induction on the
syntactical structure of $\phi \in \call_{\rm BRM}$ as follows:
\cws{0}{\begin{array}{rcll}
\calt_{\rm BE}(\true, A, \varrho_{n}) & \!\! = \!\! & \true & \\
\calt_{\rm BE}(M, A, \varrho_{n}) & \!\! = \!\! & \bigwedge\limits_{a_{i} \in \supp(M)}^{} \left(
\bigwedge\limits_{k = 1}^{M(a_{i})} \eilbdiam{x_{i, k}}{} \true \land \bigwedge\limits_{h =
1}^{\sharp(a_{i}, \varrho_{n}) - M(a_{i})} \lnot\eilbdiam{z_{i, h}}{} \true \right) & \\
& & \land \bigwedge\limits_{b \in A \setminus \supp(M)}^{} \lnot (y : b) \eilbdiam{y}{} \true &
\hspace{-1.0cm} \textrm{with } y \textrm{ fresh} \\
\calt_{\rm BE}(\lnot\phi', A, \varrho_{n}) & \!\! = \!\! & \lnot\calt_{\rm BE}(\phi', A, \varrho_{n}) & \\
\calt_{\rm BE}(\phi_{1} \land \phi_{2}, A, \varrho_{n}) & \!\! = \!\! & \calt_{\rm BE}(\phi_{1}, A,
\varrho_{n}) \land \calt_{\rm BE}(\phi_{2}, A, \varrho_{n}) & \\
\calt_{\rm BE}(\diam{a}{} \phi', A, \varrho_{n}) & \!\! = \!\! & \eildiam{x : a}{} \calt_{\rm BE}(\phi', A,
\varrho_{n} \cup \{(n + 1, (x, a))\}) & \hspace{-1.0cm} \textrm{with } x \textrm{ fresh} \\
\calt_{\rm BE}(\diam{a^{\dag}}{} \phi', A, \varrho_{n}) & \!\! = \!\! & (x : a) \eilbdiam{x}{} \calt_{\rm
BE}(\phi', A, \varrho_{n}) & \hspace{-1.0cm} \textrm{with } x \textrm{ fresh} \\
\end{array}}
where in the translation of $M$ the finite sequence $\varrho_{n}$ is such that:

	\begin{itemize}

\item $(x_{i, k}, a_{i}) \in \rge(\varrho_{n})$, with $x_{i, k} \neq x_{i, k'}$ for $k \neq k'$ and all the
identifiers $x_{i, k}$ being taken starting from the end of $\varrho_{n}$, i.e., the top of the stack.

\item $\sharp(a_{i}, \varrho_{n})$ is the number of occurrences of $a_{i}$ in $\varrho_{n}$.

\item $(z_{i, h}, a_{i}) \in \rge(\varrho_{n})$, with $z_{i, h} \notin \{ x_{i, k} \mid 1 \leq k \leq
M(a_{i}) \}$ and $z_{i, h} \neq z_{i, h'}$ for $h \neq h'$.

	\end{itemize}

	\begin{thm}\label{thm:translation_brml_eil}

Let $P \in \procs$, $\phi \in \call_{\rm BRM}$, and $\act(P)$ be the set of actions in $P$. Then
$P \models \phi$ iff $\exists \varrho_{n} \ldotp \exists \rho \in \perm(P, \calt_{\rm BE}(\phi, \act(P),
\varrho_{n})) \ldotp P \models_{\rho} \calt_{\rm BE}(\phi, \act(P), \varrho_{n})$.

		\begin{proof}

We proceed by induction on $k = \depth(\phi)$:

    			\begin{itemize}

\item If $k = 0$ then there are two cases:

				\begin{itemize}

\item If $\phi$ is $\true$ then $\calt_{\rm BE}(\true, \act(P), \varrho_{n}) = \true$ and $P$ satisfies both
formulas (the second one for all $\varrho_{n}$ and $\rho$).

\item If $\phi$ is $M$ then we divide the proof into two parts. Starting from $P \models M$, we derive that
$\brm(P) = M$, hence for all $a_{i} \in \supp(M)$ there exists $P'_{a_{i, k}} \xarrow{\theta_{a_{i, k}}}{}
P$ for $1 \leq k \leq M(a_i)$ such that $\act(\theta_{a_{i, k}}) = a_{i}$, with $P$ having no other incoming
transitions. If we consider any sequence of proved transitions $P_{1} \arrow{\theta_{1}}{} P_{2}
\arrow{\theta_{2}}{} \dots \arrow{\theta_{m}}{} P_{m + 1}$ such that $P_{1}$ is $\toinitial(P)$ and $P_{m +
1}$ is $P$, all the proof terms $\theta_{a_{i, k}}$ appear in the sequence $\theta_{1}, \dots, \theta_{m}$
because the transitions $P'_{a_{i, k}} \xarrow{\theta_{a_{i, k}}}{} P$ are all independent from each other.
\\
In the construction of $\varrho_{n}$, for all $a_{i} \in \supp(M)$ we have to consider all the
$a_{i}$-transitions along any path from $\toinitial(P)$ to $P$. Therefore, we take as $n$ the number of
proof terms $\theta_{j}$ in the sequence $\theta_{1}, \dots, \theta_{m}$ such that $\act(\theta_{j}) =
a_{i}$ for all $a_{i} \in \supp(M)$. Since $\brm(P) = M$, the total number of incoming transitions of $P$ is
$\sum\limits_{a_{i} \in \supp(M)} M(a_{i})$, while the number of proof terms $\theta_{j}$ in the sequence
$\theta_{1}, \dots, \theta_{m}$ such that $\act(\theta_{j}) = a_{i}$ and $\theta_{j} \not\in \{
\theta_{a_{i, k}} \mid 1 \leq k \le M(a_{i}) \}$ is $n - \sum\limits_{a_{i} \in \supp(M)} M(a_{i})$. We can
map all numbers between $1$ and $n - \sum\limits_{a_{i} \in \supp(M)} M(a_{i})$ to the pairs $(z_{i, h},
a_{i})$ and all numbers between $n - \sum\limits_{a_{i} \in \supp(M)} M(a_{i}) + 1$ and $n$ to the pairs
$(x_{i, k}, a_{i})$. \\
At this point, we can construct $\rho$ by mapping every $x_{i, k}$ to $\theta_{a_{i, k}}$ and every $z_{i,
h}$ to $\theta_{a_{i, h}}$ such that there is no transition $P'_{a_{i, h}} \xarrow{\theta_{a_{i, h}}}{} P$.
\\
Then $P \models_{\rho} \calt_{\rm BE}(M, \act(P), \varrho_{n}) = \bigwedge \limits_{a_{i} \in \supp(M)}
(\bigwedge\limits_{k = 1}^{M(a_{i})} \eilbdiam{x_{i, k}}{} \true \land \bigwedge\limits_{h =
1}^{\sharp(a_{i}, \varrho_{n}) - M(a_{i})} \lnot \eilbdiam{z_{i, h}}{} \true) \land \bigwedge\limits_{b \in
\act(P) \setminus \supp(M)} \lnot (y : b) \eilbdiam{y}{} \true$ as we now show by considering each of the
three main conjunctions separately:

					\begin{itemize}

\item $P \models_{\rho} \bigwedge\limits_{k = 1}^{M(a_{i})} \eilbdiam{x_{i, k}}{} \true$ because, from the
way we have constructed~$\rho$, we know that it maps every $x_{i, k}$ exactly to the proof term $\theta_{i,
k}$ such that there exists a transition $P'_{i, k} \xarrow{\theta_{a_{i, k}}}{} P$ with $\act(\theta_{a_{i,
k}}) = a_{i}$. Moreover, every $P'_{i, k}$ trivially satisfies $\true$.

\item $P \models_{\rho} \bigwedge\limits_{h = 1}^{\sharp(a_{i}, \varrho_m) - M(a_{i})} \lnot \eilbdiam{z_{i,
h}}{} \true$ because, as in the previous case, we have constructed $\rho$ in such a way that every $z_{i,
h}$ is mapped to the proof term $\theta_{a_{i, h}}$ such that $\act(\theta_{a_{i, h}}) = a_{i}$ and there is
no transition $P'_{a_{i, h}} \xarrow{\theta_{a_{i, h}}}{} P$. Hence, $P \models_{\rho} \lnot \eilbdiam{z_{i,
h}}{} \true$ for any $z_{i, h}$.

\item $P \models_{\rho} \bigwedge\limits_{b \in \act(P) \setminus \supp(M)} \lnot (y : b) \eilbdiam{y}{}
\true$ because if $b \notin \supp(M)$ then: 

						\begin{itemize}

\item either there is no $b^{\dag}. \, P' \in \sbpr(P)$ and hence $P \not\models_{\rho} (y : b)
\eilbdiam{y}{} \true$, i.e., $P \models_{\rho} \lnot (y : b) \linebreak \eilbdiam{y}{} \true$;

\item or, if such a process exists, $P$ does not have any incoming transition $P' \arrow{\theta}{} P$ such
that $\act(\theta) = b$ and hence $P \not\models_{\rho [y \mapsto \theta]} \eilbdiam{y}{} \true$, from which
it follows that $P \not\models_{\rho} (y : b) \eilbdiam{y}{} \true$, i.e., $P \models_{\rho} \lnot (y : b)
\eilbdiam{y}{} \true$.

						\end{itemize}

					\end{itemize}
				
The converse is straightforward because, if there exist $\varrho_{n}$ and $\rho \in \perm(\calt_{\rm BE}(M,
\act(P), \varrho_{n}))$ such that $P \models_{\rho} \calt_{\rm BE}(M, \act(P), \varrho_{n})$, then we can
note that the conjunctions in $\calt_{\rm BE}(M, \linebreak \act(P), \varrho_{n})$ express the fact that $P$
has exactly $\sum\limits_{a_{i} \in \supp(M)} M(a_{i})$ incoming transitions such that every $a_{i} \in
\supp(M)$ appears $M(a_{i})$ times, while there are no incoming transitions of $P$ labeled with actions not
in $\supp(M)$. Thus, we can derive that $\brm(P) = M$, i.e., $P \models M$.

				\end{itemize}

\item If $k \ge 1$ then there are four cases:

				\begin{itemize}

\item If $\phi$ is $\lnot\phi'$ then by the induction hypothesis:
\cws{4}{\hspace*{-1.0cm}\begin{array}{rcl}
P \models \lnot\phi' & \textrm{iff} & P \not\models \phi' \\
& \textrm{iff} & \exists \varrho_n \ldotp \exists \rho \ldotp P \not\models_\rho 
\calt_{\rm BE} (\phi', \act(P), \varrho_n) \\
& \textrm{iff} & \exists \varrho_n \ldotp \exists \rho \ldotp P \models_\rho 
\lnot \calt_{\rm BE} (\phi', \act(P), \varrho_n) \\
& \textrm{iff} & \exists \varrho_n \ldotp \exists \rho \ldotp P \models_\rho 
\calt_{\rm BE} (\lnot \phi', \act(P), \varrho_n) \\
\end{array}}

\item If $\phi$ is $\phi_{1} \land \phi_{2}$ then we divide the proof into two parts. Starting from $P
\models \phi_{1} \land \phi_{2}$, \linebreak by the induction hypothesis and
Lemma~\ref{lem:free_var_subst_env_mod}:
\cws{0}{\hspace*{-1.0cm}\begin{array}{rcl}
P \models \phi_{1} \land \phi_{2} & \textrm{implies} & P \models \phi_{1} \textrm{ and } P \models
\phi_{2} \\
& \textrm{implies} & \exists \varrho_{n_{1}} \ldotp \exists \rho_{1} \ldotp P \models_{\rho_{1}} \calt_{\rm
BE}(\phi_{1}, \act(P), \varrho_{n_{1}}) \textrm{ and } \\
& & \exists \varrho_{n_{2}} \ldotp \exists \rho_{2} \ldotp P \models_{\rho_{2}} \calt_{\rm BE}(\phi_{2},
\act(P), \varrho_{n_{2}}) \\
& \textrm{implies} & \exists \varrho_{n_{1}} \ldotp \exists \varrho_{n_{2}} \ldotp \exists \rho^{\sigma}
\ldotp P \models_{\rho^{\sigma}} \sigma(\calt_{\rm BE}(\phi_{1}, \act(P), \varrho_{n_{1}})) \textrm{ and }
\\
& & \hspace*{2.64cm} P \models_{\rho^{\sigma}} \sigma(\calt_{\rm BE}(\phi_{2}, \act(P), \varrho_{n_{2}})) \\
& \textrm{implies} & \exists \varrho_{n_{1}} \ldotp \exists \varrho_{n_{2}} \ldotp \exists \rho^{\sigma}
\ldotp P \models_{\rho^{\sigma}} \sigma(\calt_{\rm BE}(\phi_{1}, \act(P), \varrho_{n_{1}})) \, \land \\
& & \hspace*{3.84cm} \sigma(\calt_{\rm BE}(\phi_{2}, \act(P), \varrho_{n_{2}})) \\
& \textrm{implies} & \exists \varrho_{n} \ldotp \exists \rho^{\sigma} \ldotp P \models_{\rho^{\sigma}}
\sigma(\calt_{\rm BE}(\phi_{1}, \act(P), \varrho_{n})) \, \land \\
& & \hspace*{2.78cm} \sigma(\calt_{\rm BE}(\phi_{2}, \act(P), \varrho_{n})) \\
& \textrm{implies} & \exists \varrho_{n}^{\sigma} \ldotp \exists \rho^{\sigma} \ldotp P
\models_{\rho^{\sigma}} \calt_{\rm BE}(\phi_{1}, \act(P), \varrho_{n}^{\sigma}) \, \land \\
& & \hspace*{2.78cm} \calt_{\rm BE}(\phi_{2}, \act(P), \varrho_{n}^{\sigma}) \\
& \textrm{implies} & \exists \varrho_{n}^{\sigma} \ldotp \exists \rho^{\sigma} \ldotp P
\models_{\rho^{\sigma}} \calt_{\rm BE}(\phi_{1} \land \phi_{2}, \act(P), \varrho_{n}^{\sigma}) \\
\end{array}}
where:

					\begin{itemize}

\item $\rho^{\sigma}$ maps a set of fresh identifiers $z_{\theta}$, one for each $\theta$ appearing in any
path from $\toinitial(P)$ to $P$, to the corresponding proof term, i.e., $\rho^{\sigma}(z_{\theta}) =
\theta$.

\item $\sigma$ is defined as $\sigma(x) = z_{\rho_{1}(x)}$ for all $x \in \fid (\calt_{\rm BE}(\phi_{1},
\act(P), \varrho_{n_{1}}))$ and $\sigma(x) = z_{\rho_{2}(x)}$ for all $x \in \fid(\calt_{\rm BE}(\phi_{2},
\act(P), \varrho_{n_{2}}))$.

\item Since $P \models \phi_{1}$ and $P \models \phi_{2}$, both $\varrho_{n_{1}}$ and $\varrho_{n_{2}}$ have
to be built consistently with any path from $\toinitial(P)$ to $P$, hence they can be replaced by a single
$\varrho_{n}$ that is built in the same consistent way.   

\item $\varrho_{n}^{\sigma}$ is defined as $\varrho_{n}^{\sigma}(m) = (\sigma(x), a)$ for all $1 \leq m \leq
n$ such that $\varrho_{n}(m) = (x, a)$ and $x \in \fid(\calt_{\rm BE}(\phi, \act(P), \varrho_{n}))$. Since
all free identifiers are captured by $\varrho_{n}$, it holds that $\sigma(\calt_{\rm BE}(\phi, \act(P),
\varrho_{n})) = \calt_{\rm BE}(\phi, \act(P), \varrho_{n}^{\sigma})$.

					\end{itemize}

\noindent
As for the converse:
\cws{4}{\hspace*{-0.5cm}\begin{array}{rcl}
\exists \varrho_{n} \ldotp \exists \rho \ldotp P \models_{\rho} \calt_{\rm BE}(\phi_{1} \land \phi_{2},
\act(P), \varrho_{n}) & \textrm{implies} & \exists \varrho_{n} \ldotp \exists \rho \ldotp P \models_{\rho}
\calt_{\rm BE}(\phi_{1}, \act(P), \varrho_{n}) \, \land \\
& & \hspace*{2.4cm} \calt_{\rm BE}(\phi_{2}, \act(P), \varrho_{n}) \\
& \textrm{implies} & \exists \varrho_{n} \ldotp \exists \rho \ldotp P \models_{\rho} \calt_{\rm
BE}(\phi_{1}, \act(P), \varrho_{n}) \textrm{ and } \\
& & \hspace*{1.37cm} P \models_{\rho} \calt_{\rm BE}(\phi_{2}, \act(P), \varrho_{n}) \\
& \textrm{implies} & P \models \phi_{1} \textrm{ and } P \models \phi_{2} \\
& \textrm{implies} & P \models \phi_{1} \land \phi_{2} \\
\end{array}}

\item If $\phi$ is $\diam{a}{} \phi'$ then by the induction hypothesis:
\cws{4}{\hspace*{-0.8cm}\begin{array}{rcl}
P \models \diam{a}{} \phi' & \textrm{iff} & \textrm{there is } P \arrow{\theta}{} P' \textrm{ s.t. }
\act(\theta) = a \textrm{ and } P' \models \phi' \\
& \textrm{iff} & \textrm{there is } P \arrow{\theta}{} P' \textrm{ s.t. } \act(\theta) = a \textrm{ and } \\ 
& & \hspace*{2.0cm} \exists \varrho_{n} \ldotp \exists \rho \ldotp P' \models_{\rho [x \mapsto \theta]}
\calt_{\rm BE}(\phi', \act(P'), \varrho_{n} \cup \{ (n + 1, (x, a)) \}) \\
& \textrm{iff} & \exists \varrho_{n} \ldotp \exists \rho \ldotp P \models_{\rho} \eildiam{x : a}{}
\calt_{\rm BE}(\phi', \act(P), \varrho_{n} \cup \{ (n + 1, (x, a))\}) \\
& \textrm{iff} & \exists \varrho_{n} \ldotp \exists \rho \ldotp P \models_{\rho} \calt_{\rm BE}(\diam{a}{}
\phi', \act(P), \varrho_n) \\
\end{array}}

\item If $\phi$ is $\diam{a^{\dag}}{} \phi'$ then by the induction hypothesis:
\cws{8}{\hspace*{-0.8cm}\begin{array}{rcl}
P \models \diam{a^{\dag}}{} \phi' & \textrm{iff} & \textrm{there is } P' \arrow{\theta}{} P \textrm{ s.t. }
\act(\theta) = a \textrm{ and } P' \models \phi' \\
& \textrm{iff} & \textrm{there is } P' \arrow{\theta}{} P \textrm{ s.t. } \act(\theta) = a \textrm{ and } \\
& & \hspace*{3.7cm} \exists \varrho_{n} \ldotp \exists \rho \ldotp P' \models_{\rho [x \mapsto \theta]}
\calt_{\rm BE}(\phi', \act(P'), \varrho_{n}) \\
& \textrm{iff} & \textrm{there is } P' \arrow{\theta}{} P \textrm{ s.t. } \act(\theta) = a \textrm{ and } \\
& & \hspace*{3.7cm} \exists \varrho_{n} \ldotp \exists \rho \ldotp P \models_{\rho [x \mapsto \theta]}
\eilbdiam{x}{} \calt_{\rm BE}(\phi', \act(P), \varrho_{n}) \\
& \textrm{iff} & \textrm{there is } a^{\dag} . \, P'' \in \sbpr(P) \textrm{ s.t. } \apt(a^{\dag} . \, P'',
P) = \theta \textrm{ and } \\
& & \hspace*{4.0cm} P \models_{\rho [x \mapsto \apt(a^{\dag} . \, P'', P)]} \eilbdiam{x}{} \calt_{\rm
BE}(\phi', \act(P), \varrho_n) \\ 
& \textrm{iff} & \exists \varrho_{n} \ldotp \exists \rho \ldotp P \models_{\rho} (x : a) \eilbdiam{x}{}
\calt_{\rm BE}(\phi', \act(P), \varrho_{n}) \\
& \textrm{iff} & \exists \varrho_{n} \ldotp \exists \rho \ldotp P \models_{\rho} \calt_{\rm
BE}(\diam{a^{\dag}}{} \phi', \act(P), \varrho_{n}) \\
\end{array}}
\qedhere

				\end{itemize}

			\end{itemize}

		\end{proof}

	\end{thm}

%
\subsection{Translating $\call_{\rm EI}$ into $\call_{\rm BRM}$}
\label{sec:eil_into_brml}
%

The challenge of this translation is the encoding of formulas of the form $(x : a) \phi$, because
identifiers are not present in $\call_{\rm BRM}$ and the satisfaction of these formulas is not necessarily
related to actions to be done or undone in this moment. Rather, it is generically related to executed
actions. On the other hand, it is not clear how multisets would come into play. The study of this
translation is left for future work.

%
%
\section{Conclusions}
\label{sec:concl}
%
%

In this paper we have proposed an entirely new approach to characterize HHPB, both denotationally and
operationally, even in the presence of equidepth autoconcurrency. Unlike~\cite{AC20}, the focus is on
counting identically labeled events rather than uniquely identifying them, thus avoiding bijections between
events altogether thanks to a variant of forward-reverse bisimilarity that is extended with backward ready
multisets equality. \blue{Backward ready multisets naturally provide information about the concurrency of
events, which leads to easily distinguishing autoconcurrency from autocausation.} Moreover, on the
operational side, \linebreak it has turned out that proof terms naturally lend themselves to identification
purposes; \linebreak in a reversible setting like ours, they have been used for the first time
in~\cite{Aub22}. Finally, \linebreak we have logically characterized our backward-ready-multiset
forward-reverse bisimilarity with backward ready multiset logic and investigated the relationships of the
latter with event identifier logic~\cite{PU14}. Our results hold in the absence of non-local autoconflicts.

The operational characterization is particularly important for several reasons. Firstly, in addition to the
equational characterization over forward-only processes developed in~\cite{FL05}, HHPB can be axiomatized
over reversible processes by resorting to the approach of~\cite{DP92} as applied in~\cite{BEM24}, provided
that backward ready multisets are considered in place of backward ready sets. Secondly, in addition to the
logics of~\cite{PU14,BC14}, HHPB can be characterized also in terms of our backward ready multiset logic.
The latter is simpler as it does not make use of variables and binders, but the former contain fragments
that have been proven to characterize various behavioral equivalences in the true concurrency
spectrum~\cite{GG01,Fec04,PU12}.

As for future work, we would like to complete the investigation of the relationships between backward ready
multiset logic and event identifier logic, as well as to extend it to the logic of~\cite{BC14}. Another
direction to pursue is whether our results apply to configuration structures that are not stable, i.e., in
which it is not necessarily the case that causality among events can be always represented in terms of
partial orders, possibly defined locally to each configuration. HHPB has been defined over non-stable models
in~\cite{Gla06,BGPS22} and a logical characterization for it has been provided in~\cite{BGPS22}, which is a
conservative extension of the one in~\cite{BC14}. Finally,
we plan to understand what needs to be added to backward-ready-multiset forward-reverse bisimilarity so as
to characterize HHPB exactly, i.e., also in the presence of non-local autoconflicts. In case of success, we
will then study the verification of the resulting bisimilarity by taking into account, as far as HHPB is
concerned, the undecidability result over finite labeled transition systems extended with an independence
relation between transitions of~\cite{JNS03} and the polynomial-time algorithm over basic parallel processes
of~\cite{FJLS10}.

\medskip
\noindent
\textbf{Acknowledgments.}
We are grateful to Rob van Glabbeek for bringing to our attention the process examined in
Example~\ref{ex:autoconc_autocaus}, as well as the anonymous reviewers for their comments and suggestions.
This research has been supported by the PRIN 2020 project \emph{NiRvAna -- Noninterference and Reversibility
Analysis in Private Blockchains}, the PRIN 2022 project \emph{DeKLA -- Developing Kleene Logics and Their
Applications}, and the INdAM-GNCS 2024 project \emph{MARVEL -- Modelli Composizionali per l'Analisi di
Sistemi Reversibili Distribuiti}.

\bibliographystyle{alphaurl}
\bibliography{biblio}

\end{document}